\documentclass[aps,prc,floatfix,nofootinbib,preprintnumbers,superscriptaddress,longbibliography,notitlepage,twocolumn]{revtex4-1}
\usepackage[dvipsnames,svgnames,x11names]{xcolor}
\usepackage{epsfig}
\usepackage{graphicx}
\usepackage{stmaryrd}
\usepackage{amssymb,tabularx,dcolumn}
\usepackage{supertabular,ltxtable}
\usepackage{longtable}
\usepackage{amsmath}
\usepackage{amstext}
\usepackage{float}
\usepackage{color}
\usepackage{bm}
\usepackage{hyperref}
\hypersetup{breaklinks=true,colorlinks=true,linkcolor=blue,citecolor=blue,filecolor=magenta,urlcolor=cyan}
\usepackage{mathtools}
\usepackage{multirow}
\usepackage{makecell}
\usepackage[utf8]{inputenc}
\usepackage{tabu}
\usepackage{braket}
\usepackage{subfigure} 
\usepackage{enumitem}
\usepackage[english]{babel}

\definecolor{pastelgray}{rgb}{0.81, 0.81, 0.77}
\definecolor{beaublue}{rgb}{0.9, 0.9, 0.93}

\begin{document}

\title{A Statistical Analysis for the Neutrinoless Double-Beta Decay Matrix element of \texorpdfstring{$^{48}$}{}Ca}

\author{M. Horoi}
\affiliation{Department of Physics, Central Michigan University, Mount Pleasant, MI 48859, USA}

\author{A. Neacsu}
\affiliation{International Center for Advanced Training and Research in Physics (CIFRA), Magurele, Romania}

\author{S. Stoica}
\affiliation{International Center for Advanced Training and Research in Physics (CIFRA), Magurele, Romania}

\date{\today}

\begin{abstract}
Neutrinoless double beta decay ($0\nu\beta\beta$) nuclear matrix elements (NME) are the object of many theoretical calculation methods, and are very important for analysis and guidance of a large number of experimental efforts. However, there are large discrepancies between the NME values provided by different methods. In this paper we propose a statistical analysis of the  $^{48}$Ca $0\nu\beta\beta$ NME using the interacting shell model, emphasizing the range of the NME probable values and its correlations with observables that can be obtained from the existing nuclear data. Based on this statistical analysis with three independent effective Hamiltonians we propose a common probability distribution function for the $0\nu\beta\beta$ NME, which has a range of (0.45 - 0.95) at 90\% confidence level of, and a mean value of 0.68.
\end{abstract}
\maketitle

\section{Introduction} \label{intro}

The study of the double-beta decay (DBD) is currently a hot research topic since it is viewed as one of the most promising approaches to clarify important, as yet unknown, properties of neutrinos and to explore physics beyond the Standard Model (SM) \cite{Avignone2008,Vergados2012}. Two scenarios are possible for this process to occur:  i) two-neutrino double-beta ($2\nu\beta\beta$) transitions (with emission of two electrons/positrons and two anti-neutrinos/neutrinos), which conserve the lepton number and are allowed by the SM and ii) double-beta decay transitions without emission of neutrinos ($0\nu\beta\beta$), which violate the lepton number conservation and are only allowed by theories beyond SM (BSM).

Neutrinoless DBD was not experimentally detected so far, but its measurement would provide us with important information about lepton number violating (LNV) processes; neutrino properties (neutrino absolute mass scale and mass hierarchy, neutrino nature as Dirac or Majorana fermion, number of neutrino flavours); CP and Lorentz simmetries violation; constraining of different BSM mechanisms that may contribute to this decay mode, etc. The most common mechanism investigated is the light left-handed (LH) Majorana neutrinos exchange between two nucleons, but once a LNV operator is introduced in the Lagrangian, several other mechanisms are also allowed, such as: the exchange of light and  heavy neutrinos in left-right symmetric models, the exchange of supersymmetric particles, DBD with the emission of Majorons, etc. 

The DBD half-life equations can be expressed, in a good approximation, as a product of some factors. The $2\nu\beta\beta$ half-life is a product of a phase space factor (PSF), which depends on the atomic charge and energy released in the decay, and a nuclear matrix element (NME) related to the nuclear structure of the parent and daughter nuclei. The $0\nu\beta\beta$ half-life, besides the PSF and NME factors, also contains a LNV factor, related to the particular BSM mechanism that may contribute to the decay. If several mechanisms are considered, the inverse half-life can be written as a sum of all the individual contributions and their interference terms \cite{Doi1985,Vergados2012,Rodejohann2012,Deppisch2012,HoroiNeacsu2016prd,Neacsu2016ahep-dist,Ahmed2017}. Using the experimental limits of the $0\nu\beta\beta$ decay half-lives and the theoretical values of PSF and NME, one can constrain the LNV parameters and the associated BSM scenarios, usually under the assumption that only one mechanism contributes at one time. 

 There is currently significant progress in the DBD experiments (in terms of the amount of source material, decreasing background and improvement in the detection techniques), leading to the expectation that the next generation of experiments will be able to cover the entire region of the neutrino inverted mass hierarchy \cite{EngelMenendez2017}. Concurrently, the progress of the theoretical methods now provides us with accurate PSF values for all the double-beta decay modes and transitions. \cite{Kotila2012, StoicaMirea2013, MireaPahomi2015}. 
 Thus, at present, the uncertainty in the DBD calculations remains mostly the NME evaluation. 
 
 There are several nuclear structure methods for the NME calculation, the most used being: Shell model methods \cite{Caurier1990,Caurier1996, Caurier2005, HoroiStoicaBrown2007, HoroiStoica2010, Horoi2013, HoroiBrown2013, SenkovHoroi2014, NeacsuHoroi2015, NeacsuHoroi2016,18ho035502}, pnQRPA methods \cite{SuhonenCivitarese1998, Simkovic1999, Stoica2001, Rodin2006, KortelainenSuhonnen2007, Faessler2012, SimkovicRodin2013}, IBA methods \cite{Barea2009, Barea2013}, Energy Density Functional method \cite{Rodriguez2010}, PHFB \cite{Rath2013}, Coupled-Cluster method (CC)~\cite{Novario2021}, in-medium generator coordinate method (IM-GCM) \cite{Yao2020}, and valence-space in-medium similarity renormalization group method (VS-IMSRG)~\cite{Belley2021}. Each of these methods have their strengths and weakness, largely discussed over time in the literature, and the current situation is that there are still significant differences between NME values calculated with different methods, and sometimes, even between NME values calculated with the same methods (see for example the review \cite{EngelMenendez2017}).   
 For the  $2\nu\beta\beta$ decay the NME are products of two Gamow-Teller (GT) transition amplitudes, and most of the nuclear methods overestimate them, in comparison with experiment. This drawback is often treated by introducing a quenching factor that multiplies the GT operator and reduces its strength. This is equivalent to using a quenched axial vector constant, instead of its bare value $g_A=1.27$. 
 
 For the $0\nu\beta\beta$ decay the NME calculation is more complicated, since besides the GT transitions, other transitions may contribute as well. Also, the NME values calculated by different methods may differ by factors of 3-4 for most relevant isotopes, and up to 7-8 in the case if $^{48}$Ca (see e.g. Fig. 5 of Ref.~\cite{EngelMenendez2017}, and Refs. \cite{Yao2020,Novario2021}). Uncertainties in the NME values are further amplified when predicting half-lives, since they enter at the power of two in the lifetime formulas.  In addition, there is no measured lifetime for this decay mode to compared with, and these uncertainties in the NME computation reflect in the interpretation of the DBD data and planning the performances of the DBD experiments. 
 
 The shell model-based methods have some advantages such as the inclusion of all correlations between nucleons around the Fermi surface, preserving all symmetries of the nuclear many-body problem, and the use of widely tested nucleon-nucleon (NN) interactions. For different mass regions of nuclei, one uses several different effective NN effective Hamiltonians that are appropriate for the corresponding model spaces. 
 These effective Hamiltonians are usually obtained by starting with a theoretical Bruekner G-Matrix Hamiltonian that is further fine-tuned to describe the experimental energy levels for a large number of nuclei that can be investigated in the corresponding model spaces.
These effective Hamiltonians are described by a small number of single particle energies and a finite number of two-body matrix elements. As a by-product, the wave functions produced by these Hamiltonian can be used to describe and predict observables, such as the  electromagnetic transition probabilities, Gamow-Teller transitions probabilities, nucleon occupation probabilities, spectroscopic factors, etc, using relative simple changes of the transition operators in terms of effective charges and quenching factors. These effective charges and quenching factors are calibrating to the existing data. For $0\nu\beta\beta$ NME such calibrations are not yet possible due to the lack of data. However, different existing effective Hamiltonians for nuclei envolved in a given $0\nu\beta\beta$ decay produce smaller ranges of the NME.  In addition, some recent ab-initio methods, such as IMSRG \cite{Yao2020,Belley2021}, build on the modern advances in the shell model by providing ab-initio derived effective Hamiltonians and effective transition operators, and they can provide some guidance for calibrating the shell model $0\nu\beta\beta$ NME. 

 It would be thus interesting to study the robustness of the  $0\nu\beta\beta$ NME to  small changes of the parameters of different effective shell model Hamiltonians and to examine how the NME changes are correlated with other observables. 
In this work, we propose a statistical analysis of $0\nu\beta\beta$ NME of $^{48}$Ca calculated with the interacting shell-model using three independent effective Hamiltonians (FPD6, GXPF1A, KB3G), emphasizing the range of the NME probable values and their correlations with several observables that can be compared to existing nuclear data. Based on this statistical analysis we propose a common probability distribution function for the $0\nu\beta\beta$ NME. We apply our analysis to $^{48}$Ca, which is the lightest DBD isotope and thus more accessible to ab-initio calculations. We only consider in this work the standard light LH neutrino exchange mass mechanism, which is most likely to contribute to the $0\nu\beta\beta$ decay process. 

The paper is organized as follows. In section II the calculation methods of the observables and the statistical model are presented.
Then, in section III we present the results and discussions on their relevance, and in section IV we end with conclusions and outlook. Finally we included an Appendix with a short presentation of the  Gram-Charlier A series that we used in our statistical model.

\section{The Statistical Model} \label{model}

\begin{center}
\begin{table}[!ht]
    \centering
    \begin{tabular}{crrrrr} \hline
~ & Exp. & Error & FPD6 & GXPF1A & KB3G \\ \hline
$0\nu\beta\beta$ NME & N/A & N/A & 0.79 & 0.559 & 0.693 \\  
$2\nu\beta\beta$ NME & 0.035 \cite{Barabash2020} & 0.003 & 0.062 & 0.050 & 0.045 \\  
$^{48}$Ca B(E2)$\uparrow$ & 0.008\cite{Pritychenko2016} & 0.001 & 0.007 & 0.006 & 0.05 \\  
$^{48}$Ca 2+ & 3.832 \cite{NNDC_A48-2022} & 0.15 & 3.658 & 3.735 & 4.238 \\  
$^{48}$Ca 4+ & 4.503 \cite{NNDC_A48-2022} & 0.15 & 4.134 & 4.264 & 4.231 \\  
$^{48}$Ca 6+ & 7.953 \cite{NNDC_A48-2022} & 0.15 & 7.396 & 7.705 & 7.831 \\  
$^{48}$Ca Occ(Nf5) & 0.032$^*$ & 0.395$^\#$ & 0.117 & 0.032 & 0.112 \\  
$^{48}$Ca Occ(Nf7) & 7.892$^*$ & 0.395$^\#$ & 7.693 & 7.892 & 7.795 \\  
$^{48}$Ca Occ(Np1) & 0.009$^*$ & 0.395$^\#$ & 0.029 & 0.009 & 0.024 \\  
$^{48}$Ca Occ(Np3) & 0.067$^*$ & 0.395$^\#$ & 0.161 & 0.067 & 0.070 \\  
$^{48}$Ca $\rightarrow$ $^{48}$Sc GT & 1.09 \cite{GreweGT2007}& 0.28 & 1.01 & 1.226 & 0.051 \\  
$^{48}$Ti B(E2)$\uparrow$ & 0.063\cite{Pritychenko2016} & 0.003 & 0.064 & 0.052 & 0.052 \\  
$^{48}$Ti 2+ & 0.984 \cite{NNDC_A48-2022} & 0.150 & 1.118 & 1.010 & 0.985 \\  
$^{48}$Ti 4+ & 2.296 \cite{NNDC_A48-2022} & 0.150 & 2.492 & 2.168 & 2.214 \\  
$^{48}$Ti 6+ & 3.333 \cite{NNDC_A48-2022} & 0.150 & 3.425 & 2.922 & 3.046 \\  
$^{48}$Ti Occ(Nf5) & 0.168$^*$ & 0.277$^\#$ & 0.310 & 0.168 & 0.263 \\  
$^{48}$Ti Occ(Nf7) & 5.535$^*$ & 0.277$^\#$ & 5.253 & 5.535 & 5.416 \\  
$^{48}$Ti Occ(Np1) & 0.048$^*$ & 0.277$^\#$ & 0.068 & 0.048 & 0.061 \\  
$^{48}$Ti Occ(Np3) & 0.248$^*$ & 0.277$^\#$ & 0.369 & 0.248 & 0.260 \\  
$^{48}$Ti Occ(Pf5) & 0.032$^*$ & 0.092$^\#$ & 0.101 & 0.032 & 0.097 \\  
$^{48}$Ti Occ(Pf7) & 1.839$^*$ & 0.092$^\#$ & 1.672 & 1.839 & 1.763 \\  
$^{48}$Ti Occ(Pp1) & 0.010$^*$ & 0.092$^\#$ & 0.031 & 0.010 & 0.021 \\  
$^{48}$Ti Occ(Pp3) & 0.119$^*$ & 0.092$^\#$ & 0.196 & 0.119 & 0.120 \\  
$^{48}$Ti $\rightarrow$ $^{48}$Sc GT & 0.014\cite{GreweGT2007} & 0.005 & 0.050 & 0.032 & 0.056 \\  \hline
    \end{tabular}
    \label{experimental-data}
    \caption{Experimental data, experimental errors, and the calculated values using 3 effective Hamiltonians for the observables analyzed. The data for occupation probabilities ($^*$) is not available and it was replaced with the GXPF1A results and errors ($^\#$) of 5\% of the highest nucleon species occupation.}
\end{table}
\end{center}

We plan to investigate the effect of small, random variation of the shell model effective Hamiltonian on the neutrinoless double beta decay NME of $^{48}$Ca, and the NME correlations with other calculated observables, such as $2^+$ energies, B(E2)$\uparrow$ values, $2\nu\beta\beta$ matrix elements, Gamow-Teller transition probabilities, neutron and proton occupation probabilities, etc. 

To achieve that goal we selected a number of often used effective Hamiltonians describing nuclei around $^{48}$Ca in the $fp$-shell ($0f_{7/2}$, $0f_{5/2}$, $1p_{3/2}$, and $1p_{1/2}$ orbitals for both protons and neutrons), and added small random contributions to their two-body matrix elements (TBME). For this project we only considered the FPD6 Hamiltonian \cite{fpd6}, the KB3G Hamiltonian \cite{Caurier2005}, and GXPF1A Hamiltonian \cite{Honma2004,Honma2005} as starting effective Hamiltonians.  In order to maintain the magicity of $^{48}$Ca we decided to keep the single particle (s.p.) energies in the perturbed effective Hamiltonians the same as in the starting Hamiltonians.

One important decision to be made about the random contributions to the starting Hamiltonians is the choice of their maximum amplitude (range). In this work we were guided by the analysis of the USDA/USDB effective Hamiltonians \cite{BrownRichter2006} where one starts with an underlying G-matrix and modifies linear combinations of two-body matrix elements in a fine-tuning procedure \cite{BrownRichter2006} until the root mean square (RMS) deviation of the calculated energies vs the experimental ones shows some signs of convergence. In this fine-tuning process one would not want to change the TBME too much from the original G-matrix values, because the over-fitted TBME could result in unitary changes of the s.p. wave functions that may produce slightly better energies, but incorrect observables. For USDA, for example, the RMS deviation of the TBME was about 300 keV, while a small improvement in the overall energies given by USDB resulted in an additional change of 100 keV if the RMS deviation of the TBME. This analysis suggests that an additional RMS of about 100 keV would not dramatically change the quality of the TBME in the sd-shell and we extended this choice to the fp-shell. An analysis of the TBME for all three starting Hamiltonians listed above indicates that a $\pm 10\%$ range for the random contributions would suffice. 

In the analysis we included as observables the $0\nu\beta\beta$ NME, the $2\nu\beta\beta$ NME, the Gamow-Teller probability to reach the first $1^+$ state in $^{48}$Sc from the ground states (g.s.) of the parent, $^{48}$Ca, and of the daughter, $^{48}$Ti, the energies of the $2^+$, $4^+$ and $6^+$ states of the parent and daughter, the B(E2)$\uparrow$ transition probabilities to the first $2^+$ state of parent and daughter, the neutron occupation probabilities of the $pf$ states of the parent, and neutron and proton occupation probabilities of the $pf$ states of the daughter nucleus. The experimental occupation probabilities for the nuclei relevant for the $^{48}$Ca $0\nu\beta\beta$ decay are not available, but we include synthetic (calculated) values in the analysis because the corresponding occupation probabilities are available for other nuclei of interest for $0\nu\beta\beta$ decay~\cite{Schiffer2008,Kay2009,Kay2013}, and it might be interesting to see if they have any correlations with the $0\nu\beta\beta$ NME. All in all, there are 24 observables included in our statistical analysis.

The main goals are: (i) for each starting effective Hamiltonian find correlations between $0\nu\beta\beta$ NME and the other observables that are accessible experimentally; (ii) find theoretical ranges for each observables; (iii) establish the shape of different distributions for each observables and starting Hamiltonians; (iv) use this information to find weights of contributions  from different starting Hamiltonians to the "optimal" distribution of the $0\nu\beta\beta$ NME; (v) find an "optimal" value of the $0\nu\beta\beta$ NME and its predicted probable range (theoretical error). One should mention that similar studies for other observables were recently proposed~\cite{PhysRevC.101.054308}.

The $0\nu\beta\beta$ NME is related to the half-life of the respective process \cite{HoroiStoica2010} by

\begin{equation}
\left( T^{0\nu}_{1/2} \right)^{-1}
= G_{0\nu} (E_0, Z) g_A^4 \mid M_{0\nu}\mid^2 \mid <\eta_l>\mid^2 \ ,
\label{t0n}
\end{equation}
\noindent
where $G_{ 0\nu}$ and $M_{0\nu}$ are the PSF and nuclear matrix elements  for the $0\nu$ decay, $g_A$ is the axial vector coupling constant, and $<\eta_l> \equiv <m_{\beta\beta}>/m_e c^2$ is a BSM parameter associated with the light neutrino exchange mechanisms. Here we only consider the contribution from the light LH neutrino exchange mechanism, which is likely to contribute to the $0\nu\beta\beta$ decay. The methodology of calculating the $0\nu\beta\beta$ NME, $M_{0\nu}$,  within the shell model was extensively described elsewhere \cite{HoroiStoica2010,Horoi2013,18ho035502} and it will not be repeated here. Suffices to say that it includes a short range correlation function that can be viewed as an effective modification of the bare operator (see below).

The $2\nu\beta\beta$ NME is related to the half-life of the respective process \cite{HoroiStoicaBrown2007} by

\begin{equation}
 \left( T^{2\nu}_{1/2} \right)^{-1} = G_{2\nu} (E_0, Z) g_A^4\mid m_ec^2 M_{2\nu}\mid^2 \ .
\label{t2n}
\end{equation}

\noindent
Here, $G_{2\nu}$ is the appropriate PSF, and $M_{2\nu}$ can be calculated with
\begin{equation}
  M_{2\nu} = \sum_k \frac{q^2\left<0^+_f \mid \sigma \tau^- \mid 1^+_k\right>\left<1^+_k \mid \sigma \tau^- \mid 0^+_i\right>}{E_k-E_0} \ ,
\label{m2n}
\end{equation}
where the summation is on the $1^+_k$ states in $^{48}$Sc and $E_0=Q_{\beta\beta}/2+\Delta M(^{48}Sc-^{48}Ca)$

Often, the shell model calculations of the $0\nu\beta\beta$ NME are described as using the "bare" transition operator. This characterization is unfortunate, while the transition operator (see e.g. Eq. (7-12) of Ref. \cite{HoroiStoica2010}) contains the bare operator from the underlying theory of $0\nu\beta\beta$ decay, 
modified by a phenomenological effective short-range correlation function, $1+f(r)$, which is quenching the $0\nu\beta\beta$ NME. Therefore, the short-range modification of the bare operator acts as an effective operator. In practice the parameters of an effective operator need to be calibrated to the data. Given that the short-range correlator has radial dependence, its calibration has been only done relative to some ab-initio results. The standard Miller-Spencer short-range correlator \cite{MillerSpencer1976,HoroiStoica2010} produces the highest quenching of the NME, while the CD-Bonn parameterization of the short-range correlator \cite{PhysRevC.79.055501} produces little to no quenching. A direct renormalization of the $0\nu\beta\beta$ NME by a similarity renormalization group (SRG) evolution of the NME of the bare operator from 200 to 10 major harmonic oscillator shells using CDBonn two body wave functions, indicates that using a phenomenological CDBonn parametrization of the short-range correlator is a reasonable approach \cite{HoroiAPS2016}. More recent ab-initio calculations of the $0\nu\beta\beta$ NME using the $N^3LO$ Hamiltonian provides more quenched values, more consistent with the shell model results base on Miller-Spencer parametrization of the short-range correlator. In an effort to calibrate the effective operator used in shell model calculations to the latest ab-initio results we used the Miller-Spencer correlator in this study.

The other observables used in this study including the excited state energies, the GT strengths to the first $1^+$ state $^{48}$Sc, the B(E2)$\uparrow$ to the first $2^+$ state in the parent and daughter, as well as the s.p. occupation probabilities, are calculated in the standard way. Here we use in all cases the same effective charges ($e_p=1.5$ and $e_{n}=0.5)$) for the B(E2)$\uparrow$, and the same quenching factor ($q=0.74$) for the the GT strengths and $M_{2\nu}$.

\begin{widetext}
\begin{center}
\begin{table}[!ht]
    \centering
    \begin{tabular}{|c|r|r|r|r|r|r|r|r|r|r|r|r|}
    \hline
~ & $0\nu\beta\beta$ & $2\nu\beta\beta$ & $^{48}$Ca & $^{48}$Ca & $^{48}$Ca& $^{48}$Ca & $^{48}$Ca Occ& $^{48}$Ca Occ& $^{48}$Ca Occ& $^{48}$Ca Occ& $^{48}$Ca $\rightarrow$& $^{48}$Ti \\
~ & NME & NME & B(E2)$\uparrow$&2+&4+&6+& (Nf5)& (Nf7) & (Np1) & (Np3) & $^{48}$Sc GT& B(E2)$\uparrow$ \\ \hline
$0\nu\beta\beta$ NME & 1.00 & 0.90 & 0.43 & 0.22 & 0.24 & 0.18 & 0.62 & -0.58 & 0.30 & 0.41 & 0.12 & -0.06 \\  \hline
$2\nu\beta\beta$ NME & 0.90 & 1.00 & 0.38 & 0.28 & 0.30 & 0.17 & 0.44 & -0.47 & 0.25 & 0.37 & 0.10 & -0.18\\ \hline
$^{48}$Ca B(E2)$\uparrow$ & 0.43 & 0.38 & 1.00 & -0.35 & -0.27 & -0.31 & 0.55 & -0.91 & 0.53 & 0.92 & -0.17& 0.32 \\ \hline
$^{48}$Ca 2+ & 0.23 & 0.28 & -0.35 & 1.00 & 0.96 & 0.49 & -0.06 & 0.32 & -0.23 & -0.42 & 0.51 & -0.38\\ \hline
$^{48}$Ca 4+ & 0.24 & 0.30 & -0.27 & 0.96 & 1.00 & 0.44 & -0.01 & 0.26 & -0.17 & -0.36 & 0.54 & -0.33\\ \hline
$^{48}$Ca 6+ & 0.18 & 0.17 & -0.31 & 0.49 & 0.44 & 1.00 & -0.22 & 0.25 & -0.19 & -0.20 & 0.26 & -0.22\\ \hline
$^{48}$Ca Occ(Nf5) & 0.62 & 0.44 & 0.55 & -0.06 & -0.01 & -0.22 & 1.00 & -0.77 & 0.38 & 0.42 & -0.04 & 0.12\\ \hline
$^{48}$Ca Occ(Nf7) & -0.58 & -0.47 & -0.91 & 0.32 & 0.26 & 0.25 & -0.77 & 1.00 & -0.58 & -0.90 & 0.17 & -0.30\\ \hline
$^{48}$Ca Occ(Np1) & 0.30 & 0.25 & 0.53 & -0.23 & -0.17 & -0.19 & 0.38 & -0.58 & 1.00 & 0.48 & -0.11 & 0.19\\ \hline
$^{48}$Ca Occ(Np3) & 0.41 & 0.37 & 0.92 & -0.42 & -0.36 & -0.20 & 0.42 & -0.90 & 0.48 & 1.00 & -0.22 & 0.34\\ \hline
$^{48}$Ca $\rightarrow$ $^{48}$Sc GT & 0.12 & 0.10 & -0.17 & 0.51 & 0.54 & 0.26 & -0.04 & 0.17 & -0.11 & -0.22 & 1.00 & -0.27\\ \hline
$^{48}$Ti B(E2)$\uparrow$ & -0.06 & -0.18 & 0.32 & -0.38 & -0.34 & -0.22 & 0.12 & -0.30 & 0.19 & 0.34 & -0.27 & 1.00\\ \hline
$^{48}$Ti 2+ & 0.80 & 0.89 & 0.37 & 0.32 & 0.32 & 0.20 & 0.47 & -0.47 & 0.24 & 0.35 & 0.23 & -0.24\\ \hline
$^{48}$Ti 4+ & 0.78 & 0.85 & 0.35 & 0.35 & 0.35 & 0.25 & 0.45 & -0.45 & 0.23 & 0.34 & 0.22 & -0.19\\ \hline
$^{48}$Ti 6+ & 0.75 & 0.80 & 0.42 & 0.28 & 0.28 & 0.24 & 0.48 & -0.52 & 0.26 & 0.42 & 0.03 & 0.13\\ \hline
$^{48}$Ti Occ(Nf5) & 0.14 & -0.03 & 0.39 & -0.16 & -0.11 & -0.36 & 0.75 & -0.53 & 0.25 & 0.24 & -0.00 & 0.27\\ \hline
$^{48}$Ti Occ(Nf7) & 0.12 & 0.32 & -0.51 & 0.47 & 0.42 & 0.37 & -0.37 & 0.53 & -0.33 & -0.49 & 0.20 & -0.82\\ \hline
$^{48}$Ti Occ(Np1) & -0.13 & -0.29 & 0.27 & -0.34 & -0.29 & -0.24 & 0.10 & -0.26 & 0.56 & 0.25 & -0.02 & 0.65\\ \hline
$^{48}$Ti Occ(Np3) & -0.25 & -0.38 & 0.43 & -0.51 & -0.47 & -0.26 & 0.03 & -0.37 & 0.21 & 0.50 & -0.28 & 0.88\\ \hline
$^{48}$Ti Occ(Pf5) & 0.22 & 0.01 & 0.20 & 0.00 & 0.03 & -0.08 & 0.56 & -0.35 & 0.15 & 0.12 & 0.21 & 0.27\\ \hline
$^{48}$Ti Occ(Pf7) & 0.21 & 0.41 & -0.16 & 0.24 & 0.20 & 0.16 & -0.10 & 0.16 & -0.11 & -0.16 & -0.01 & -0.85\\ \hline
$^{48}$Ti Occ(Pp1) & -0.11 & -0.29 & 0.13 & -0.19 & -0.16 & -0.14 & 0.06 & -0.13 & 0.30 & 0.12 & 0.16 & 0.60\\ \hline
$^{48}$Ti Occ(Pp3) & -0.33 & -0.49 & 0.11 & -0.28 & -0.25 & -0.16 & -0.09 & -0.06 & 0.05 & 0.14 & -0.09 & 0.89\\ \hline
$^{48}$Ti $\rightarrow$ $^{48}$Sc GT & 0.15 & 0.40 & 0.23 & -0.03 & -0.02 & -0.24 & 0.15 & -0.20 & 0.13 & 0.19 & -0.55 & 0.03 \\ \hline
\end{tabular}
    \label{correlations-matrix}
    \caption{Correlation matrix for the 24 observables described in the text 
    using the FPD6 starting Hamiltonian (continues in Table III). 
    }
\end{table}

\begin{table}[!ht]
    \centering
    \begin{tabular}{|c|r|r|r|r|r|r|r|r|r|r|r|r|}
    \hline
& $^{48}$Ti & $^{48}$Ti & $^{48}$Ti & $^{48}$Ti Occ & $^{48}$Ti Occ& $^{48}$Ti Occ& $^{48}$Ti Occ& $^{48}$Ti Occ& $^{48}$Ti Occ& $^{48}$Ti Occ& $^{48}$Ti Occ & $^{48}$Ti $\rightarrow$\\
& 2+ & 4+ & 6+ &(Nf5)&(Nf7)&(Np1)&(Np3)&(Pf5)&(Pf7)&(Pp1)&(Pp3)& $^{48}$Sc GT \\ \hline
$0\nu\beta\beta$ NME & 0.80 & 0.78 & 0.75 & 0.14 & 0.12 & -0.13 & -0.25 & 0.22 & 0.21 & -0.11 & -0.33 & 0.15 \\ \hline
$2\nu\beta\beta$ NME & 0.89 & 0.85 & 0.80 & -0.03 & 0.32 & -0.29 & -0.38 & 0.01 & 0.41 & -0.28 & -0.48 & 0.39 \\ \hline
$^{48}$Ca B(E2)$\uparrow$ & 0.37 & 0.35 & 0.42 & 0.39 & -0.51 & 0.27 & 0.43 & 0.19 & -0.16 & 0.13 & 0.11 & 0.23  \\ \hline
$^{48}$Ca 2+ & 0.32 & 0.35 & 0.27 & -0.16 & 0.47 & -0.34 & -0.51 & 0.00 & 0.24 & -0.19 & -0.28 & -0.03  \\ \hline
$^{48}$Ca 4+ & 0.32 & 0.35 & 0.28 & -0.11 & 0.42 & -0.29 & -0.47 & 0.03 & 0.20 & -0.16 & -0.25 & -0.02  \\ \hline
$^{48}$Ca 6+ & 0.20 & 0.25 & 0.24 & -0.36 & 0.37 & -0.24 & -0.26 & -0.07 & 0.16 & -0.14 & -0.16 & -0.24  \\ \hline
$^{48}$Ca Occ(Nf5) & 0.47 & 0.45 & 0.48 & 0.75 & -0.37 & 0.10 & 0.03 & 0.56 & -0.10 & 0.06 & -0.09 & 0.15  \\ \hline
$^{48}$Ca Occ(Nf7) & -0.47 & -0.45 & -0.52 & -0.53 & 0.53 & -0.26 & -0.37 & -0.35 & 0.16 & -0.13 & -0.06 & -0.20  \\ \hline
$^{48}$Ca Occ(Np1) & 0.24 & 0.23 & 0.26 & 0.25 & -0.33 & 0.56 & 0.21 & 0.15 & -0.11 & 0.30 & 0.05 & 0.13  \\ \hline
$^{48}$Ca Occ(Np3) & 0.35 & 0.34 & 0.42 & 0.24 & -0.49 & 0.25 & 0.50 & 0.12 & -0.16 & 0.12 & 0.14 & 0.19  \\ \hline
$^{48}$Ca $\rightarrow$ $^{48}$Sc GT & 0.23 & 0.22 & 0.03 & 0.00 & 0.20 & -0.02 & -0.28 & 0.21 & -0.01 & 0.16 & -0.09 & -0.55  \\ \hline
$^{48}$Ti B(E2)$\uparrow$ & -0.24 & -0.19 & 0.13 & 0.27 & -0.82 & 0.65 & 0.88 & 0.27 & -0.85 & 0.60 & 0.89 & 0.03  \\ \hline
$^{48}$Ti 2+ & 1.00 & 0.96 & 0.85 & 0.16 & 0.25 & -0.24 & -0.42 & 0.20 & 0.36 & -0.19 & -0.51 & 0.28  \\ \hline
$^{48}$Ti 4+ & 0.96 & 1.00 & 0.89 & 0.16 & 0.21 & -0.21 & -0.36 & 0.22 & 0.29 & -0.15 & -0.44 & 0.24  \\ \hline
$^{48}$Ti 6+ & 0.85 & 0.89 & 1.00 & 0.20 & -0.01 & -0.09 & -0.09 & 0.22 & 0.08 & -0.07 & -0.18 & 0.35  \\ \hline
$^{48}$Ti Occ(Nf5) & 0.16 & 0.16 & 0.20 & 1.00 & -0.64 & 0.34 & 0.22 & 0.81 & -0.42 & 0.35 & 0.18 & -0.05  \\ \hline
$^{48}$Ti Occ(Nf7) & 0.25 & 0.21 & -0.01 & -0.64 & 1.00 & -0.74 & -0.88 & -0.55 & 0.86 & -0.66 & -0.80 & 0.12  \\ \hline
$^{48}$Ti Occ(Np1) & -0.24 & -0.21 & -0.09 & 0.34 & -0.74 & 1.00 & 0.66 & 0.41 & -0.75 & 0.88 & 0.67 & -0.22  \\ \hline
$^{48}$Ti Occ(Np3) & -0.42 & -0.36 & -0.09 & 0.22 & -0.88 & 0.66 & 1.00 & 0.19 & -0.82 & 0.56 & 0.90 & -0.10  \\ \hline
$^{48}$Ti Occ(Pf5) & 0.20 & 0.22 & 0.22 & 0.81 & -0.55 & 0.41 & 0.19 & 1.00 & -0.57 & 0.56 & 0.26 & -0.30  \\ \hline
$^{48}$Ti Occ(Pf7) & 0.36 & 0.29 & 0.08 & -0.42 & 0.86 & -0.75 & -0.82 & -0.57 & 1.00 & -0.81 & -0.94 & 0.31  \\ \hline
$^{48}$Ti Occ(Pp1) & -0.19 & -0.15 & -0.07 & 0.35 & -0.66 & 0.88 & 0.56 & 0.56 & -0.81 & 1.00 & 0.67 & -0.35  \\ \hline
$^{48}$Ti Occ(Pp3) & -0.51 & -0.44 & -0.18 & 0.18 & -0.80 & 0.67 & 0.90 & 0.26 & -0.94 & 0.67 & 1.00 & -0.23  \\ \hline
$^{48}$Ti $\rightarrow$ $^{48}$Sc GT & 0.28 & 0.24 & 0.35 & -0.05 & 0.12 & -0.22 & -0.10 & -0.30 & 0.31 & -0.35 & -0.23 & 1.00 \\ \hline
    \end{tabular}
    \label{correlations-cont}
    \caption{Correlation matrix: continuation of Table II. 
    }
\end{table}
\end{center}
\end{widetext}

\section{Results} \label{results}
The experimental data used in this study listed in Table 
I and in the legends of the rightmost column in Tables \ref{tab:parent-fpd6}/\ref{tab:daughter-fpd6} - \ref{tab:parent-kb3g}/\ref{tab:daughter-kb3g} are taken from Ref. \cite{NNDC_A48-2022} (excitation energies of the $2^+$, $4^+$ and $6^+$ states of $^{48}$Ca and $^{48}$Ti in MeV), Ref. \cite{Barabash2020} ($2\nu\beta\beta$ NME in MeV$^{-1}$), Ref. \cite{Pritychenko2016} ( B(E2)$\uparrow$ in $e^2b^2$), and Ref. \cite{GreweGT2007} for the GT transition probabilities to the first excited $1^+$ state in $^{48}$Sc. The experimental errors for the excitation energies are very small, and for the calculation  the $\chi^2$ we use the typical theoretical RMS value of 150 keV~\cite{Honma2004}. The experimental occupation probabilities are not available, and we took as reference the GXPF1A results assuming an uniform error that we choose to be 5\% of the highest occupation for a each nucleon species in the $fp$-shell. In the Tables Occ (Nf7) designates the neutron occupation probability of the $f_{7/2}$ s.p. orbital, Occ (Pf3) designates the proton occupation probability of the $p_{3/2}$ s.p. orbital, etc.

Tables \ref{tab:parent-fpd6}/\ref{tab:daughter-fpd6} - \ref{tab:parent-kb3g}/\ref{tab:daughter-kb3g} show the main results of this study. The leftmost columns indicates the 24 observables discussed in section \ref{model}, including the $0\nu\beta\beta$ NME. The middle column shows the scatter plots of the correlation of each variable with the $0\nu\beta\beta$ NME, and the last column shows the distribution of each observable when the random term is added to the respective effective Hamiltonian. The legends in column with correlations show the standard Pearson correlator $R$, and in the last columns the legends include the mean, standard deviations, and the skewness (normalized 3rd moment) of the distributions, as well as the result for the starting interactions (FPD6, GXPF1A, and KB3G) and the experimental values when available. Tables \ref{tab:parent-fpd6}/\ref{tab:daughter-fpd6} present the results for the FPD6 effective Hamiltonian, Tables \ref{tab:parent-gxpf1a}/\ref{tab:daughter-gxpf1a} show the results for the GXPF1A effective Hamiltonian, and Tables \ref{tab:parent-kb3g}/\ref{tab:daughter-kb3g} present the results for the KB3G effective Hamiltonian. For each starting effective Hamiltonian we use 20,000 random Hamiltonians produced by the procedure described in section \ref{model}.

The results in Tables \ref{tab:parent-fpd6}/\ref{tab:daughter-fpd6} - \ref{tab:parent-kb3g}/\ref{tab:daughter-kb3g} indicate strong correlations between the $0\nu\beta\beta$ NME and the $2\nu\beta\beta$ NME. Alternative approaches of obtaining these NME, e.g. QRPA calculations, are calibrating parts of their nuclear Hamiltonian, such as the isoscalar particle-particle interaction $g_{pp}$,  to describe the experimental value of the $2\nu\beta\beta$ NME and to approximately restore the isospin symmetry, thus inducing correlation with the $2\nu\beta\beta$ NME. In the shell model approach, the Hamiltonian remains unchanged, and all symmetries are enforced. Therefore, we conclude that the strong correlations between the $0\nu\beta\beta$ NME and the $2\nu\beta\beta$ NME are genuine.

Interestingly, the correlations between the $0\nu\beta\beta$ NME and the Gamow-Teller (GT) transitions probabilities to the first $1^+$ state in $^{48}$Sc are much reduced. One explanation of this phenomena is based on the fact the distributions of the GT strength from the parent and daughter (see last column in Tables \ref{tab:parent-fpd6}/\ref{tab:daughter-fpd6} - \ref{tab:parent-kb3g}/\ref{tab:daughter-kb3g}) are asymmetric in opposite direction, thus diminishing the correlation effects. A quick look to the full correlation matrix in Table \ref{correlations-cont} shows that the GT strengths to the first $1^+$ state in $^{48}$Sc from $^{48}$Ca and $^{48}$Ti are anti-correlated with a correlation coefficient of about -0.5.

Other observables that have relatively high (anti)correlations with the $0\nu\beta\beta$ NME are the energies of the $2^+$, $4^+$ and $6^+$ states in $^{48}$Ti, and the neutron occupation probabilities in $^{48}$Ca. Overall, the correlators $R$ with the $2\nu\beta\beta$ NME are around 0.9, the ones with the energies of the $2^+$, $4^+$ and $6^+$ states in $^{48}$Ti are about 0.77, and correlators with the $0f_{5/2}$ occupation probability is about 0.6, while the occupation probability of the $0f_{7/2}$ is anti-correlated with the $0\nu\beta\beta$ NME, $R \approx -0.6$.

\begin{figure} 
\includegraphics[width=8cm]{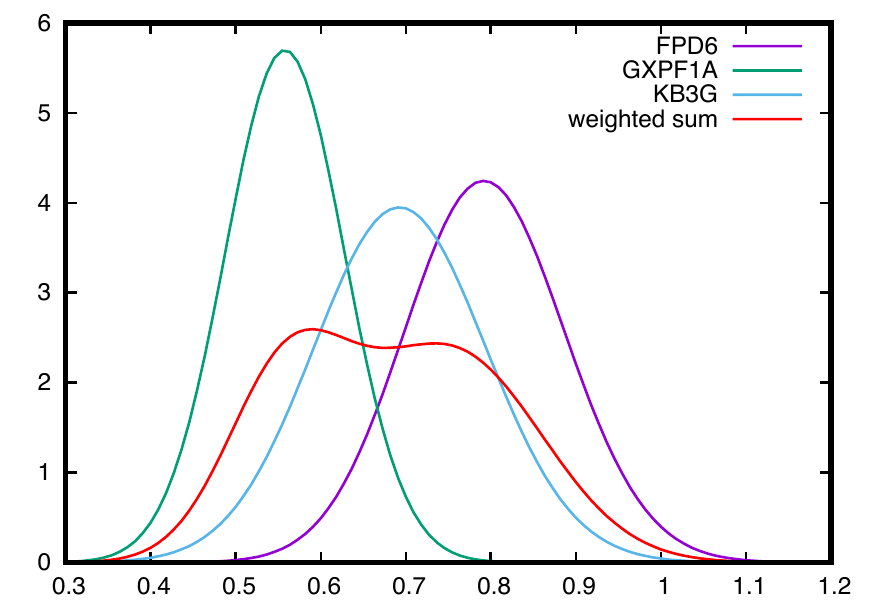}
\caption{PDF of the $0\nu\beta\beta$ NME distributions for the FPD6, GXPF1A and KB3G Hamiltonians and their weighted sum (see text for details).}
\label{nme-range}
\end{figure}

Additional interesting information can be extracted from the full correlation matrix for all 24 observables. Tables \ref{correlations-matrix} - III
show the full correlation matrix evaluated for the FPD6 starting Hamiltonian. It is interesting to analyze which other observables are correlated with those that are directly correlated with the $0\nu\beta\beta$ NME. We already discussed the correlations between the GT strengths and the $2\nu\beta\beta$ NME. In addition, one can observe that the $2^+$, $4^+$ and $6^+$ in $^{48}$Ti are correlated with the neutron $f$ and $p$ states occupancies in $^{48}$Ca, which in their turn are correlated with some of the neutron states occupancies in $^{48}$Ti. Also, some of the neutron occupation  probabilities in $^{48}$Ti are correlated to the B(E2)$\uparrow$ values. These observations highlight the importance of a reliable experimental investigation of the occupation probabilities for these nuclei.

It would be interesting to extract some information about possible range and mean value of the $0\nu\beta\beta$ NME based on this statistical analysis. First, it is clear that the value of all observables are quite stable to reasonably small changes of the effective Hamiltonian. No hints of any wild departure from the main values that would indicate some phase transitions are found. This seems to be a consequence of the preservation of nuclear many-body symmetries in the shell model. One can further try using the distributions of all available effective Hamiltonians to draw conclusions on some optimal values for the $0\nu\beta\beta$ NME and its range (error). One direct approach would be to superpose the distributions of the NME produced in Tables \ref{tab:parent-fpd6}/\ref{tab:daughter-fpd6} - \ref{tab:parent-kb3g}/\ref{tab:daughter-kb3g} withe some weighting factors $W_H$,

\begin{equation} \label{superposition}
\begin{split}
P(x)  & =  W_{FPD6} P_{FPD6}(x) + W_{GXPF1A} P_{GXPF1A}(x) \\
& +  W_{KB3G} P_{KB3G}(x) \ ,
\end{split}
\end{equation}
where $x$ is the value of the $0\nu\beta\beta$ NME. The normalized weights $W_H$ can be inferred using, for example, the likelyhood probability $\propto exp(-\chi^2/2)$,  or replacing the bare $\chi^2$ with its individual contributions weighted by the corresponding correlators $R$. Based on the data we show in Table \ref{experimental-data}, we get the following $\chi^2$ for each starting effective Hamiltonian: 7.9 for FPD6, 4.8 for GXPF1A, and 7.3 for KB3G. Unfortunately, there is no experimental data for the occupation probabilities that seem to correlate directly and indirectly with the $0\nu\beta\beta$ NME. Therefore, here we present the results of a "democratic" approach in which all $W_H$ are 0.33. Fig. \ref{nme-range} shows the probability distribution functions (PDF) for the three starting effective Hamiltonians and their weighted sum. To calculate each PDF we use the Gram-Charlier A series expansion \cite{Gram-Charlier} (see Appendix for detail), based on the first four normalized moments of the distributions presented in Tables \ref{tab:parent-fpd6}/\ref{tab:daughter-fpd6} - \ref{tab:parent-kb3g}/\ref{tab:daughter-kb3g}. Based on the results of our statistical analysis summarized in Fig. \ref{nme-range} (see "weighted sum" curve) one can infer that with 90\% confidence the $0\nu\beta\beta$ NME lies in the range between 0.45 and 0.95, with a mean value of about 0.68.

\section{Conclusion and Outlook} \label{conclusions}

In conclusion, we developed a statistical model for analyzing the distribution of the $0\nu\beta\beta$ NME of $^{48}$Ca using the interactive shell model in the $fp$-shell model space. In the analysis we started from three widely used effective Hamiltonians for the low part of the $fp$-shell, FPD6, GXPF1A and KB3G, to which we added a random contributions to the TBME of $\pm$10\%. Using sample sizes of 20,000 points we analyzed
for each starting effective Hamiltonian: (i) the correlations between $0\nu\beta\beta$ NME and the other observables that are accessible experimentally; (ii) the theoretical ranges for each observables; (iii)  the shape of different distributions for each observables and starting Hamiltonians; (iv) the weighted contributions from different starting Hamiltonians to the "optimal" distribution of the $0\nu\beta\beta$ NME; (v) an "optimal" value of the $0\nu\beta\beta$ NME and its predicted probable range (theoretical error).

We found that the $0\nu\beta\beta$ NME correlates strongly with the $2\nu\beta\beta$ NME, but much less with the Gamow-Teller strengths to the first $1^+$ state in $^{48}$Sc. We also found that the $0\nu\beta\beta$ NME exhibits reasonably strong correlations with the energies of the $2^+$, $4^+$ and $6^+$ states in $^{48}$Ti, and with the neutron occupation probabilities in $^{48}$Ca.
We also found that there are additional correlations between observables, such as the energies of the $2^+$, $4^+$ and $6^+$ states in $^{48}$Ti and the neutron occupation probabilities, as well as between B(E2)$\uparrow$ values in $^{48}$Ti and proton and neutron occupation probabilities, which can indirectly influence the $0\nu\beta\beta$ NME. Therefore, we conclude that reliable experimental values of the occupation probabilities in $^{48}$Ti and $^{48}$Ca would be useful for this analysis, potentially helpful to reduce the uncertainties of the $0\nu\beta\beta$ NME.

Based on this statistical analysis with three independent effective Hamiltonians we propose a common probability distribution function for the $0\nu\beta\beta$ NME, which has a range (theoretical error)  of (0.45 - 0.95) at 90\% confidence level, and a mean value of 0.68. We also hope that the present analysis will help ab-initio studies, such as \cite{Yao2020,Novario2021,Belley2021} to better identify correlations and further reduce the uncertainties of the $0\nu\beta\beta$ NME

\begin{table}[htbp]
\centering
\begin{tabular}{c|c|c}
\hline
Observable & Correlation & PDF\\ \hline
$0\nu\beta\beta$ NME & \includegraphics[width=0.3\linewidth]{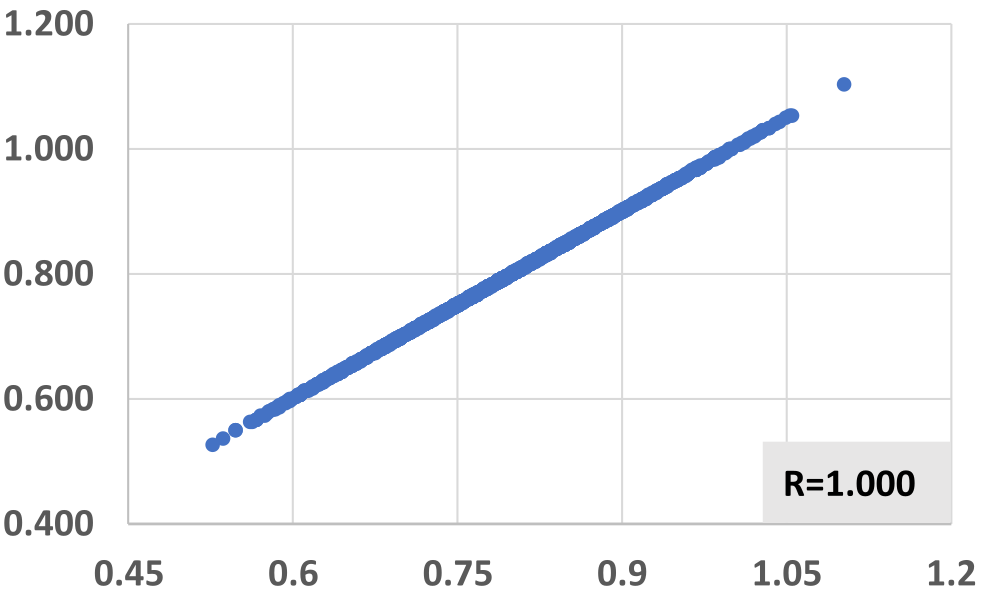} & \includegraphics[width=0.3\linewidth]{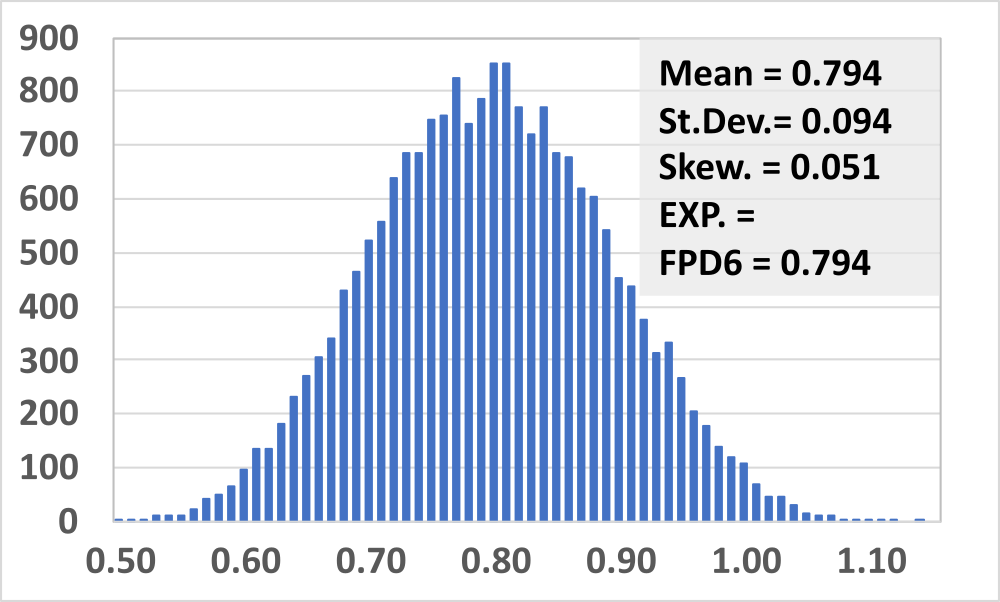} \\
\hline
$2\nu\beta\beta$ NME & \includegraphics[width=0.3\linewidth]{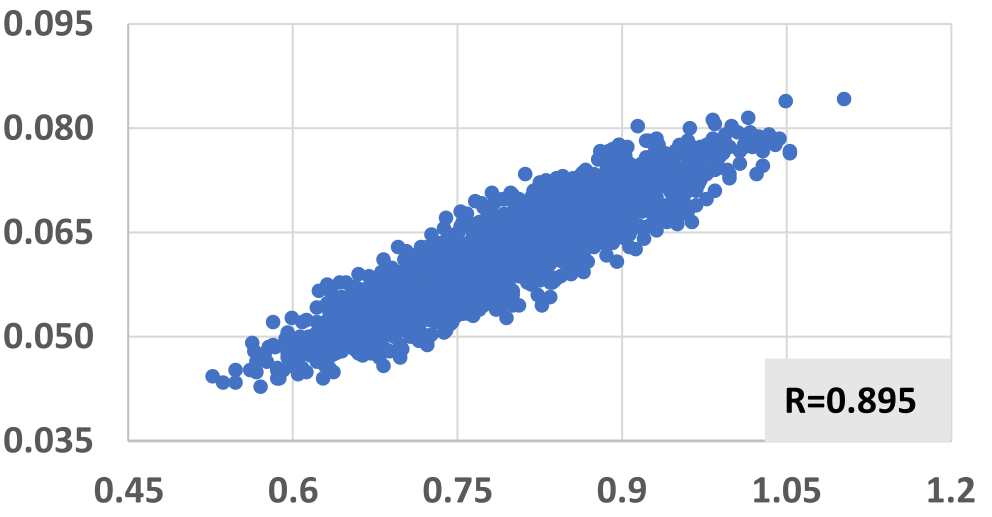} & \includegraphics[width=0.3\linewidth]{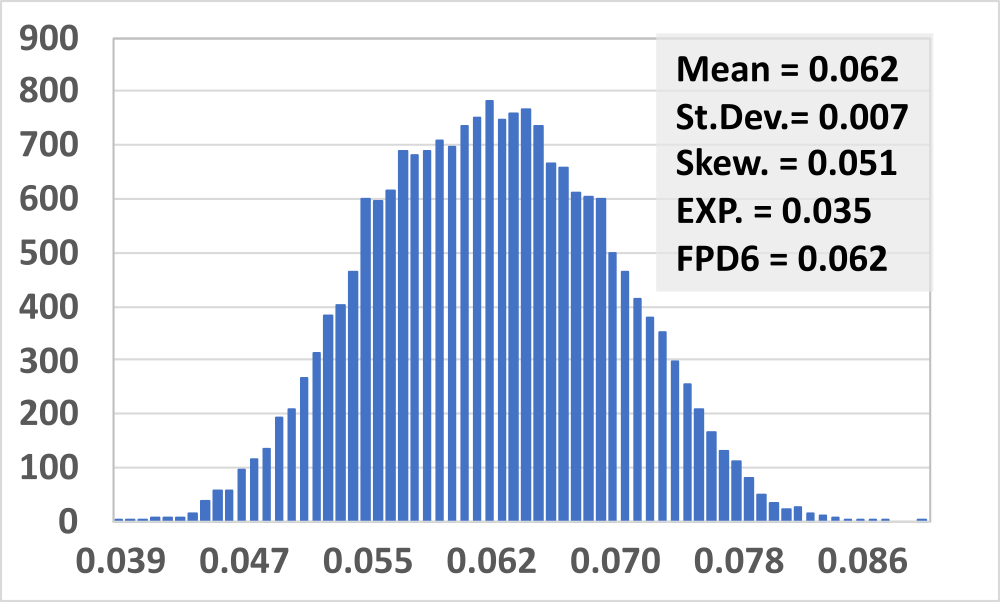}  
\\ \hline
$^{48}$Ca $\rightarrow$ $^{48}$Sc GT & \includegraphics[width=0.3\linewidth]{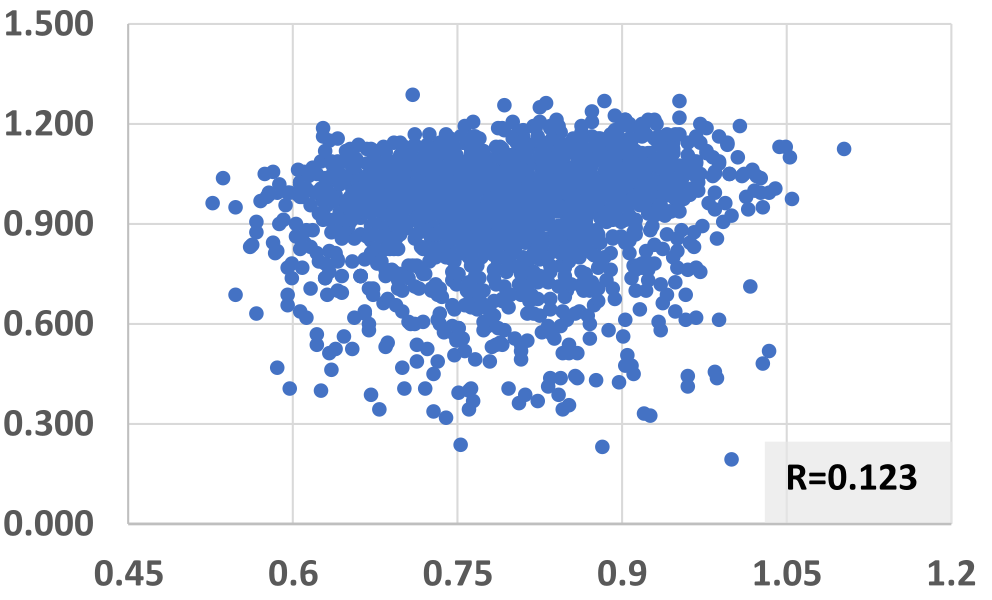} & \includegraphics[width=0.3\linewidth]{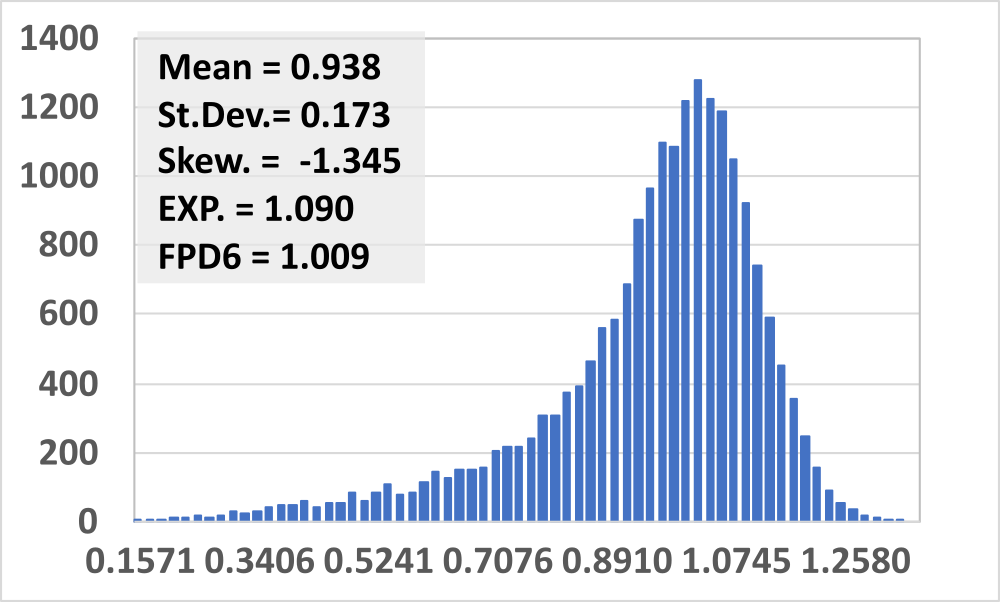}  
\\ \hline
$^{48}$Ti $\rightarrow$ $^{48}$Sc GT & \includegraphics[width=0.3\linewidth]{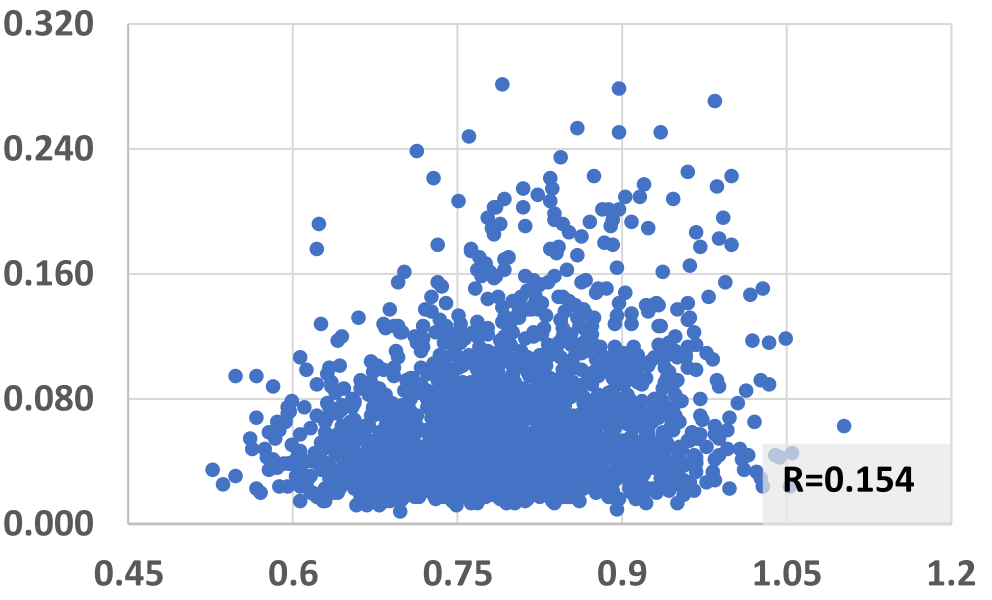} & \includegraphics[width=0.3\linewidth]{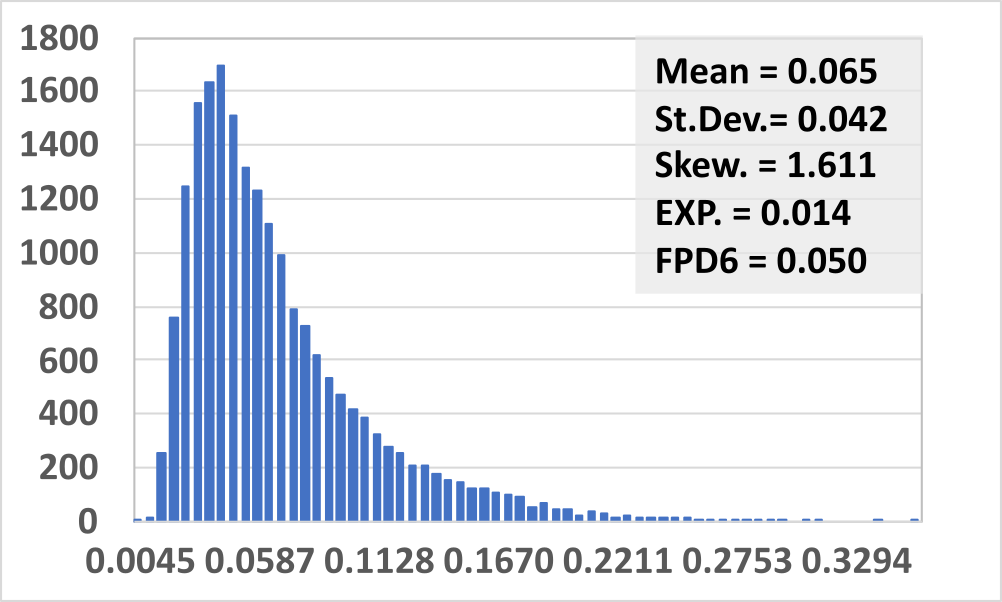}  
\\ \hline
$^{48}$Ca B(E2)$(\uparrow)$ & \includegraphics[width=0.3\linewidth]{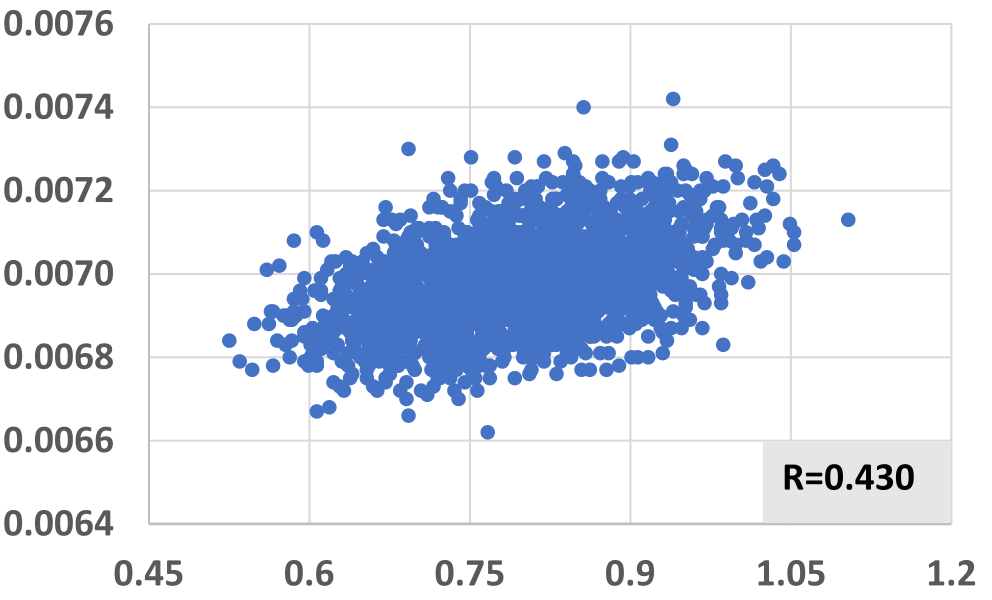} & \includegraphics[width=0.3\linewidth]{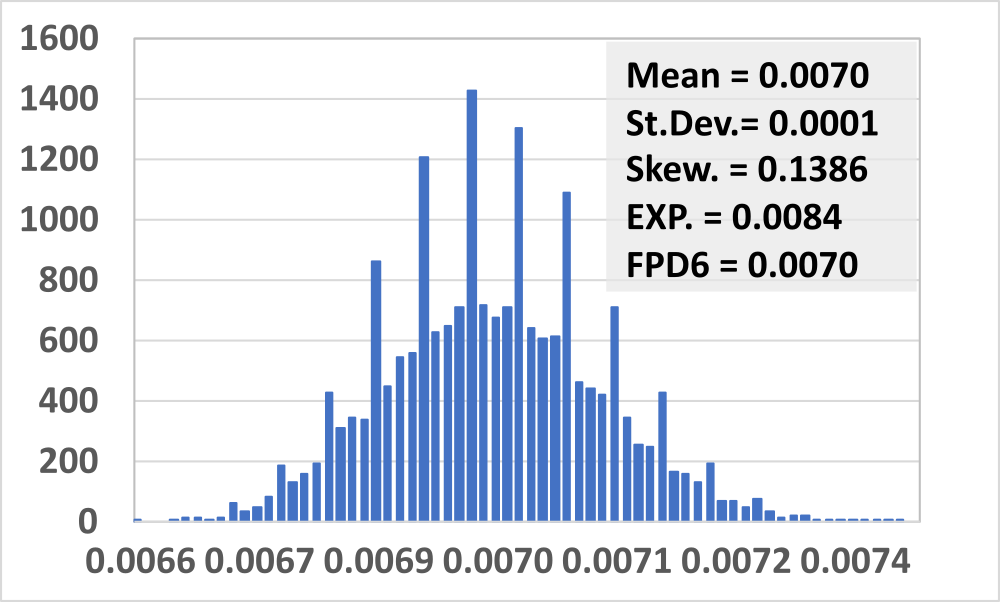} 
\\ \hline
$^{48}$Ca $2^{+}$ & \includegraphics[width=0.3\linewidth]{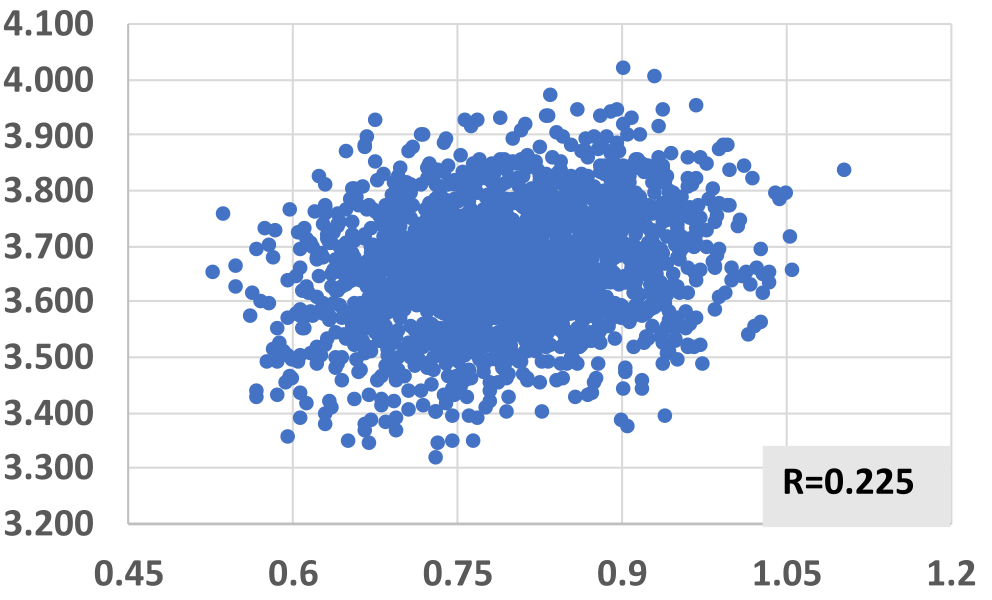} & \includegraphics[width=0.3\linewidth]{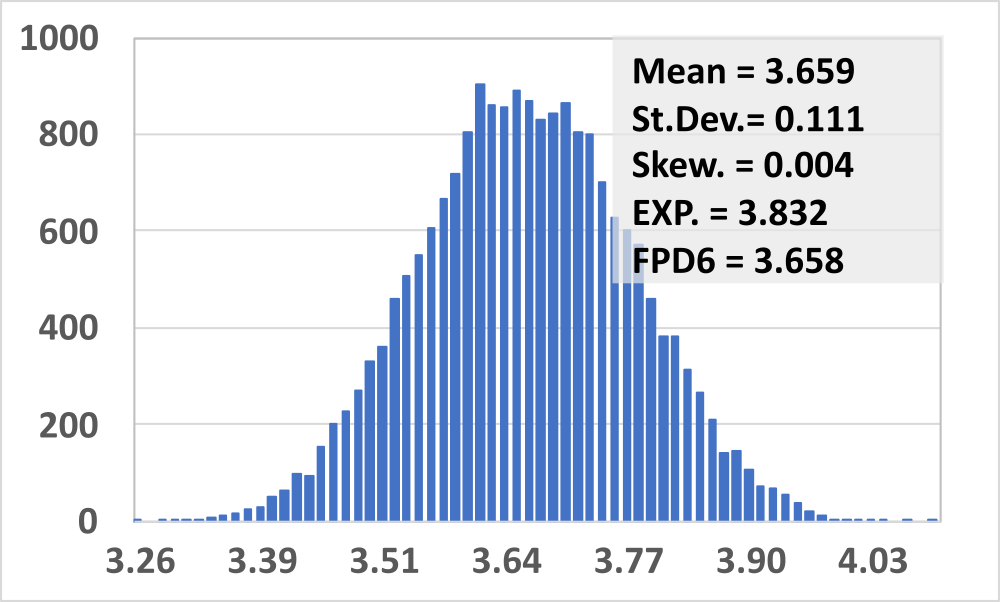}  
\\ \hline
$^{48}$Ca $4^{+}$ & \includegraphics[width=0.3\linewidth]{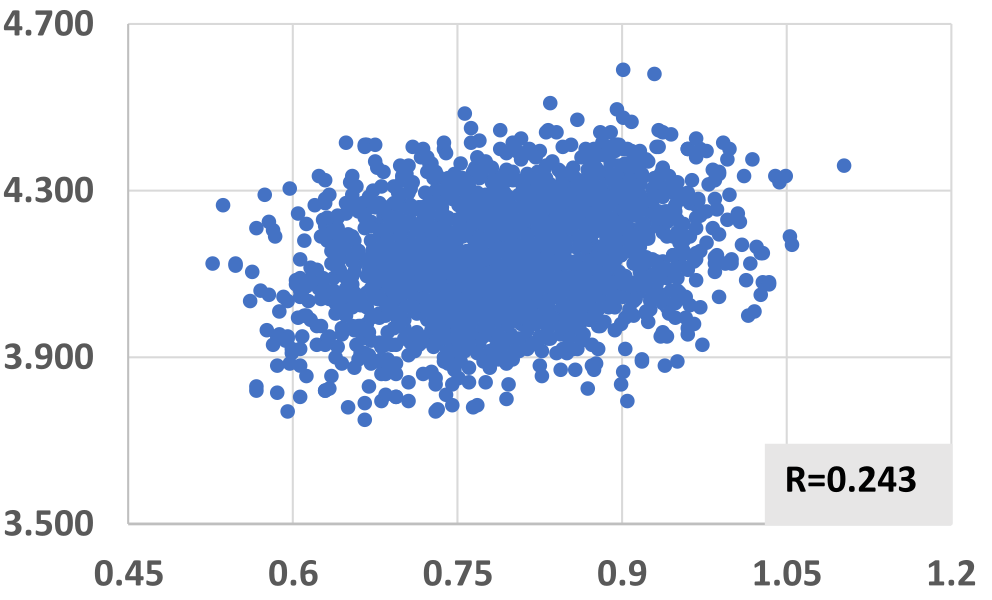} & \includegraphics[width=0.3\linewidth]{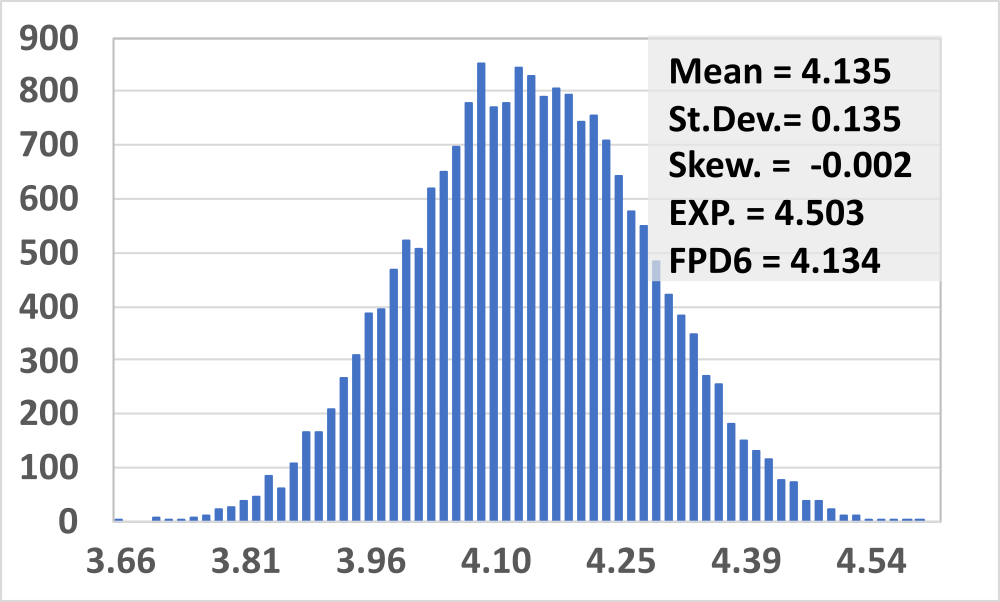}  
\\ \hline
$^{48}$Ca $6^{+}$ & \includegraphics[width=0.3\linewidth]{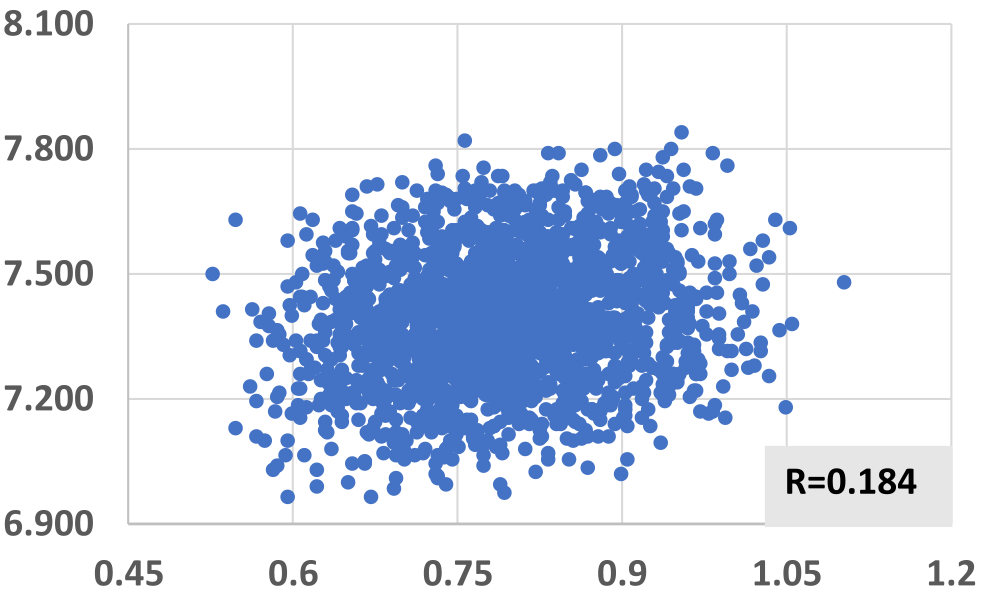} & \includegraphics[width=0.3\linewidth]{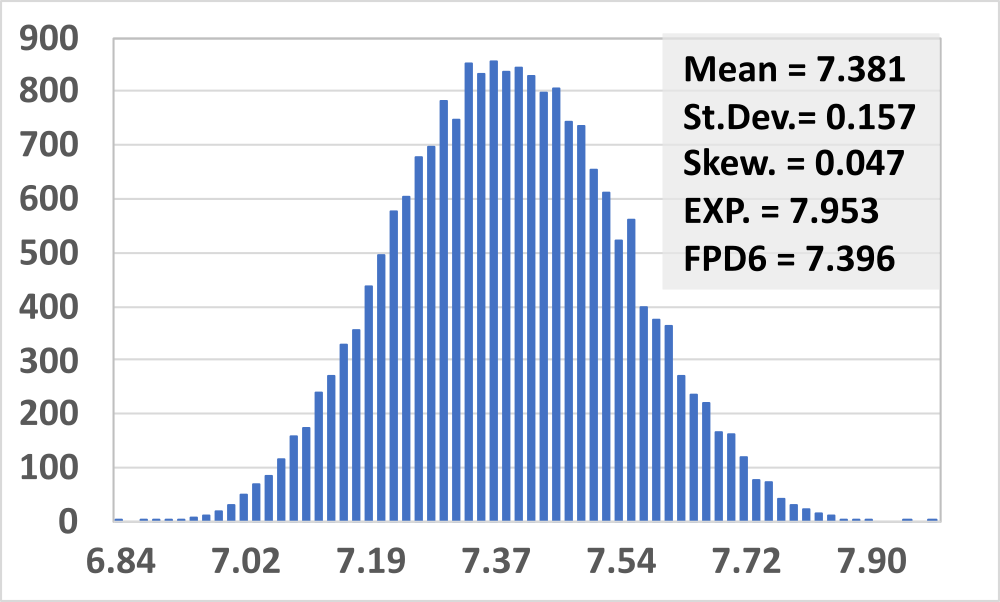}  
\\ \hline
$^{48}$Ca Occ(Nf5) & \includegraphics[width=0.3\linewidth]{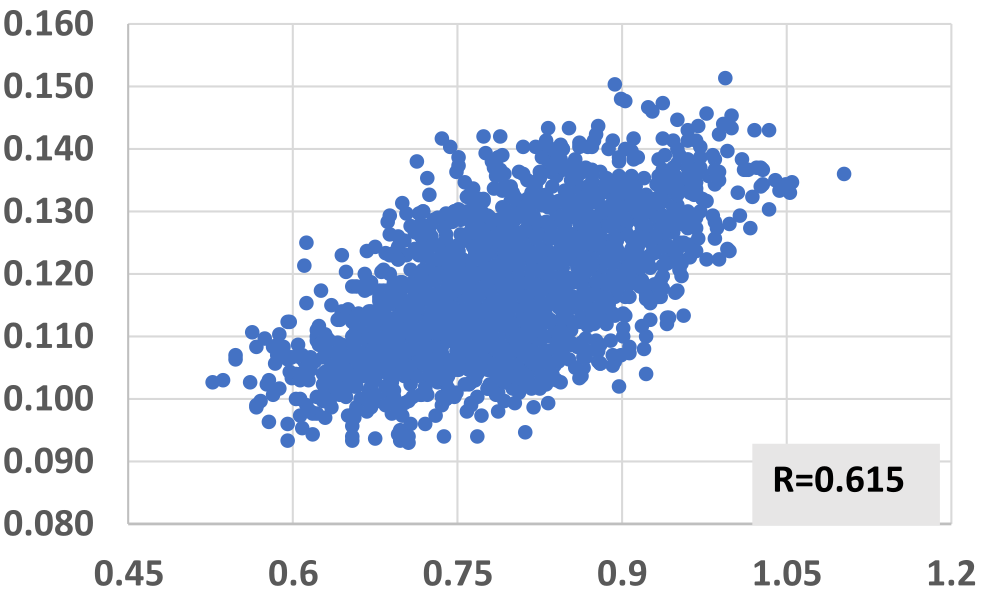} & \includegraphics[width=0.3\linewidth]{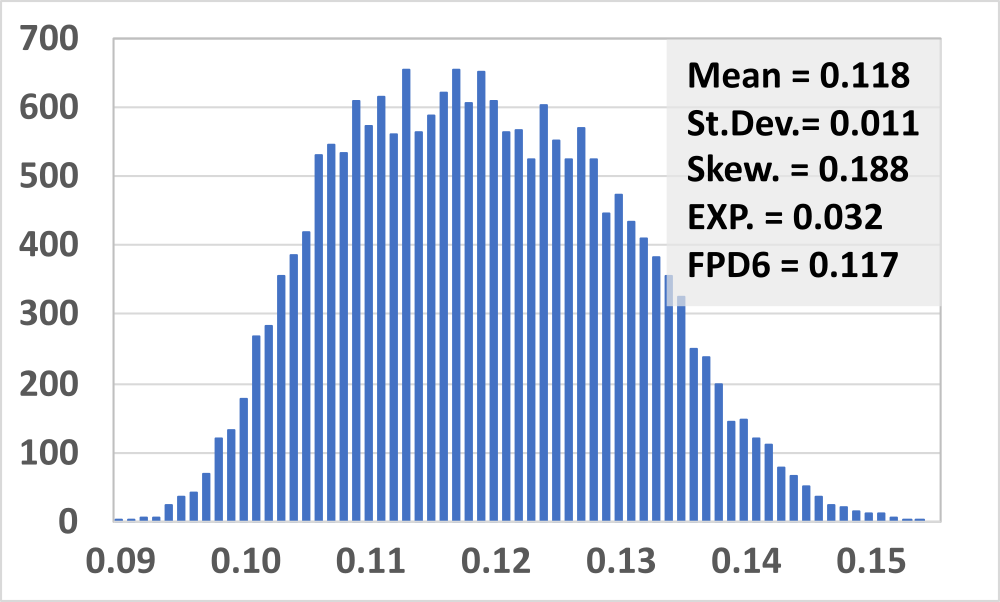}  
\\ \hline
$^{48}$Ca Occ(Nf7) & \includegraphics[width=0.3\linewidth]{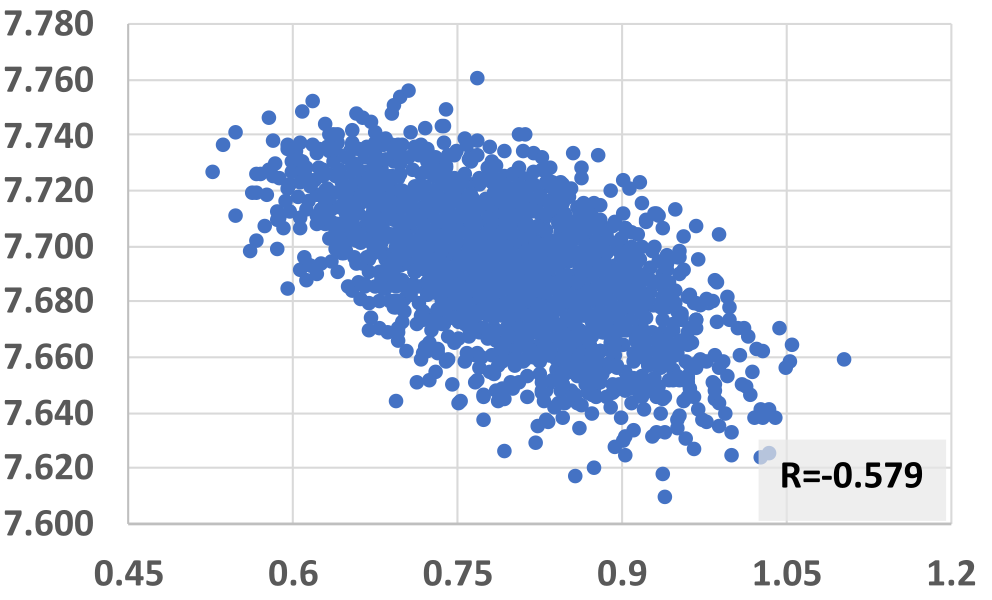} & \includegraphics[width=0.3\linewidth]{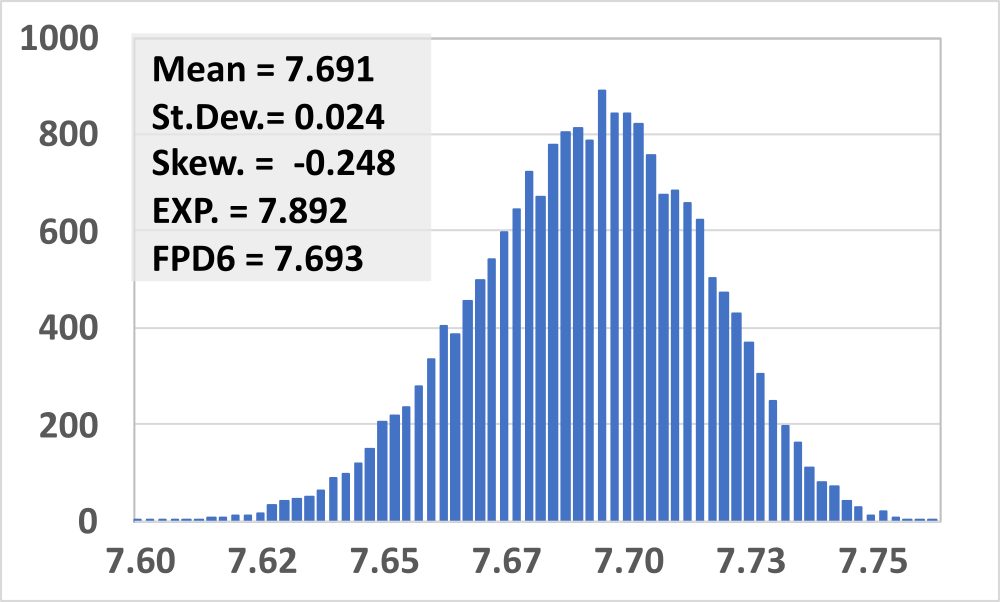}  
\\ \hline
$^{48}$Ca Occ(Np1) & \includegraphics[width=0.3\linewidth]{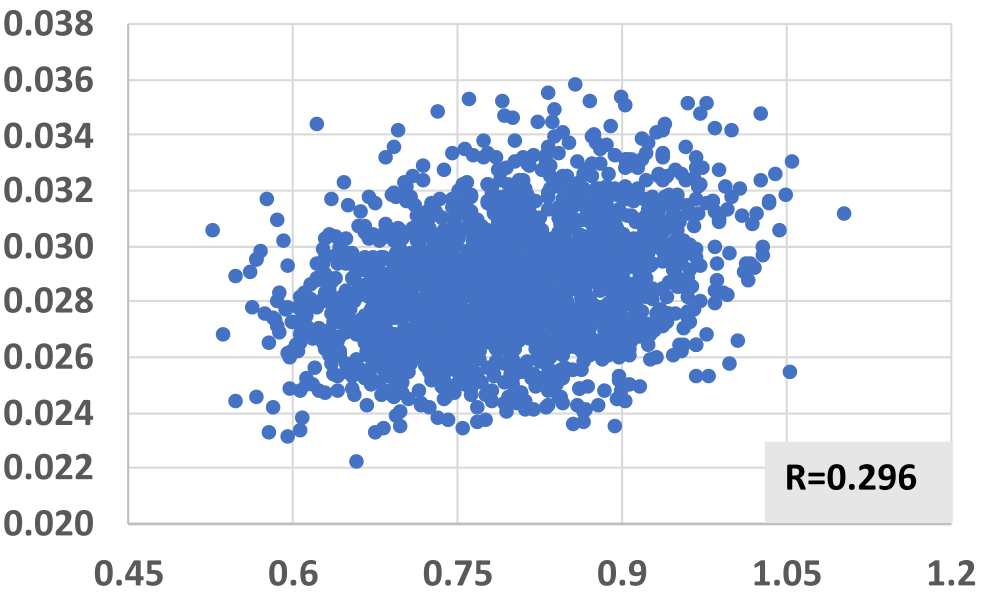} & \includegraphics[width=0.3\linewidth]{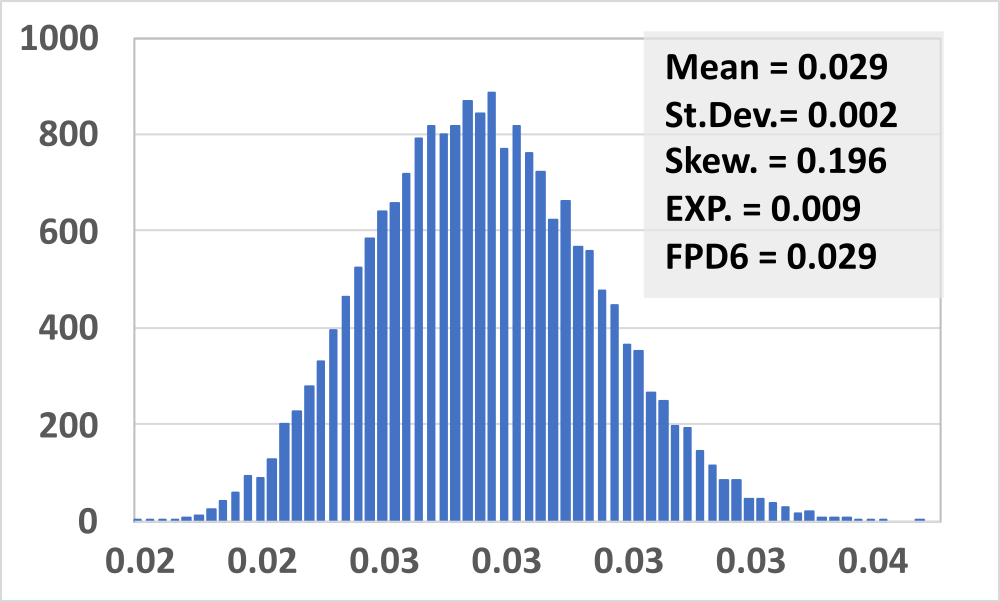}  
\\ \hline
$^{48}$Ca Occ(Np3) & \includegraphics[width=0.3\linewidth]{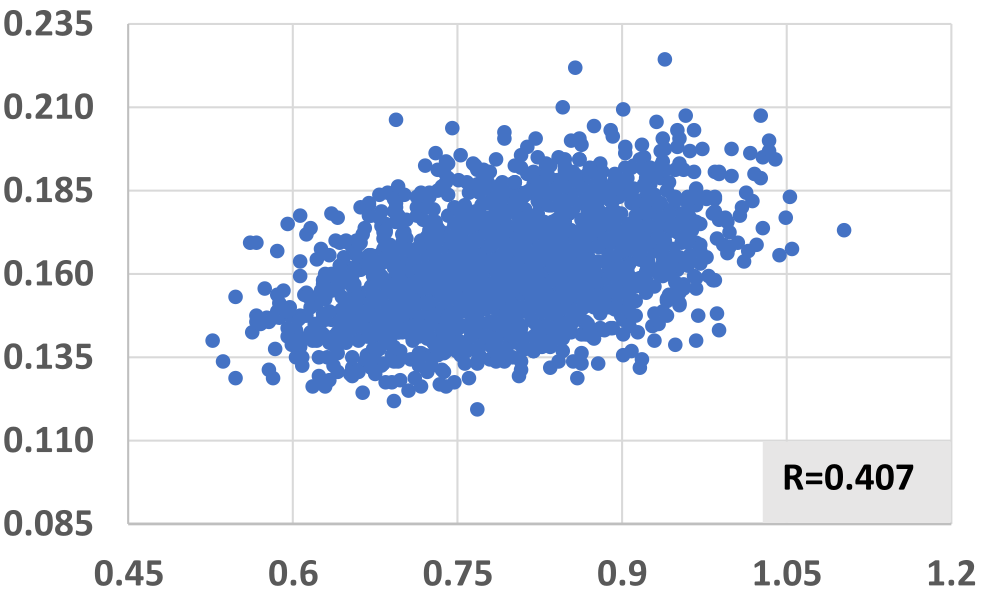} & \includegraphics[width=0.3\linewidth]{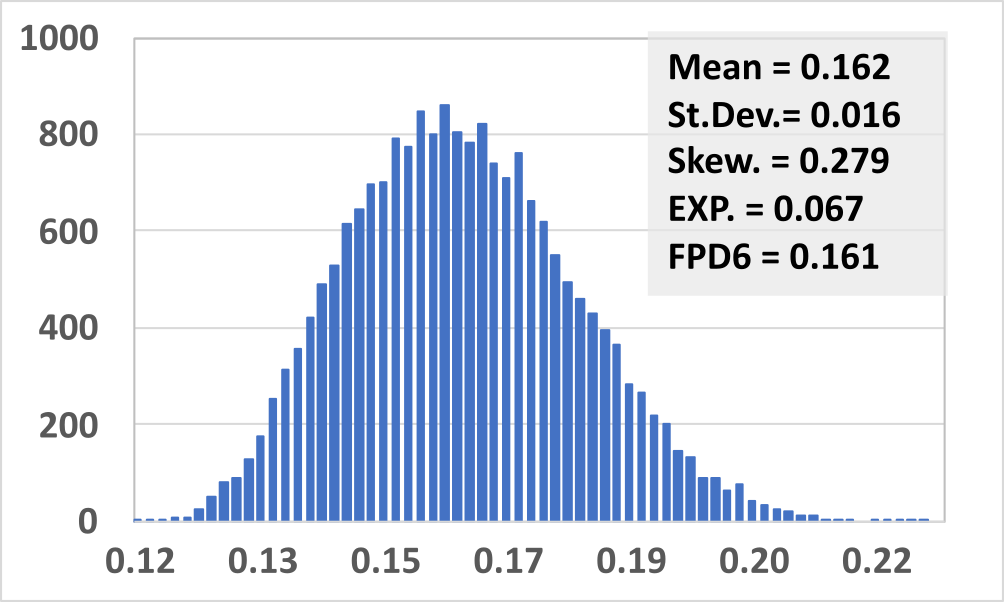}  
\\ \hline
\end{tabular}
\caption{(Color online) Correlations scattered plots and PDFs for the FPD6 starting Hamiltonian. }
\label{tab:parent-fpd6}
\end{table}
\begin{table}[htbp]
\centering
\begin{tabular}{c|c|c}
\hline
Observable & Correlation & PDF\\ \hline
$^{48}$Ti B(E2)$(\uparrow)$ & \includegraphics[width=0.3\linewidth]{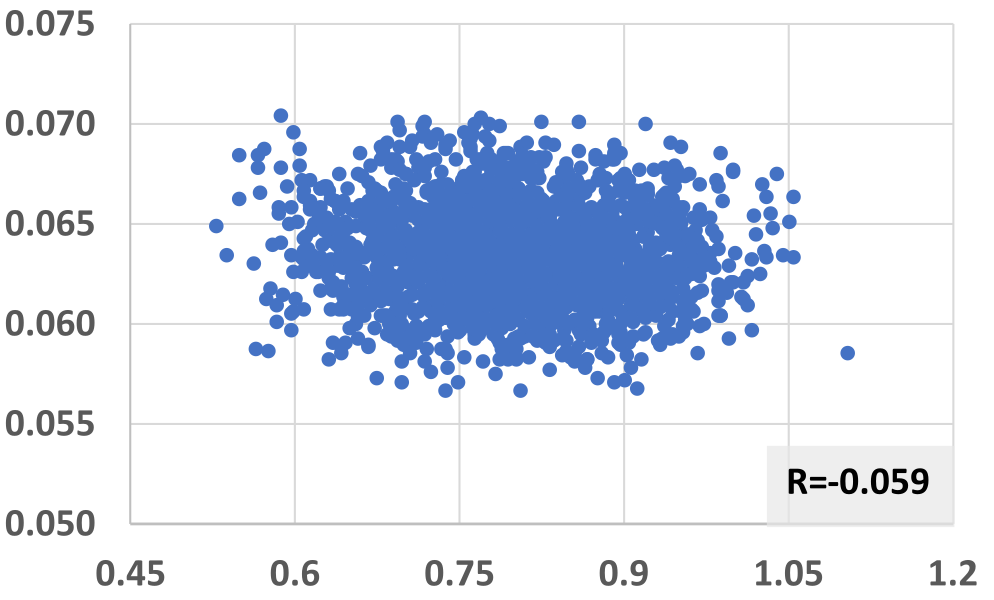} & \includegraphics[width=0.3\linewidth]{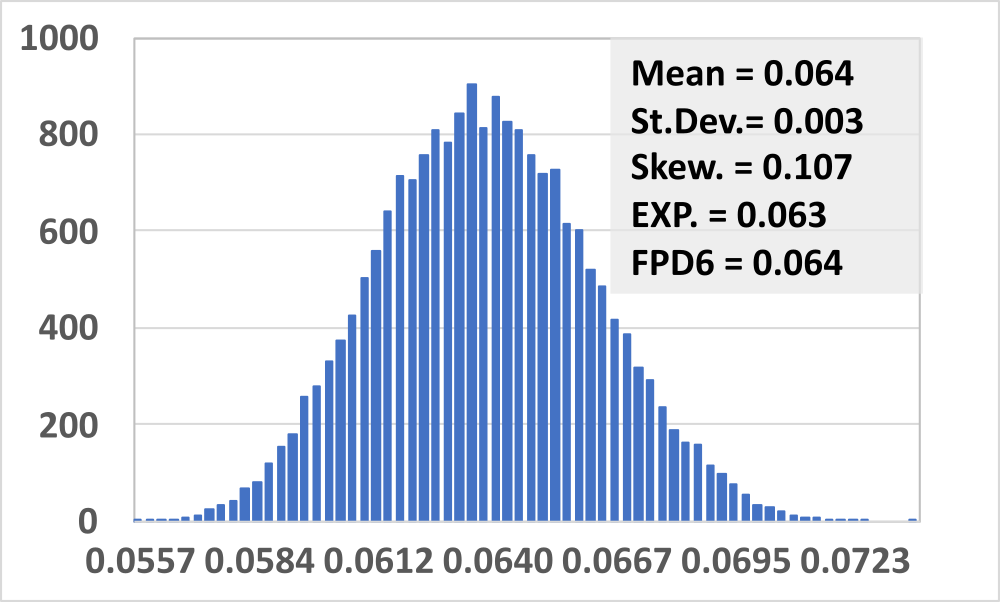} 
\\ \hline
$^{48}$Ti $2^{+}$ & \includegraphics[width=0.3\linewidth]{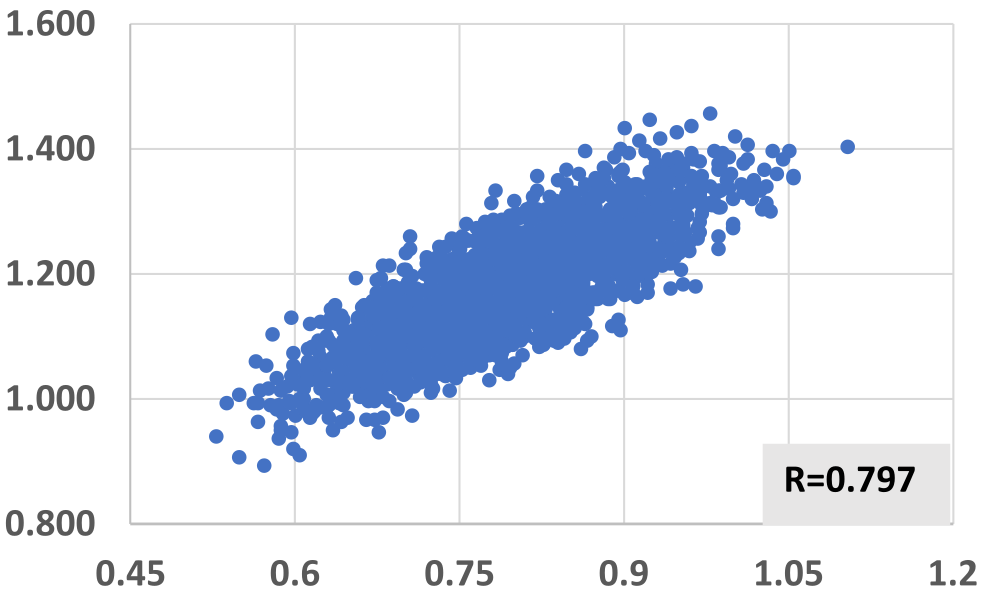} & 
 \includegraphics[width=0.3\linewidth]{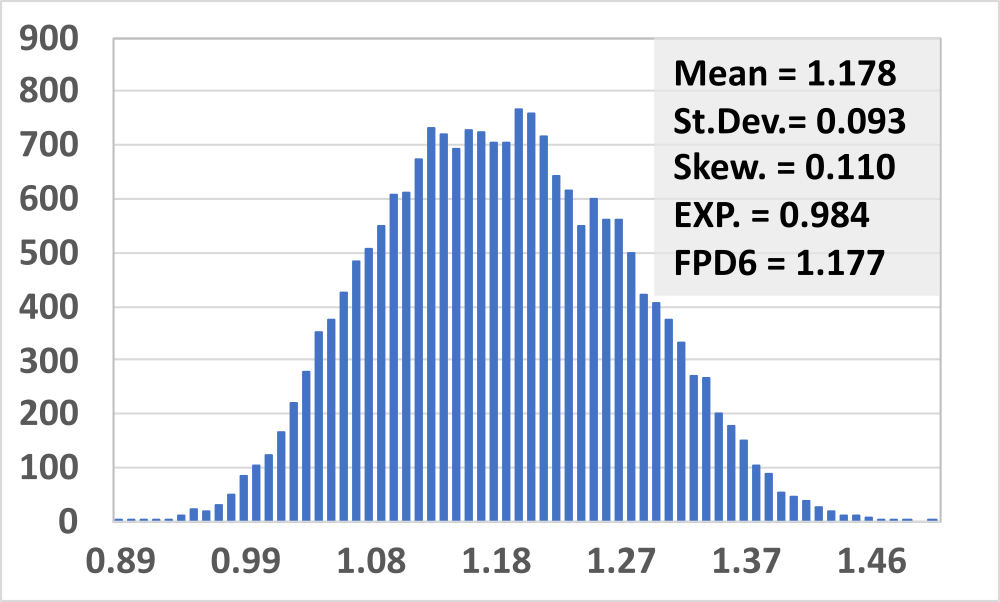} 
\\ \hline
$^{48}$Ti $4^{+}$ & \includegraphics[width=0.3\linewidth]{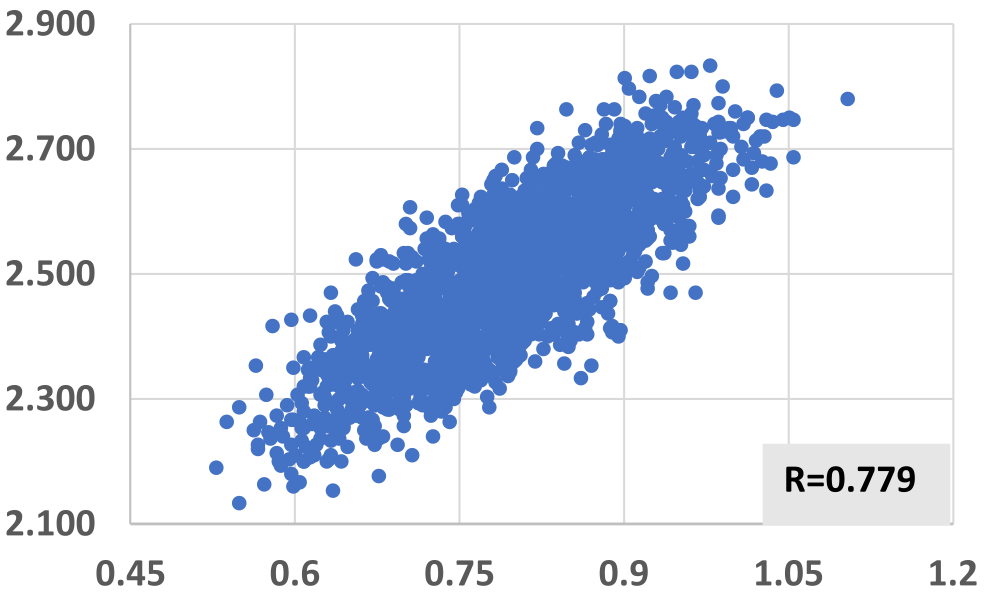} & \includegraphics[width=0.3\linewidth]{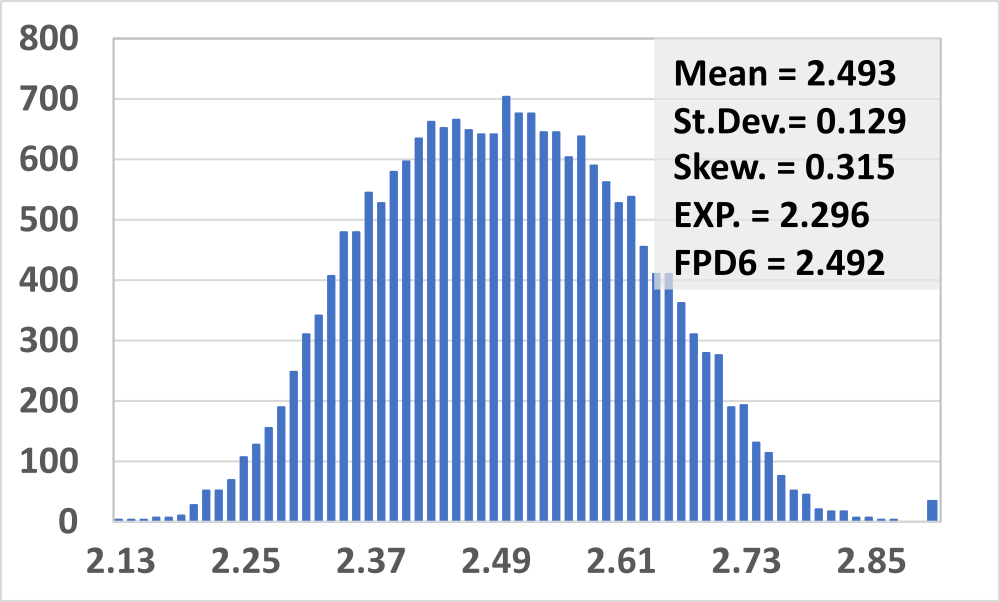}  
\\ \hline
$^{48}$Ti $6^{+}$ & \includegraphics[width=0.3\linewidth]{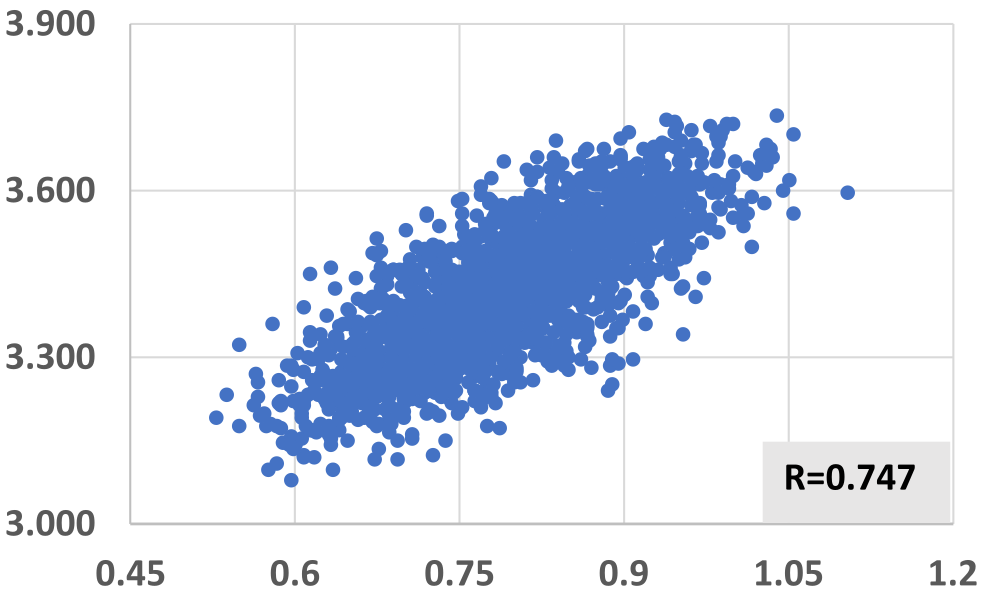} & \includegraphics[width=0.3\linewidth]{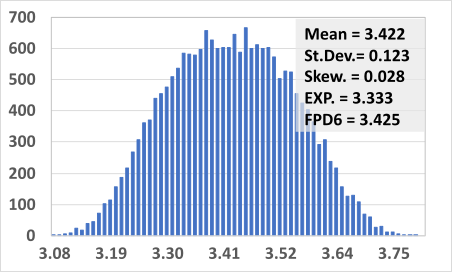}  
\\ \hline
$^{48}$Ti Occ(Nf5) & \includegraphics[width=0.3\linewidth]{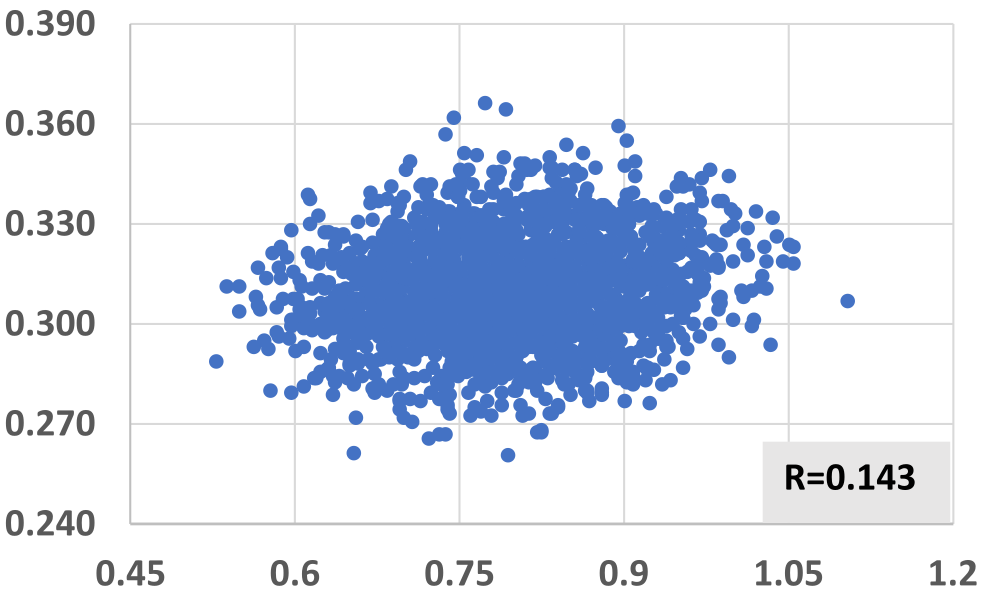} & \includegraphics[width=0.3\linewidth]{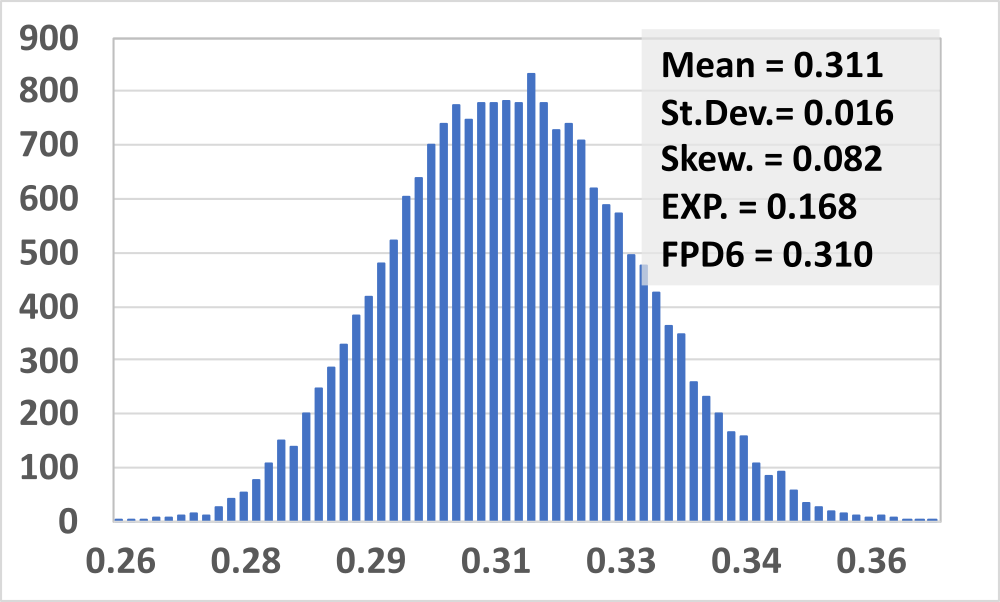}  
\\ \hline
$^{48}$Ti Occ(Nf7) & \includegraphics[width=0.3\linewidth]{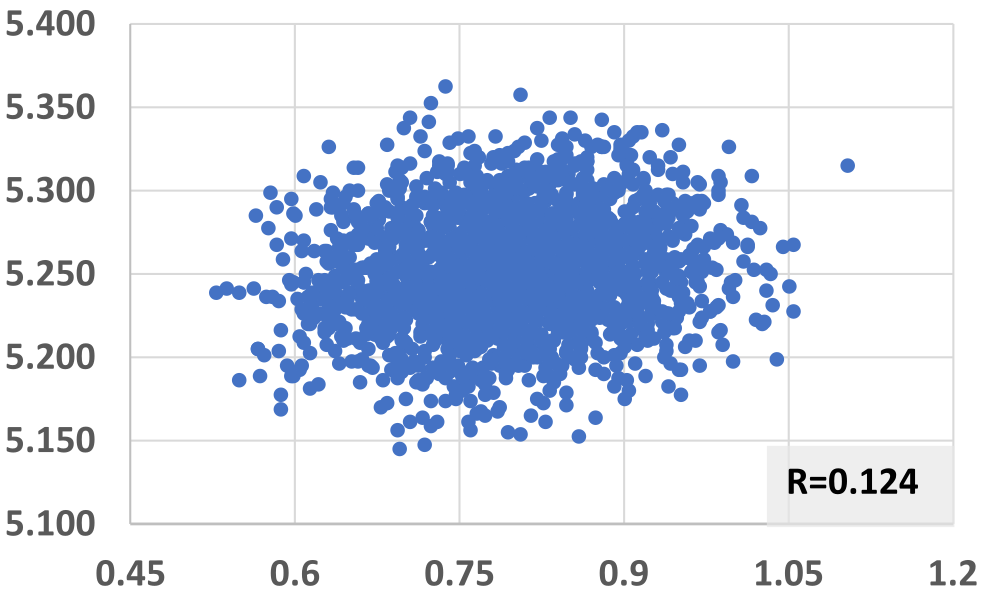} & \includegraphics[width=0.3\linewidth]{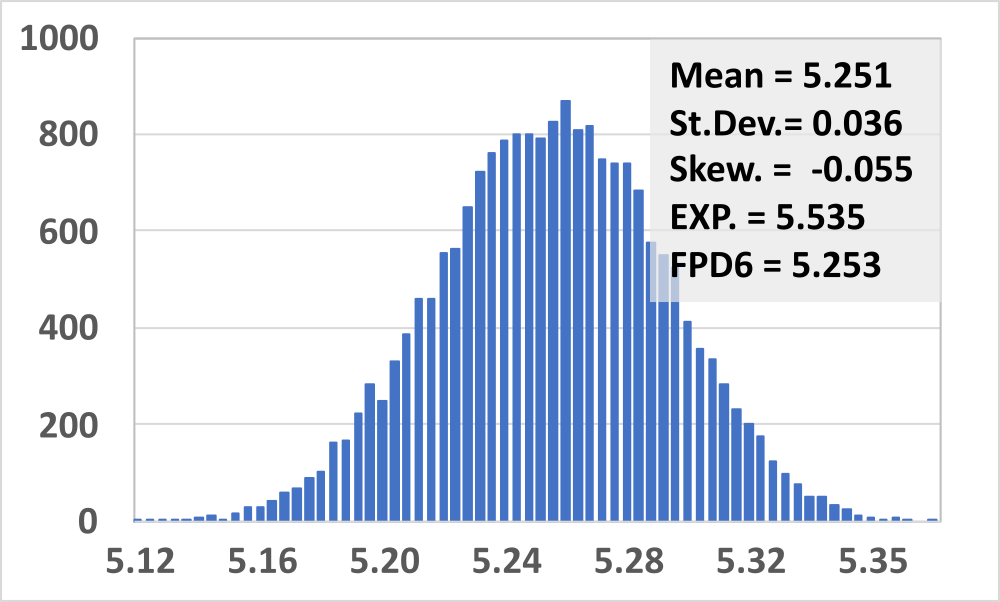}  
\\ \hline
$^{48}$Ti Occ(Np1) & \includegraphics[width=0.3\linewidth]{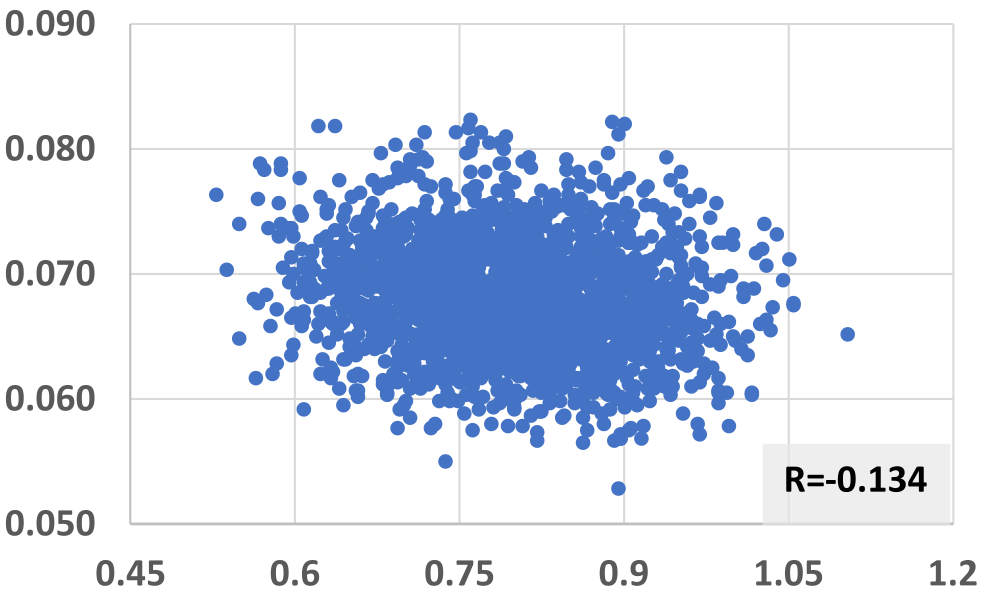} & \includegraphics[width=0.3\linewidth]{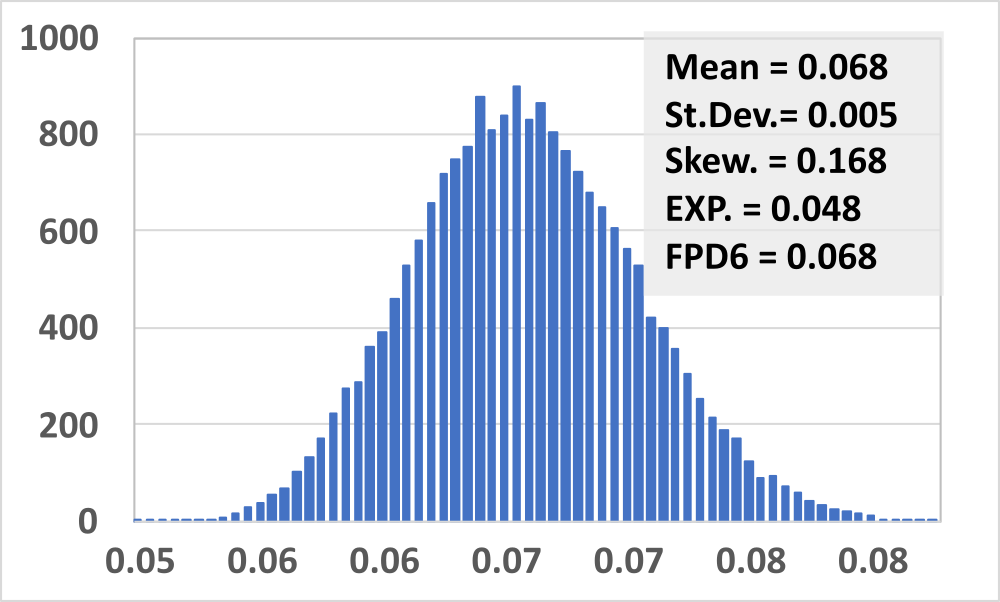}  
\\ \hline
$^{48}$Ti Occ(Np3) & \includegraphics[width=0.3\linewidth]{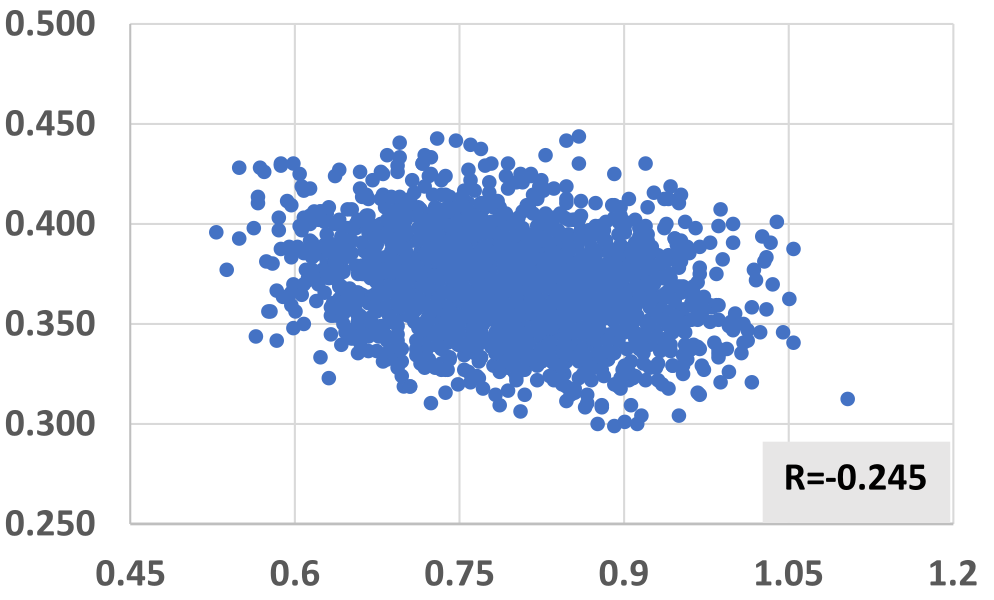} & \includegraphics[width=0.3\linewidth]{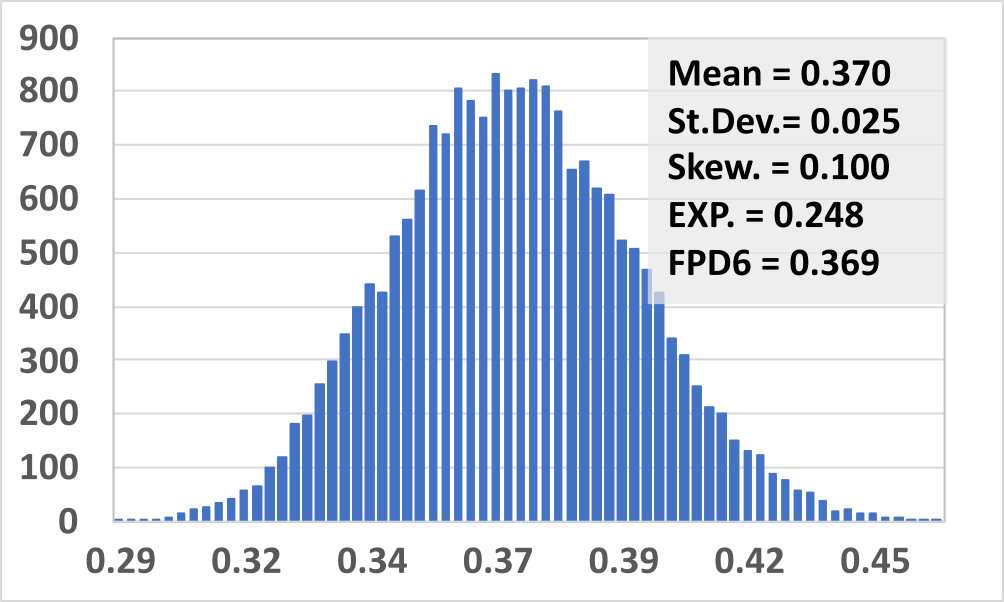}  
\\ \hline
$^{48}$Ti Occ(Pf5) & \includegraphics[width=0.3\linewidth]{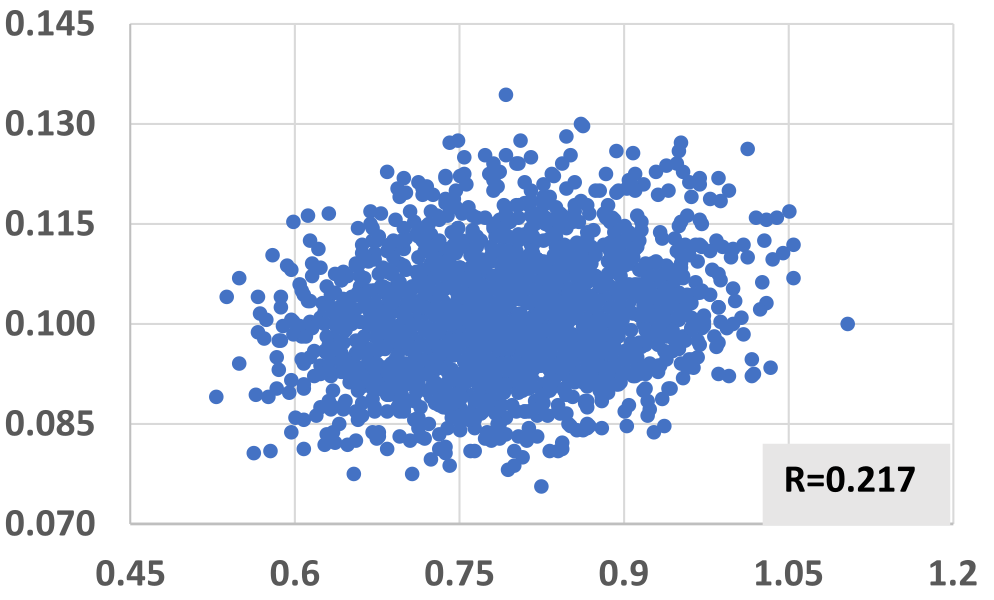} & \includegraphics[width=0.3\linewidth]{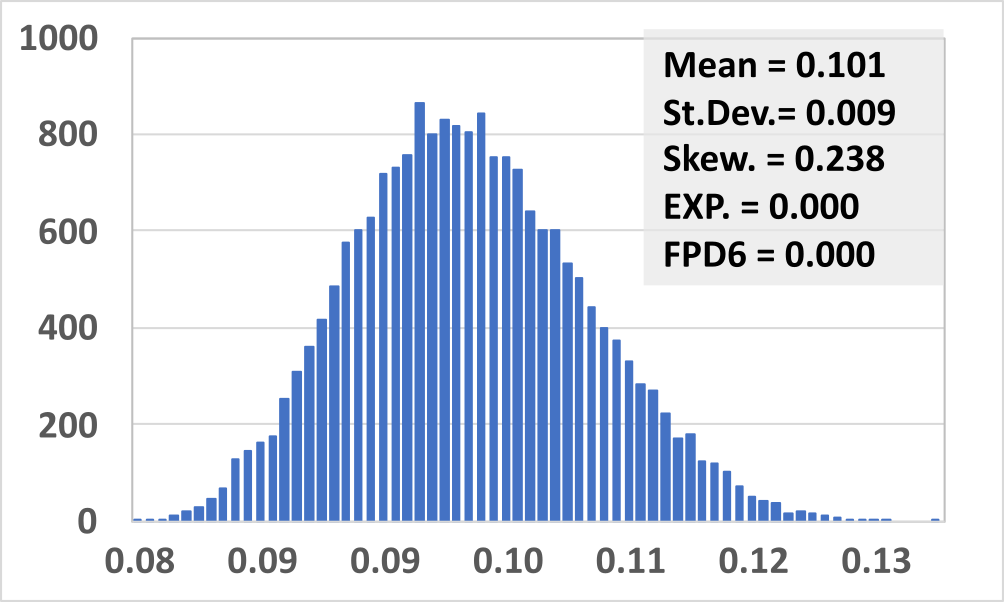}  
\\ \hline
$^{48}$Ti Occ(Pf7) & \includegraphics[width=0.3\linewidth]{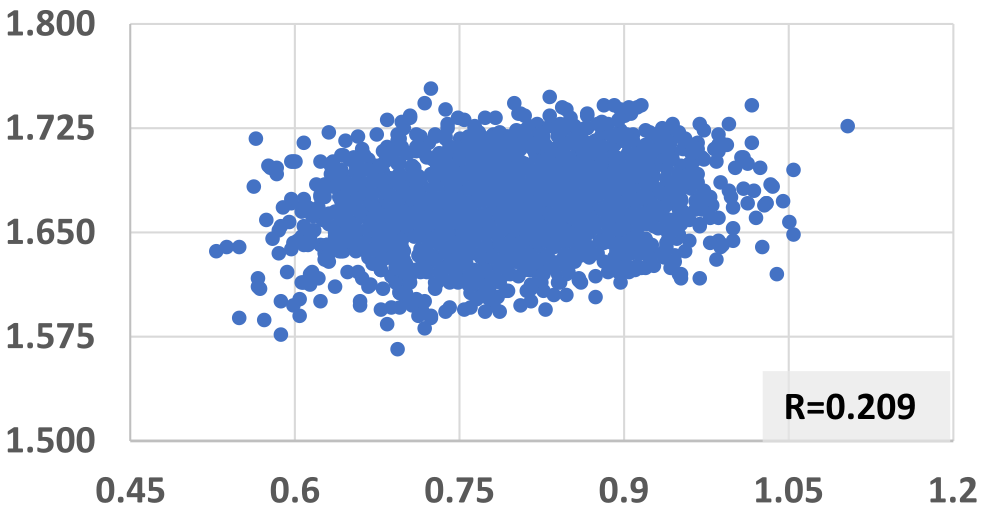} & \includegraphics[width=0.3\linewidth]{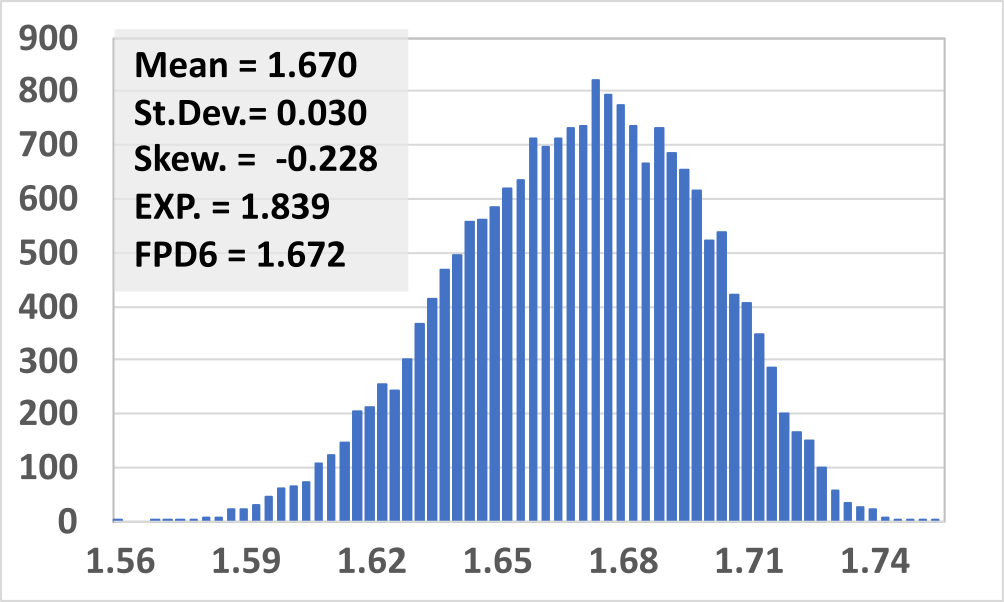}  
\\ \hline
$^{48}$Ti Occ(Pp1) & \includegraphics[width=0.3\linewidth]{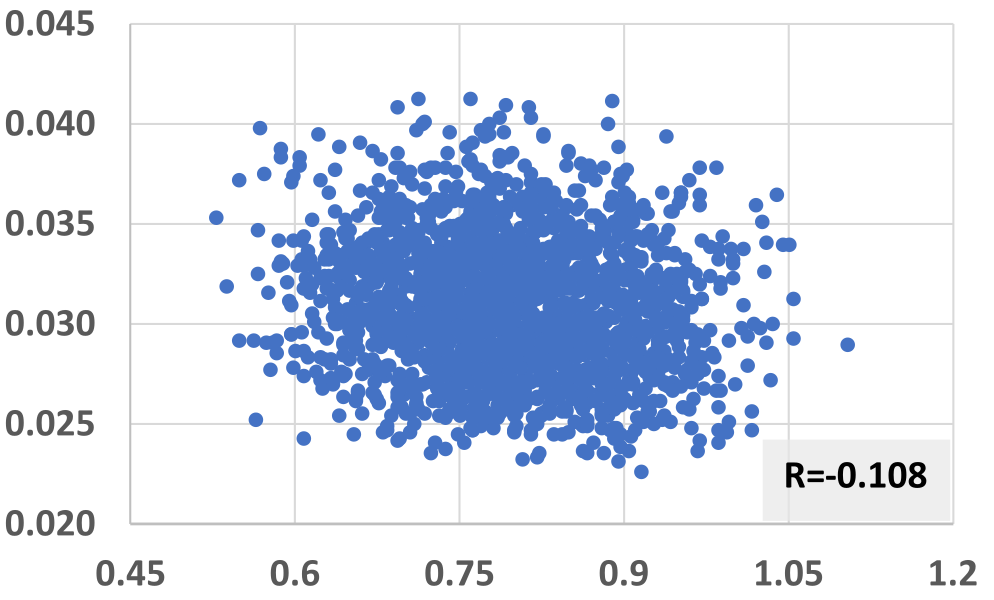} & \includegraphics[width=0.3\linewidth]{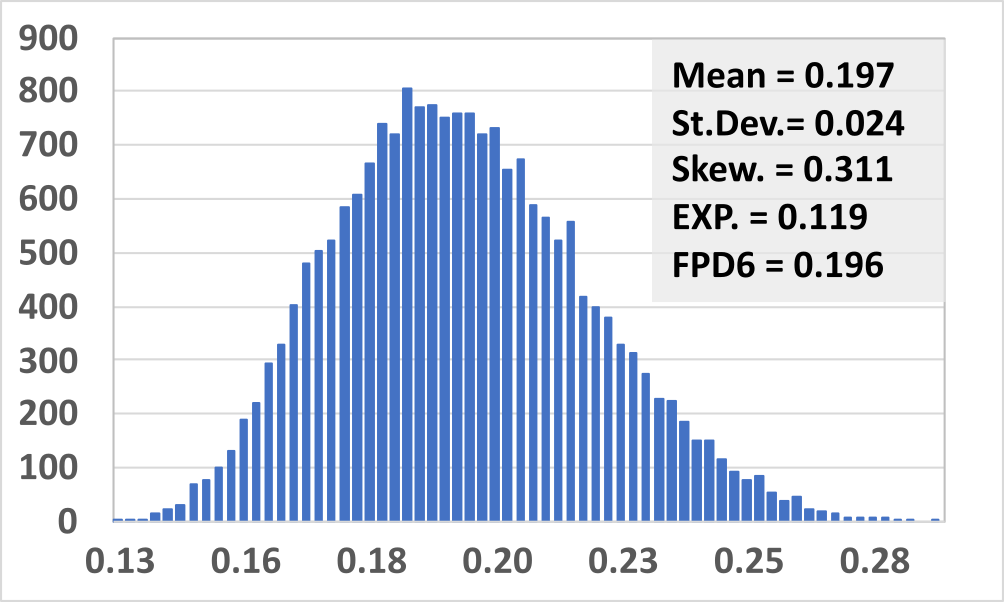}  
\\ \hline
$^{48}$Ti Occ(Pp3) & \includegraphics[width=0.3\linewidth]{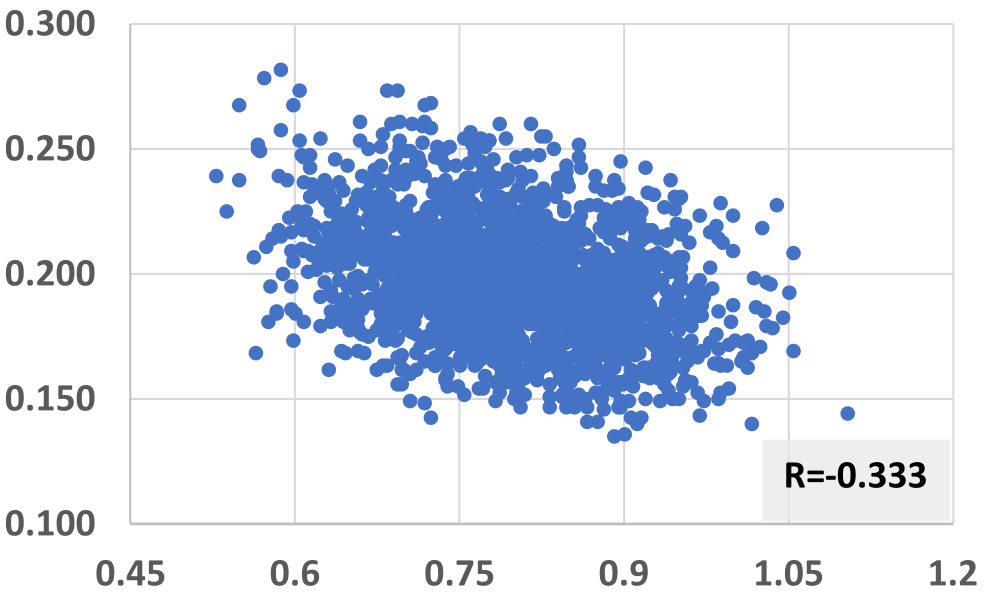} & \includegraphics[width=0.3\linewidth]{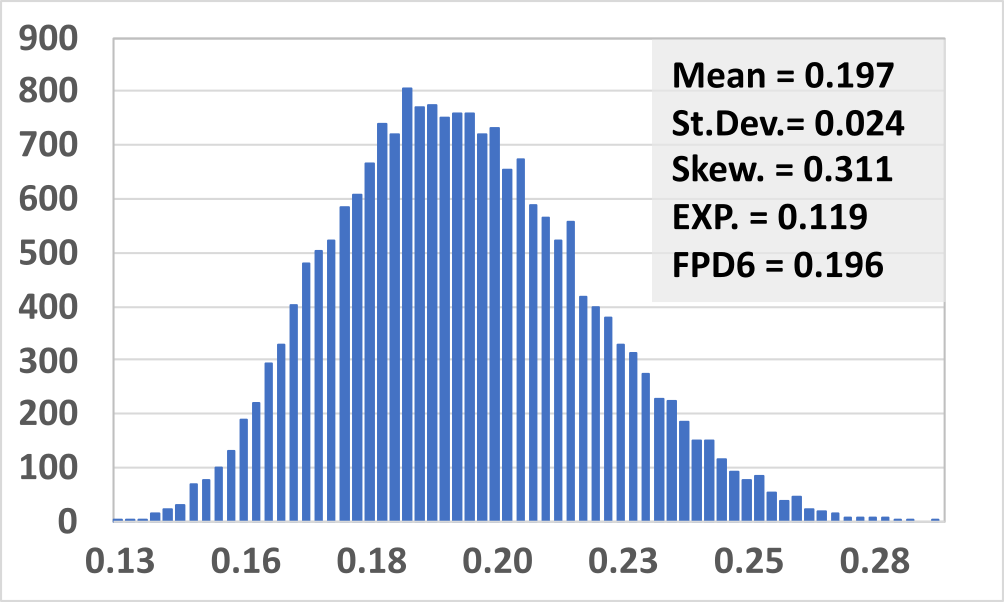}  
\\ \hline
\end{tabular}
\caption{(Color online) Continuation of Table \ref{tab:parent-fpd6}. }
\label{tab:daughter-fpd6}
\end{table}
\begin{table}[htbp]
\centering
\begin{tabular}{c|c|c}
\hline
Observable & Correlation & PDF\\ \hline
$0\nu\beta\beta$ NME & \includegraphics[width=0.32\linewidth]{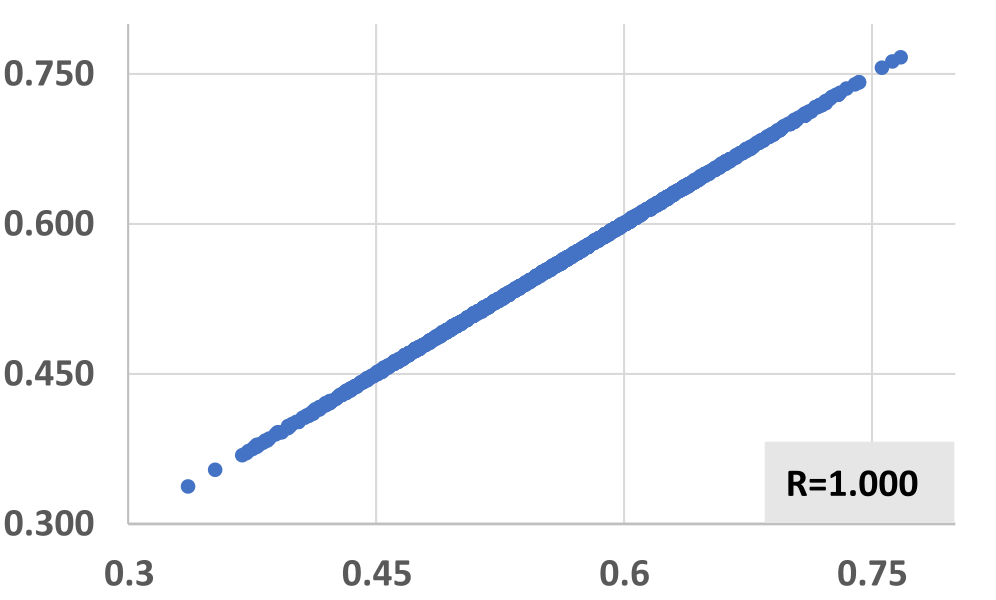} & \includegraphics[width=0.32\linewidth]{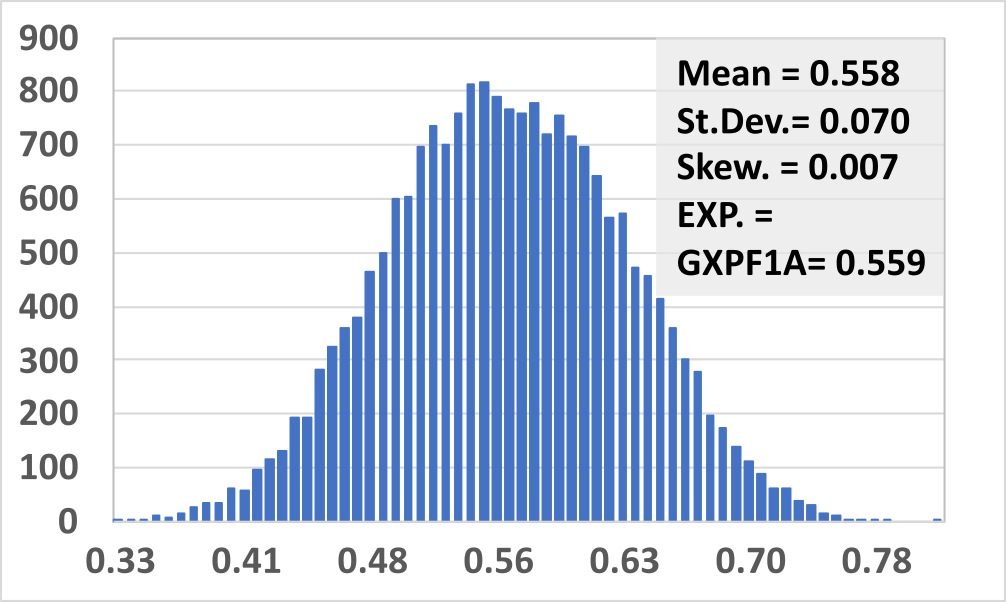} \\
\hline
$2\nu\beta\beta$ NME & \includegraphics[width=0.32\linewidth]{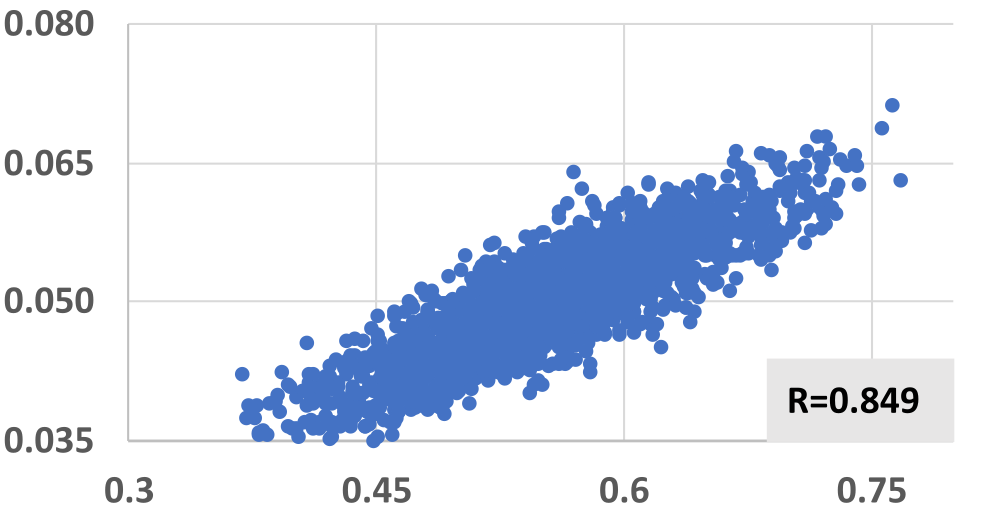} & \includegraphics[width=0.32\linewidth]{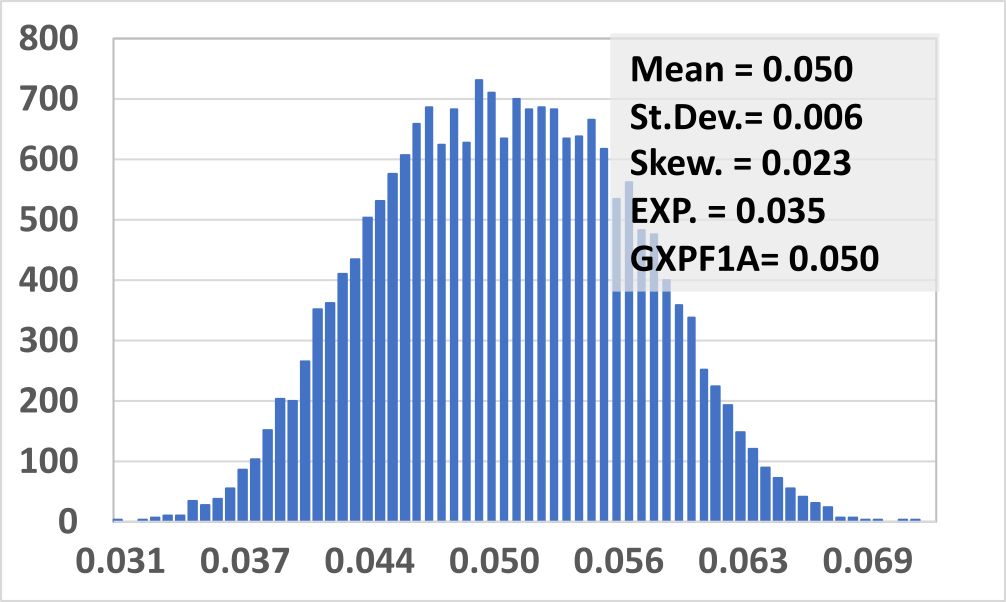}  
\\ \hline
$^{48}$Ca $\rightarrow$ $^{48}$Sc GT & \includegraphics[width=0.32\linewidth]{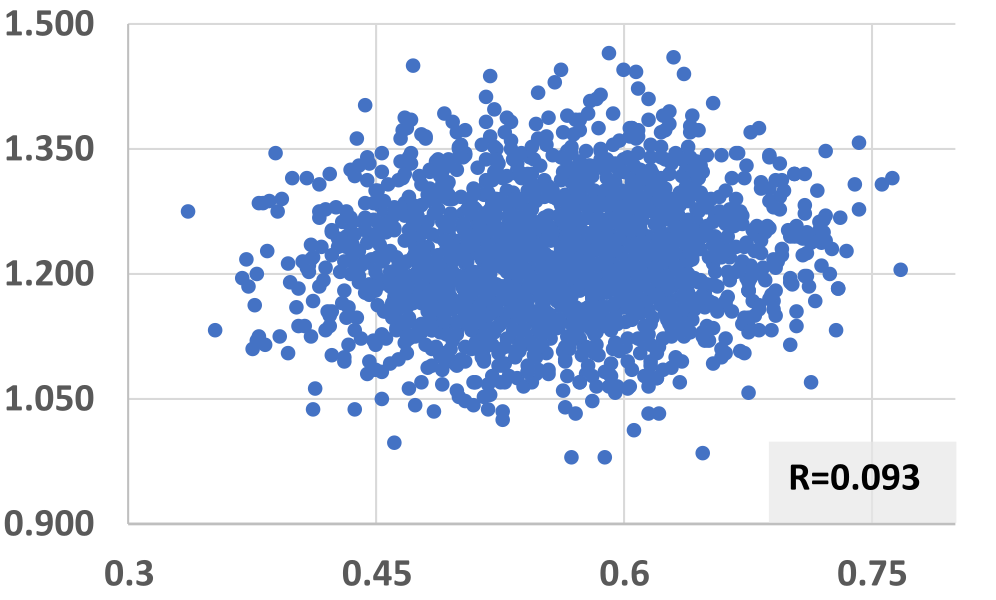} & \includegraphics[width=0.32\linewidth]{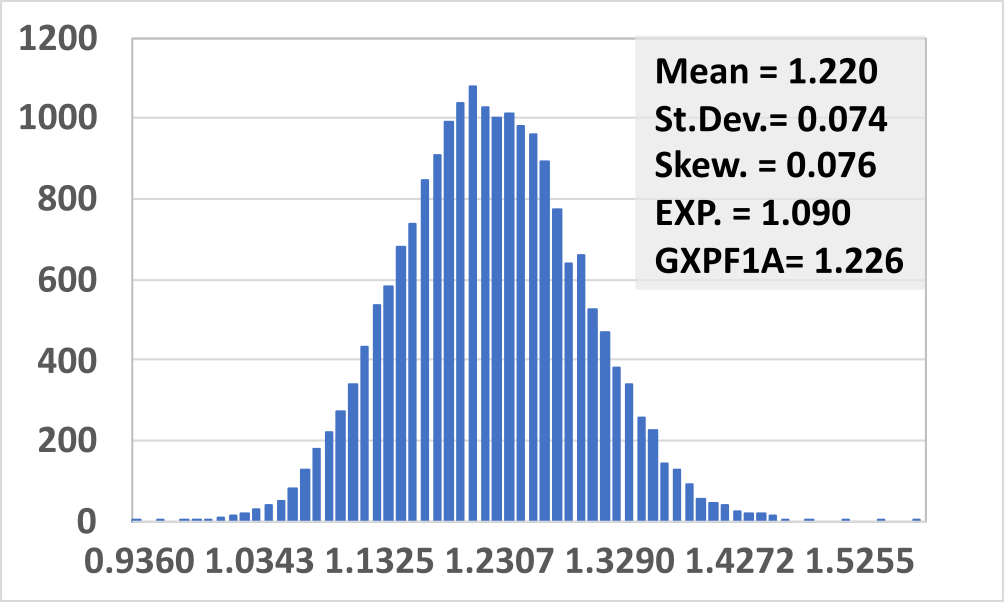}  
\\ \hline
$^{48}$Ti $\rightarrow$ $^{48}$Sc GT & \includegraphics[width=0.32\linewidth]{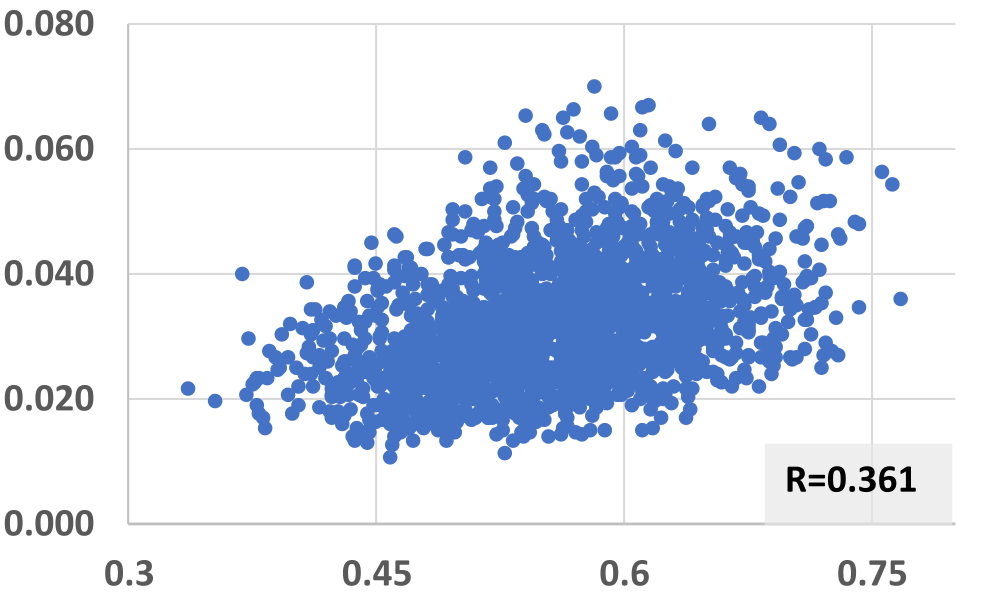} & \includegraphics[width=0.32\linewidth]{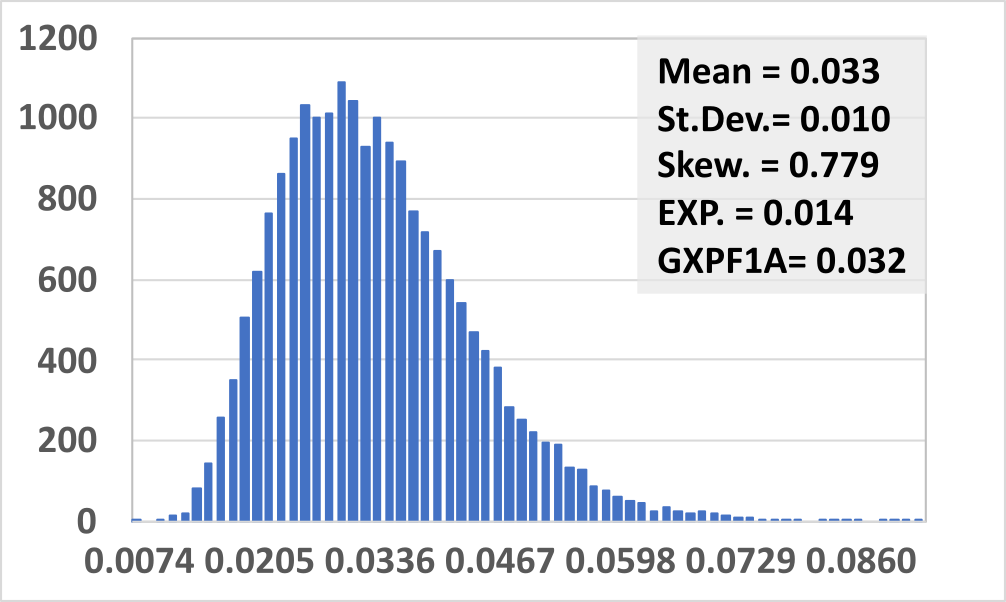}  
\\ \hline
$^{48}$Ca B(E2)$(\uparrow)$ & \includegraphics[width=0.32\linewidth]{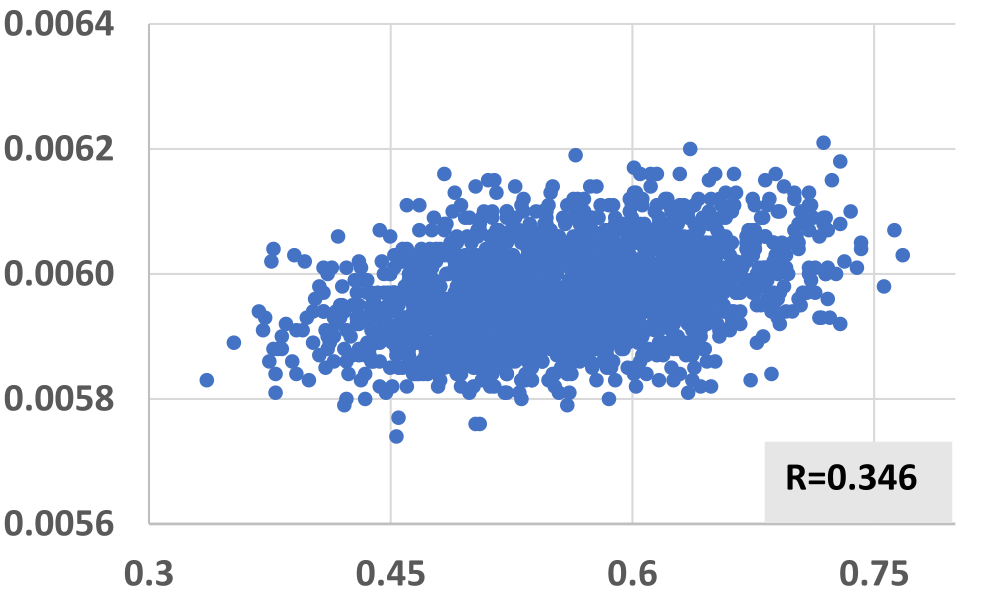} & \includegraphics[width=0.32\linewidth]{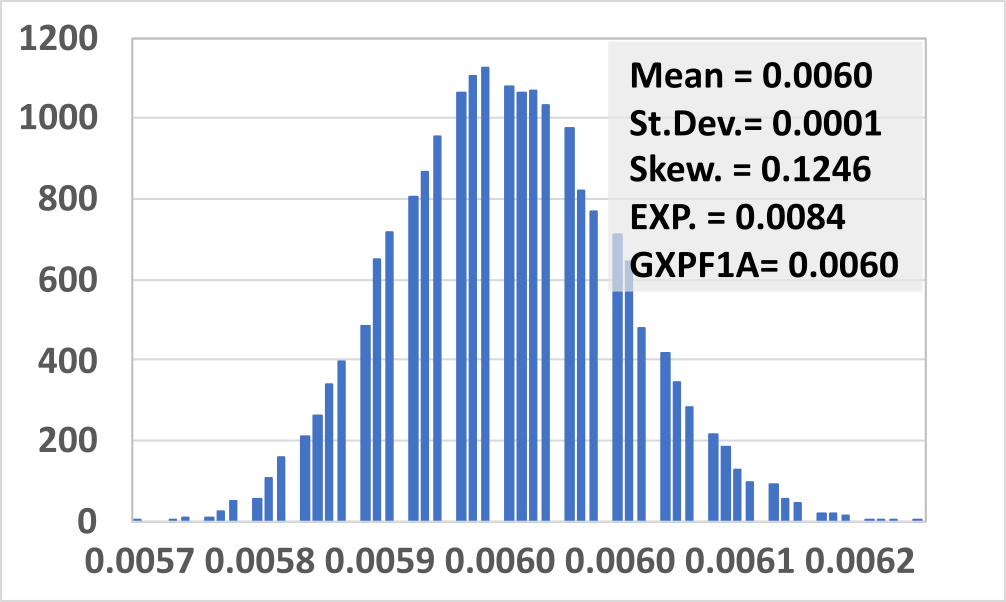} 
\\ \hline
$^{48}$Ca $2^{+}$ & \includegraphics[width=0.32\linewidth]{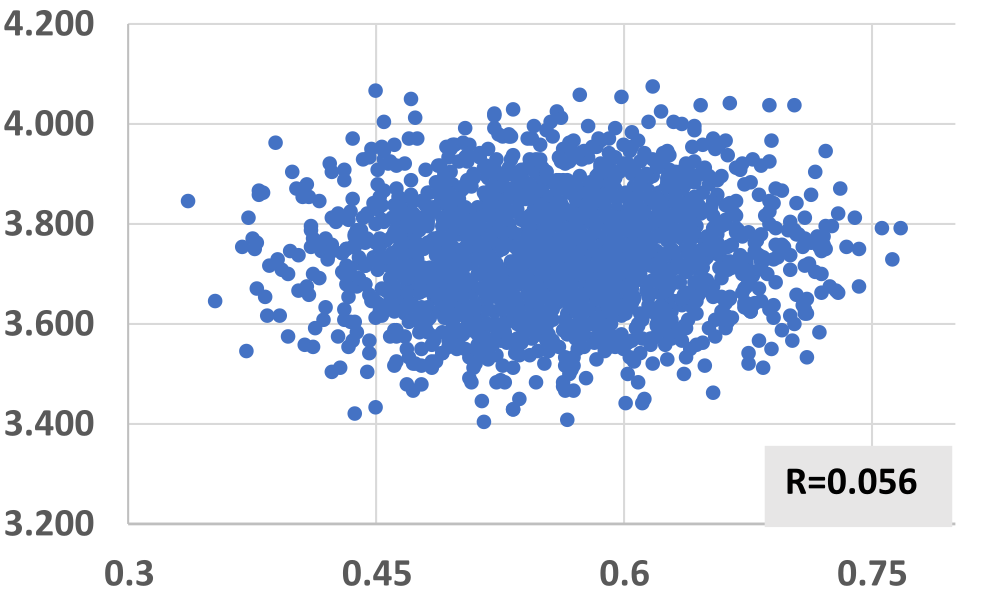} & \includegraphics[width=0.32\linewidth]{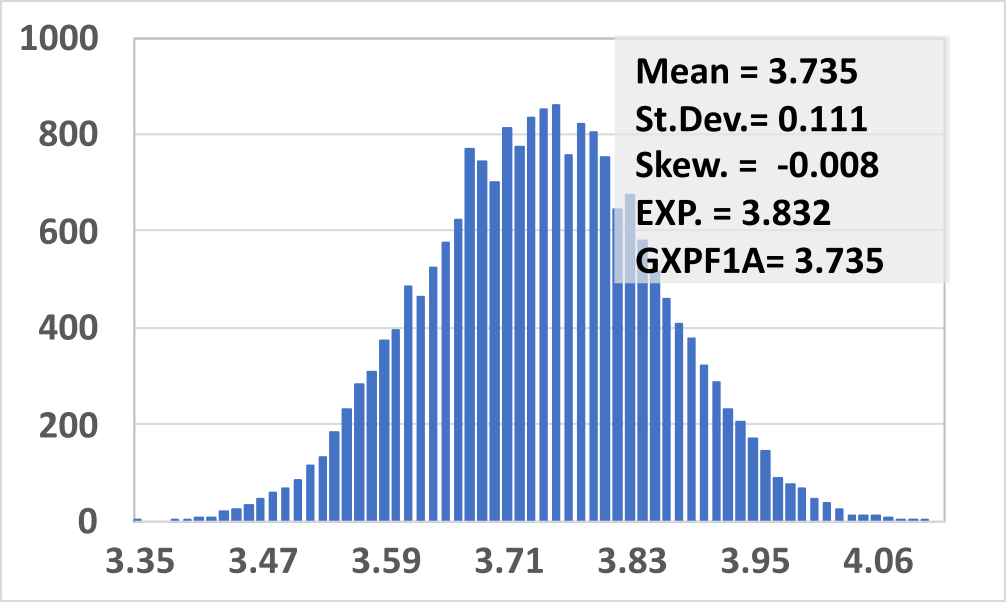}  
\\ \hline
$^{48}$Ca $4^{+}$ & \includegraphics[width=0.32\linewidth]{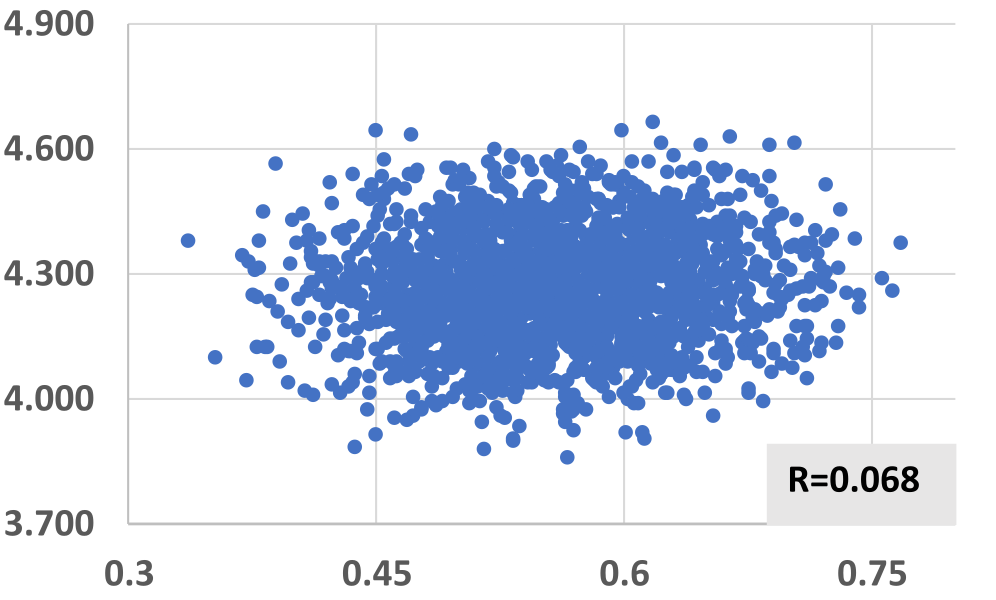} & \includegraphics[width=0.32\linewidth]{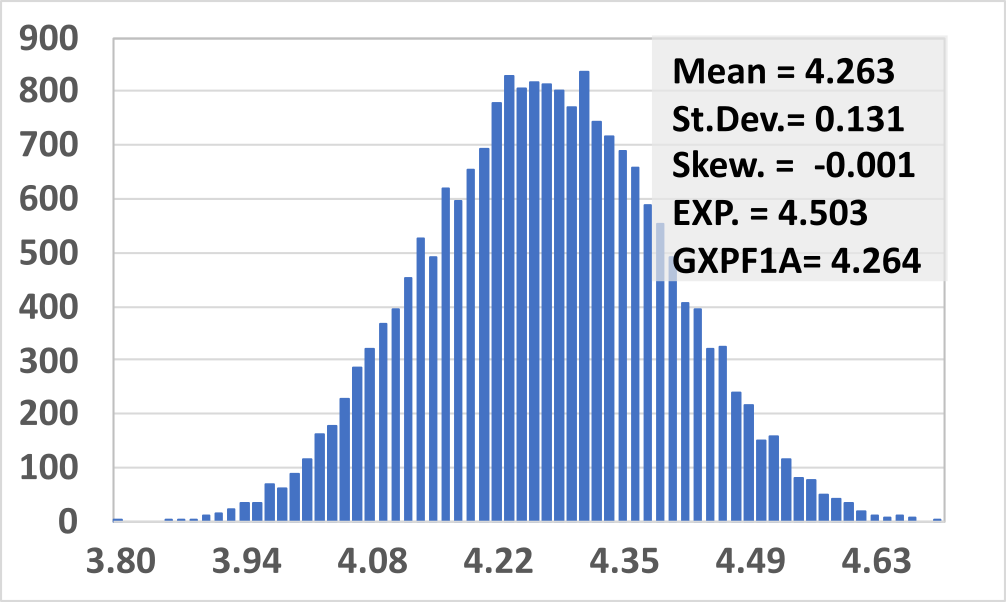}  
\\ \hline
$^{48}$Ca $6^{+}$ & \includegraphics[width=0.32\linewidth]{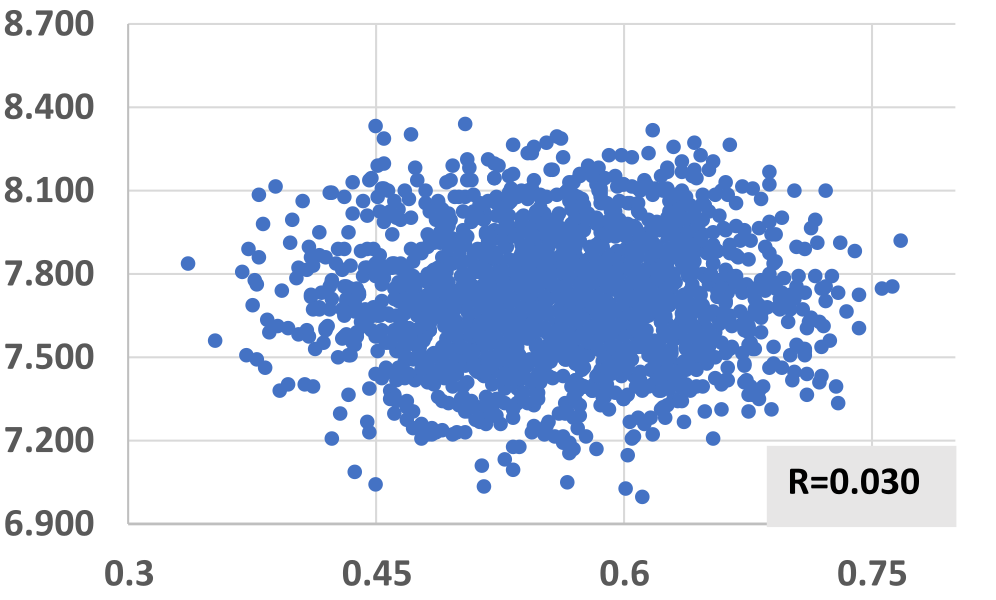} & \includegraphics[width=0.32\linewidth]{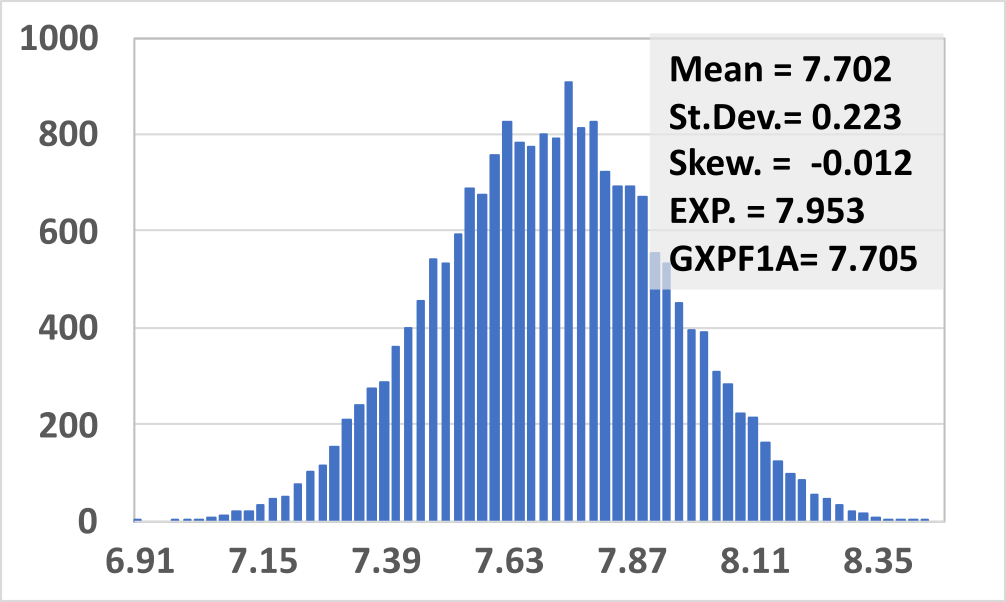}  
\\ \hline
$^{48}$Ca Occ(Nf5) & \includegraphics[width=0.32\linewidth]{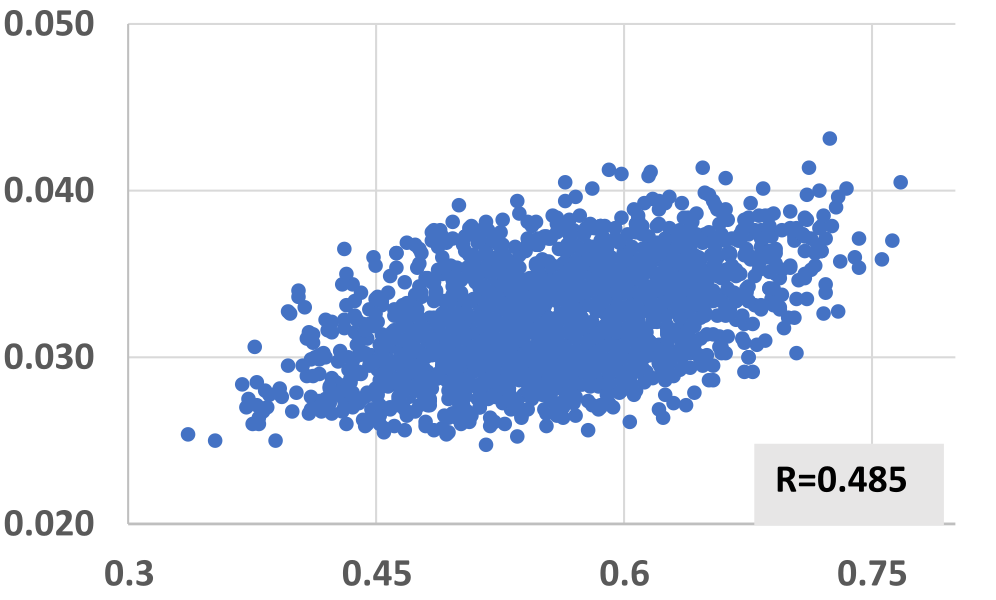} & \includegraphics[width=0.32\linewidth]{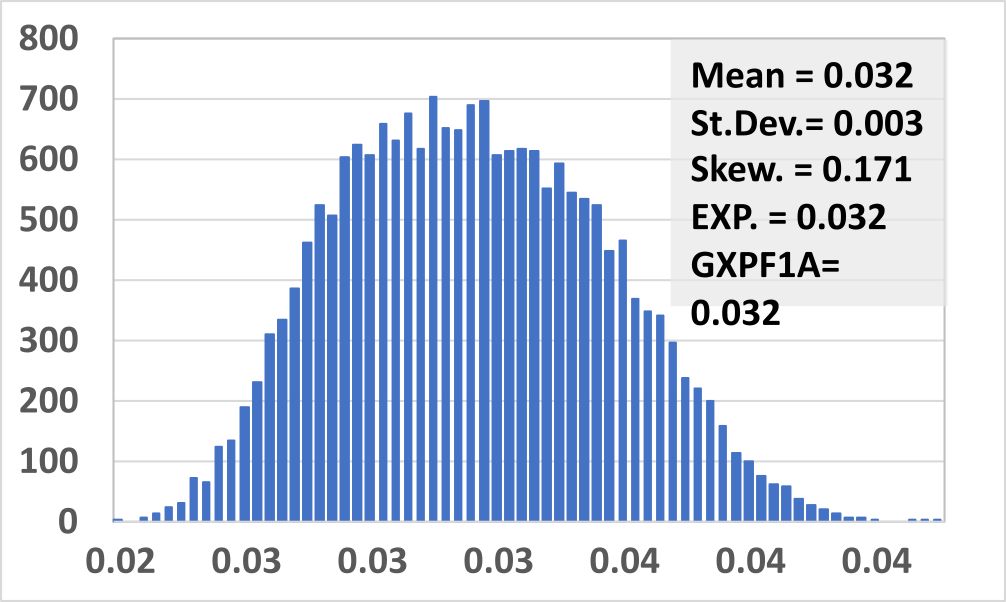}  
\\ \hline
$^{48}$Ca Occ(Nf7) & \includegraphics[width=0.32\linewidth]{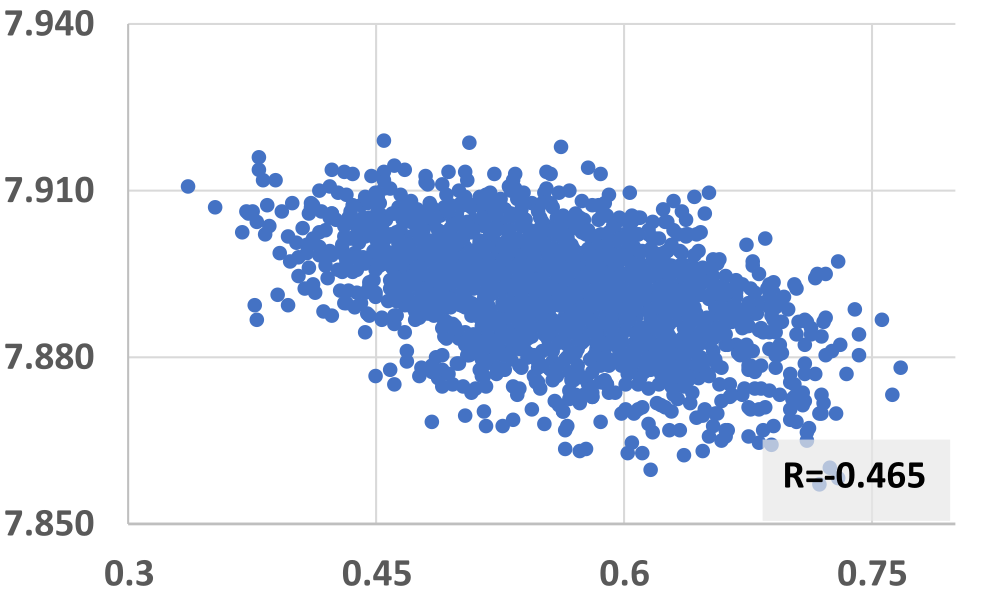} & \includegraphics[width=0.32\linewidth]{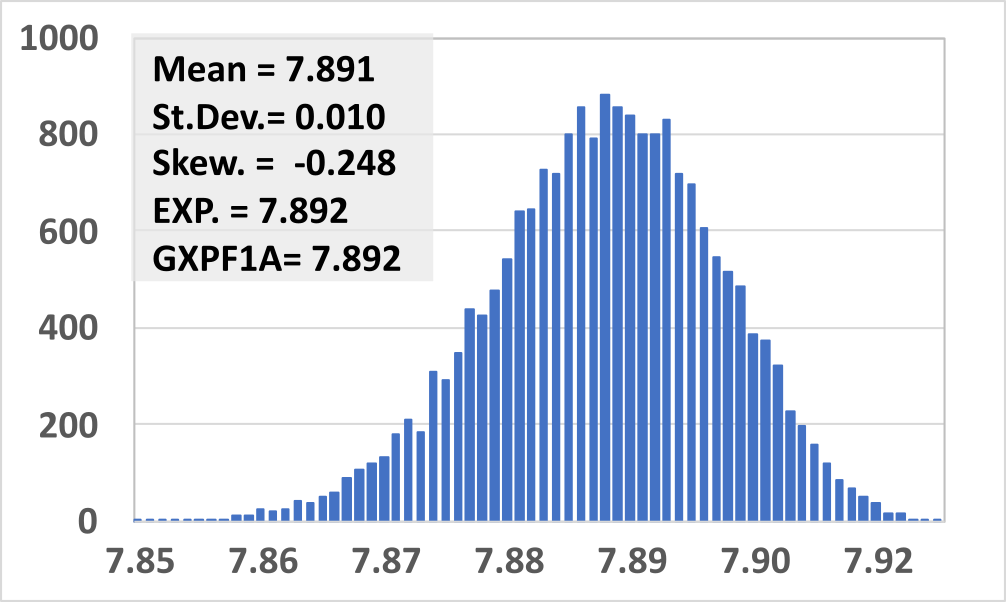}  
\\ \hline
$^{48}$Ca Occ(Np1) & \includegraphics[width=0.32\linewidth]{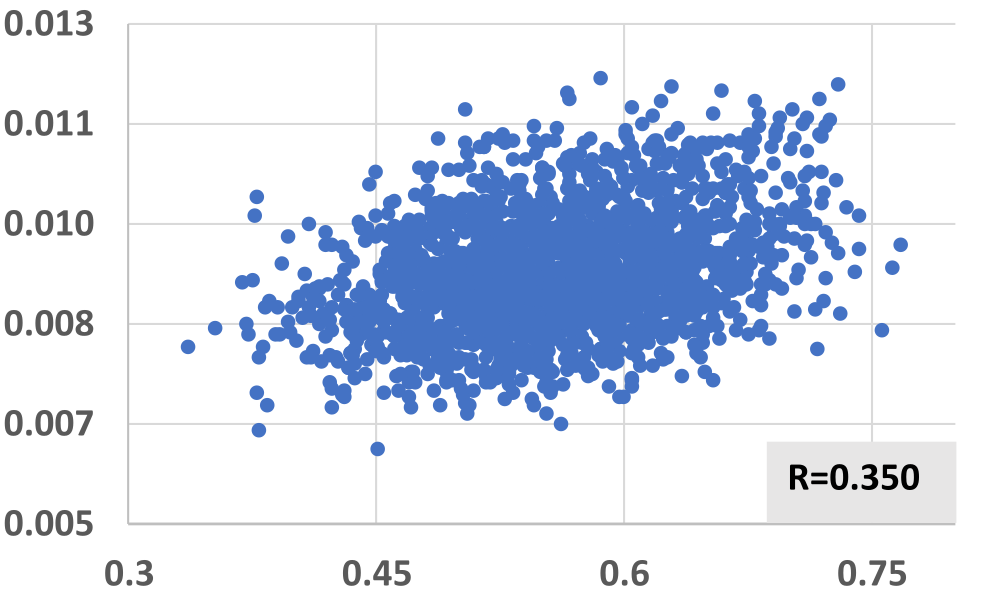} & \includegraphics[width=0.32\linewidth]{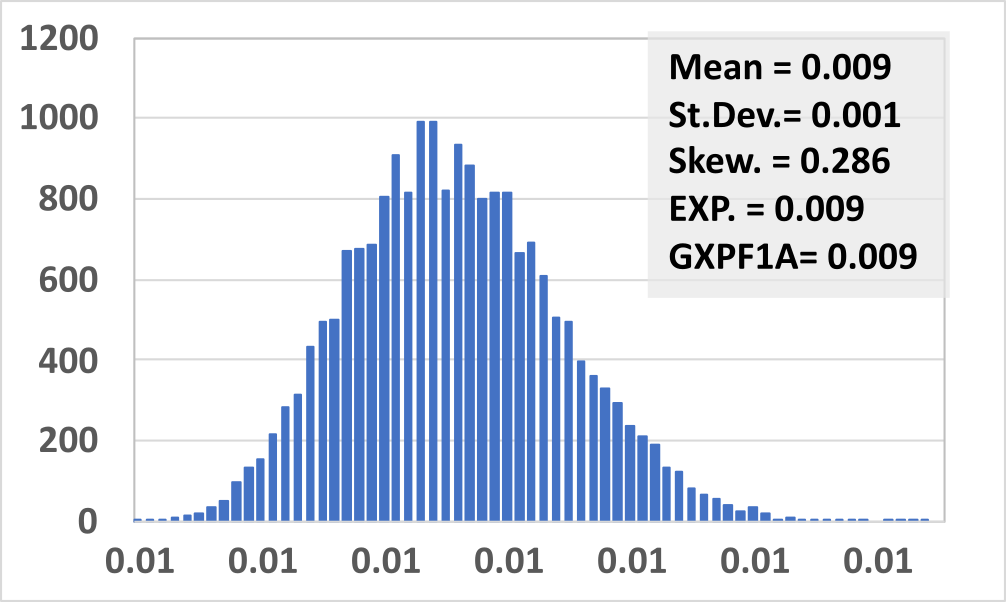}  
\\ \hline
$^{48}$Ca Occ(Np3) & \includegraphics[width=0.32\linewidth]{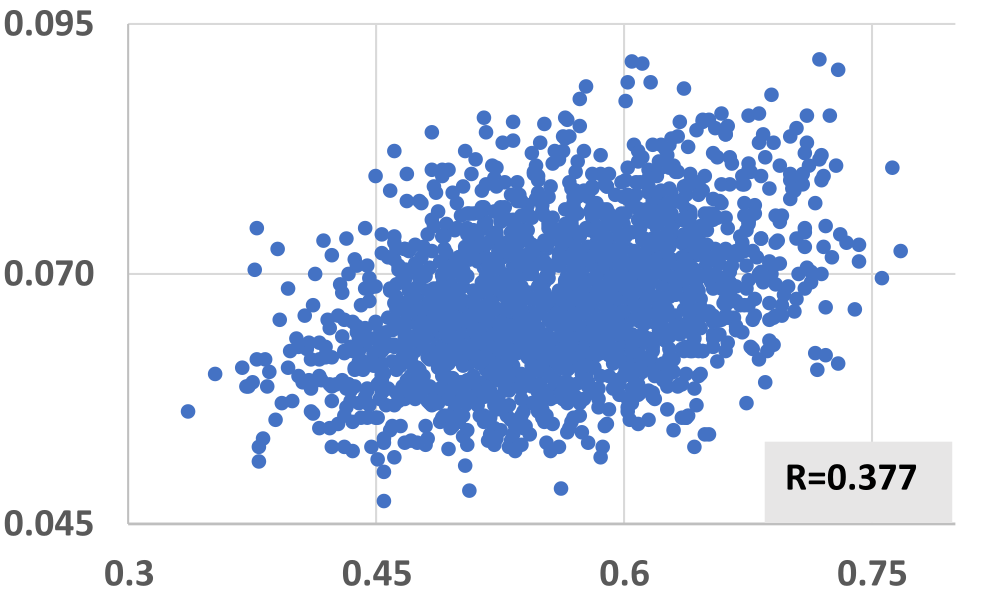} & \includegraphics[width=0.32\linewidth]{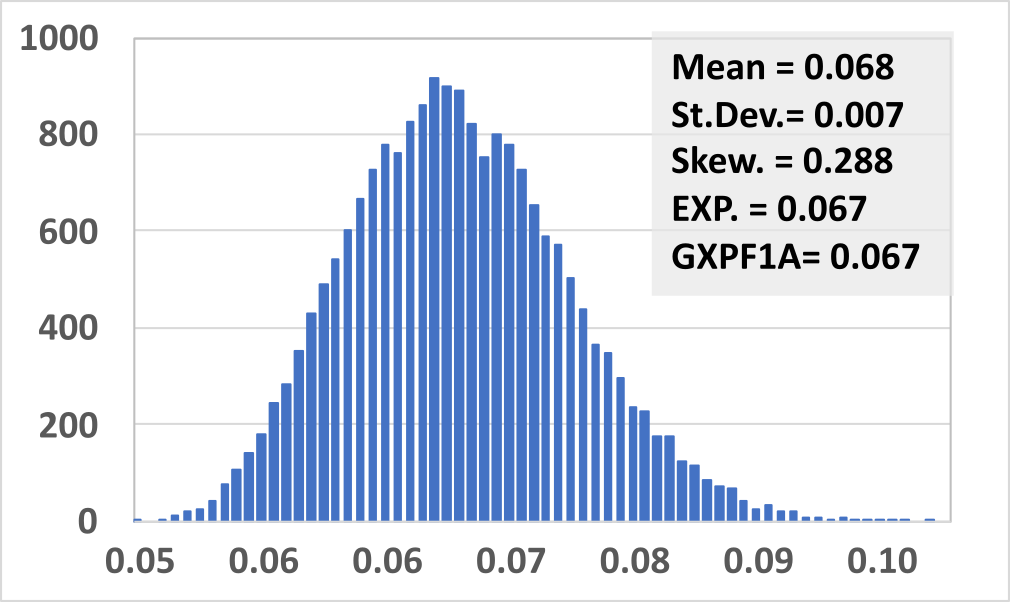}  
\\ \hline
\end{tabular}
 \caption{(Color online) Correlations scattered plots and PDFs for the GXPF1A starting Hamiltonian. }
\label{tab:parent-gxpf1a}
\end{table}

\begin{table}[htbp]
\centering
\begin{tabular}{c|c|c}
\hline
Observable & Correlation & PDF\\ \hline
$^{48}$Ti B(E2)$(\uparrow)$ & \includegraphics[width=0.32\linewidth]{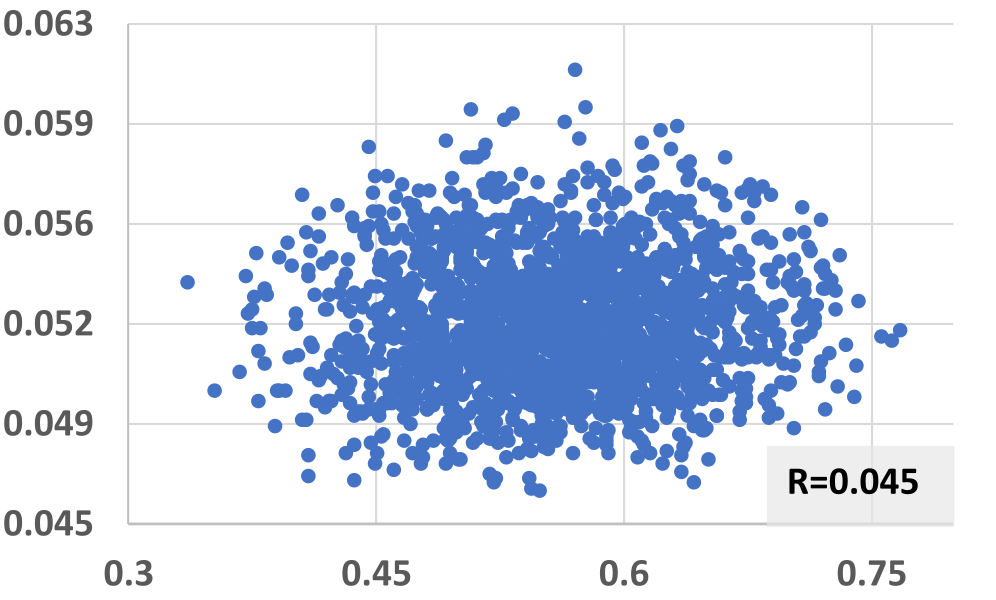} & \includegraphics[width=0.32\linewidth]{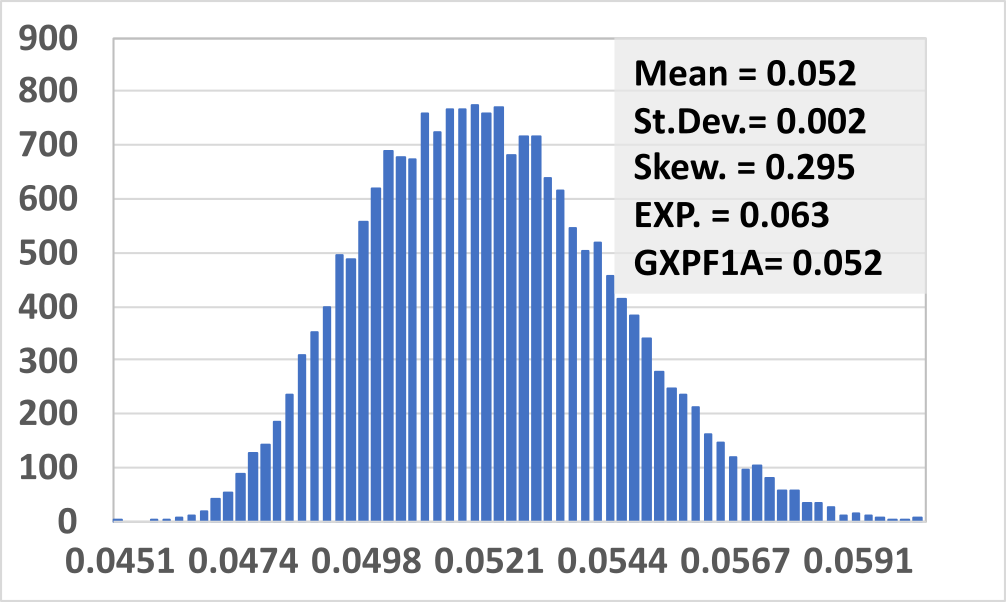} 
\\ \hline
$^{48}$Ti $2^{+}$ & \includegraphics[width=0.32\linewidth]{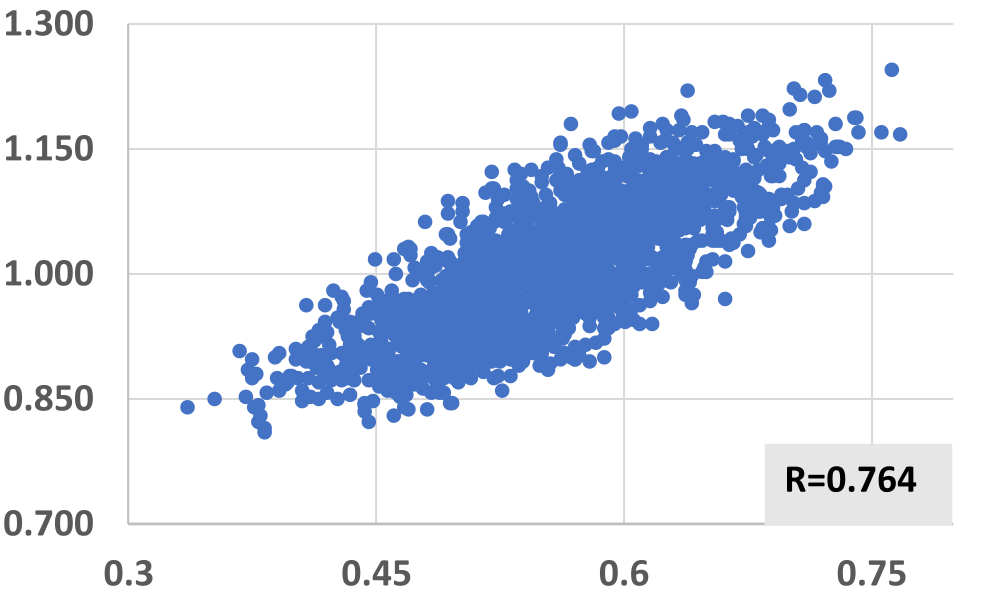} & 
 \includegraphics[width=0.32\linewidth]{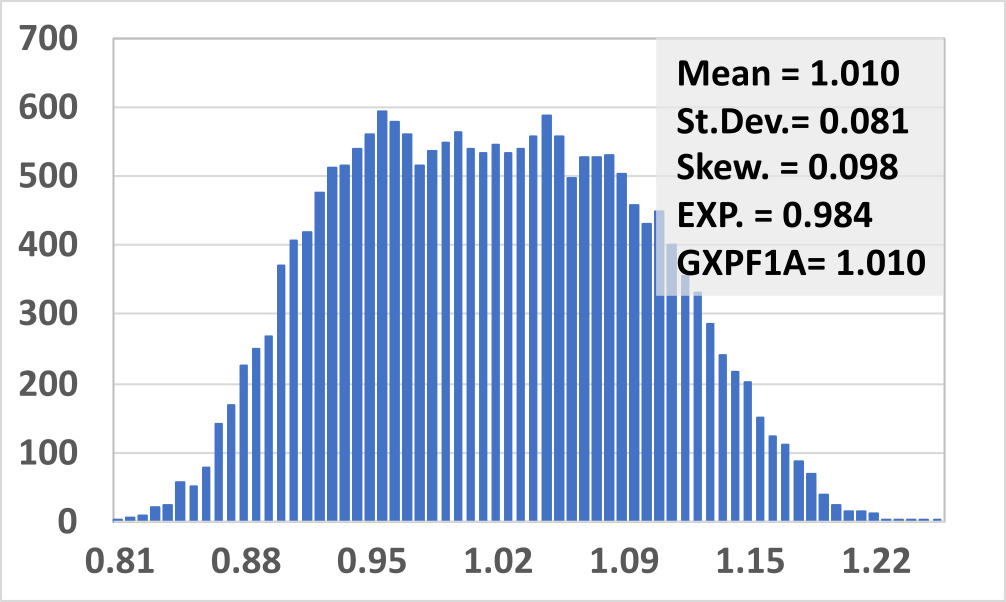} 
\\ \hline
$^{48}$Ti $4^{+}$ & \includegraphics[width=0.32\linewidth]{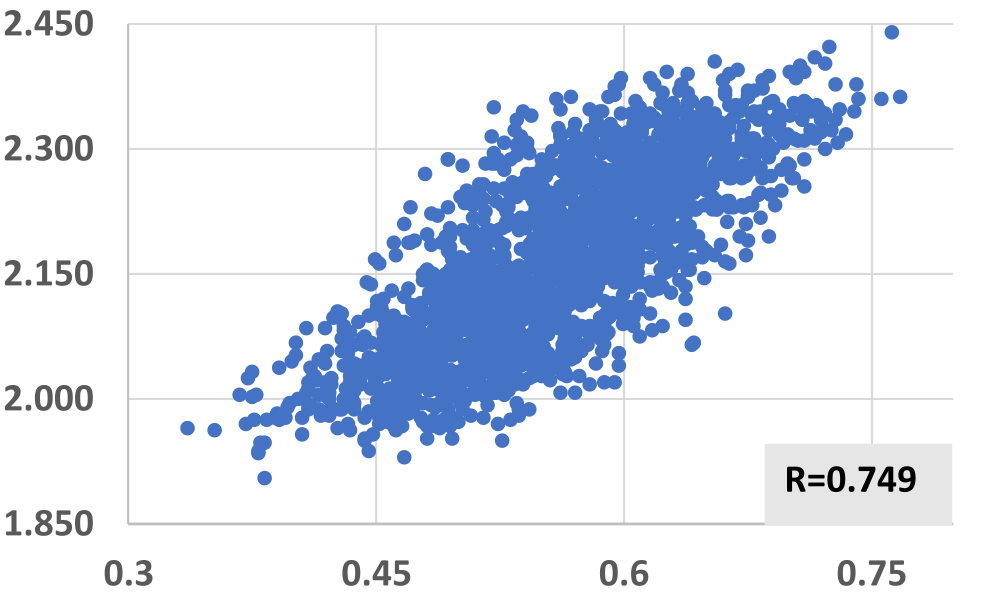} & \includegraphics[width=0.32\linewidth]{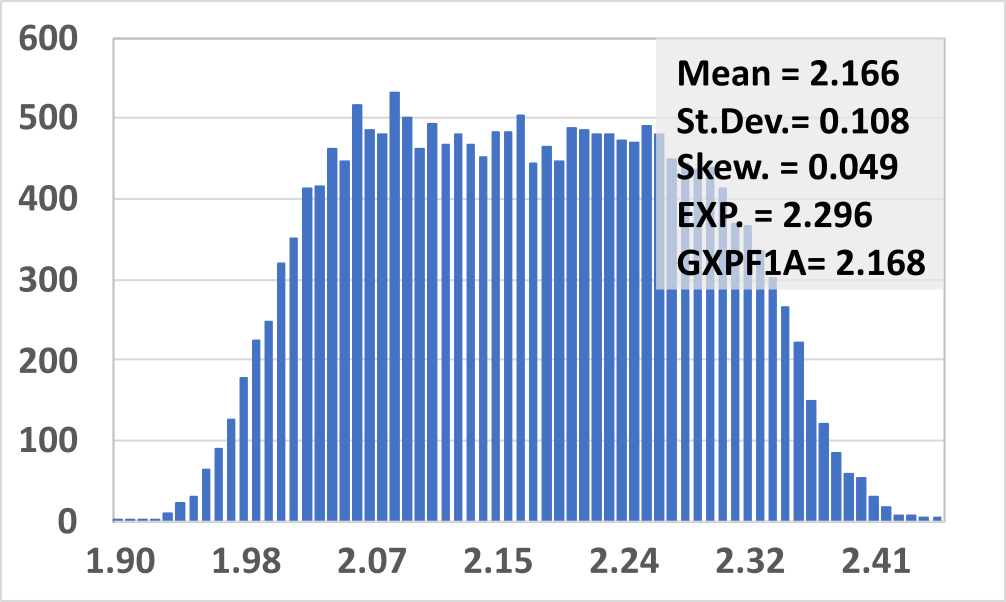}  
\\ \hline
$^{48}$Ti $6^{+}$ & \includegraphics[width=0.32\linewidth]{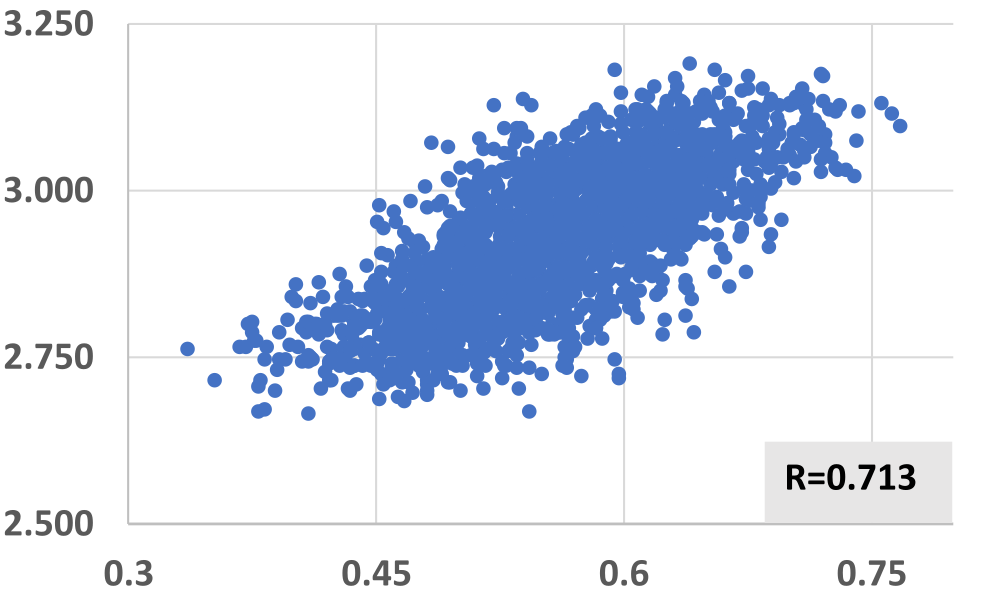} & \includegraphics[width=0.32\linewidth]{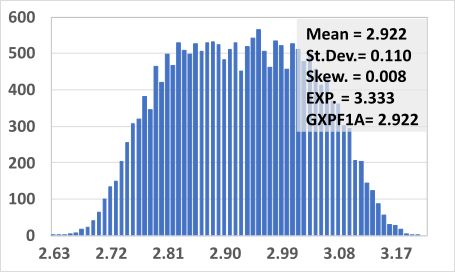}  
\\ \hline
$^{48}$Ti Occ(Nf5) & \includegraphics[width=0.32\linewidth]{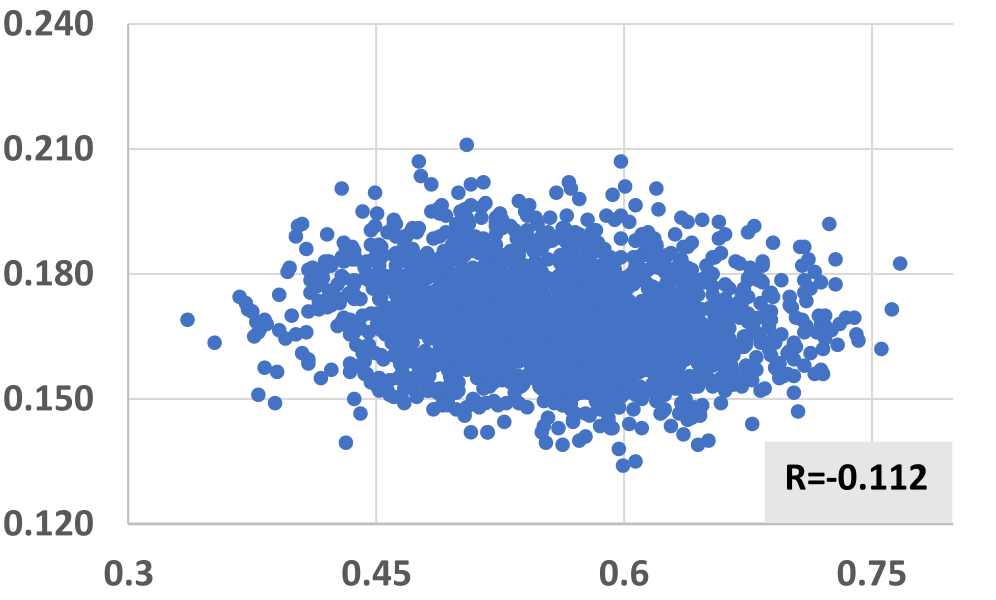} & \includegraphics[width=0.32\linewidth]{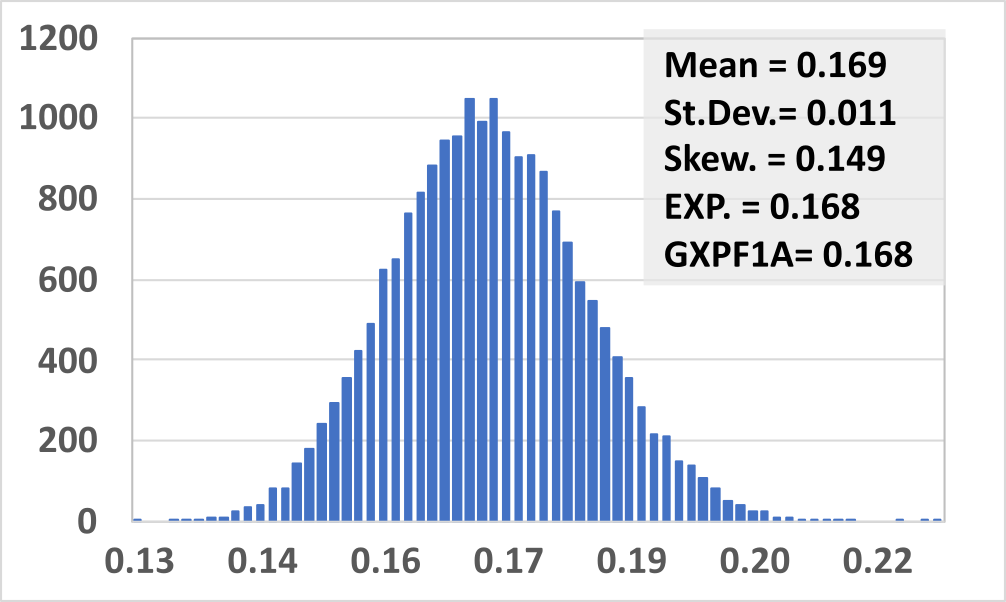}  
\\ \hline
$^{48}$Ti Occ(Nf7) & \includegraphics[width=0.32\linewidth]{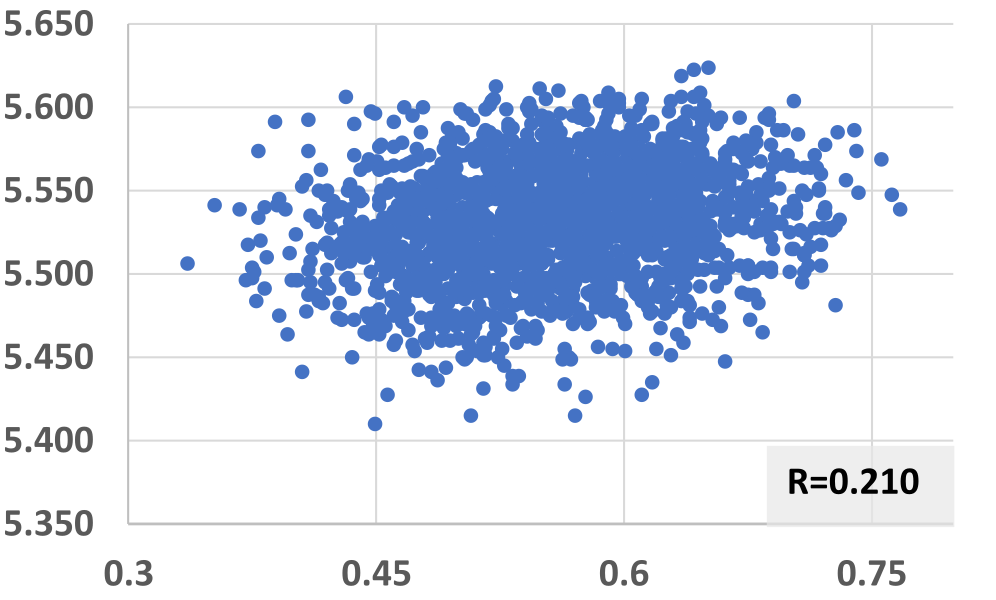} & \includegraphics[width=0.32\linewidth]{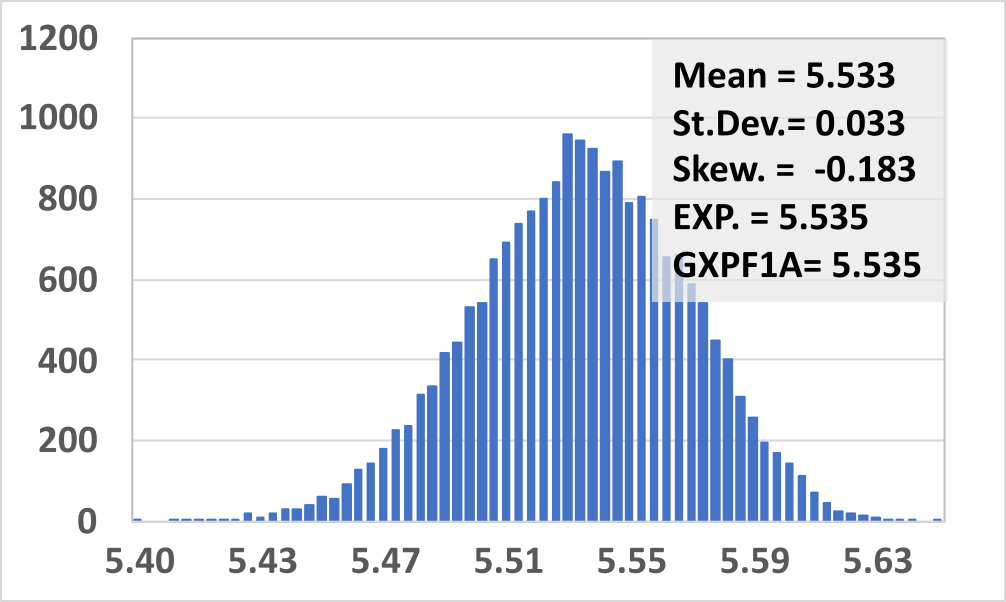}  
\\ \hline
$^{48}$Ti Occ(Np1) & \includegraphics[width=0.32\linewidth]{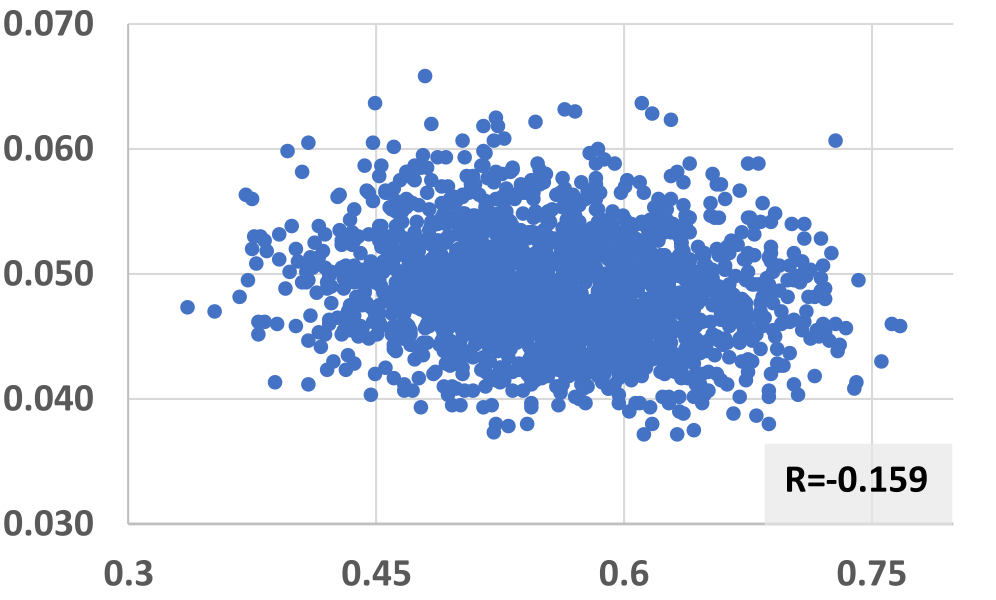} & \includegraphics[width=0.32\linewidth]{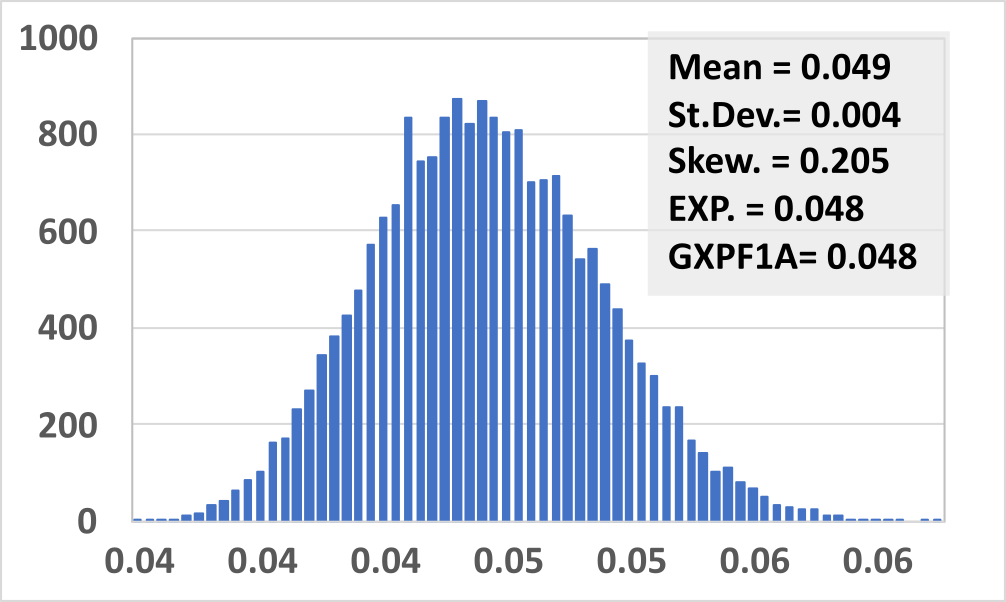}  
\\ \hline
$^{48}$Ti Occ(Np3) & \includegraphics[width=0.32\linewidth]{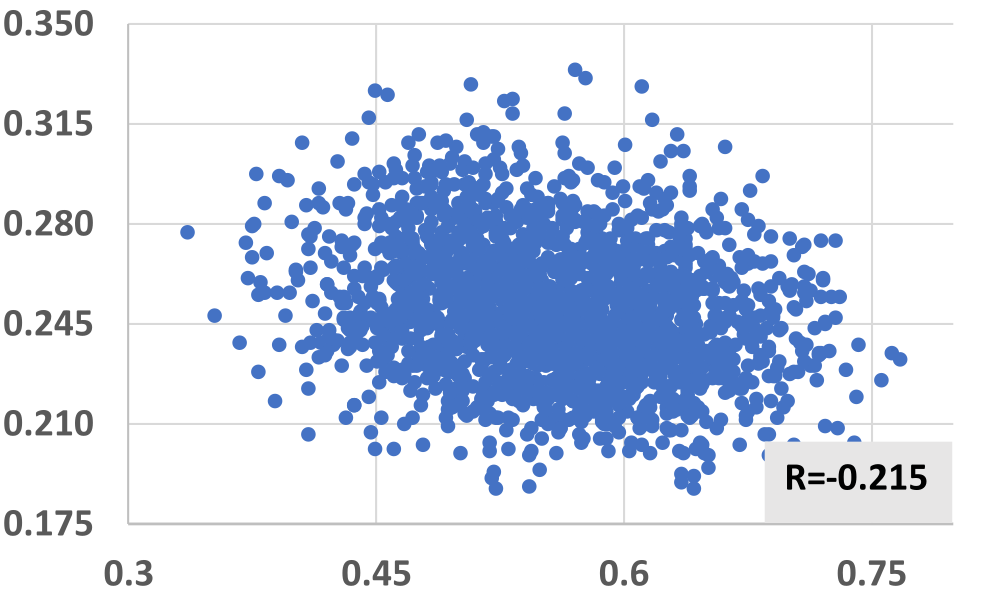} & \includegraphics[width=0.32\linewidth]{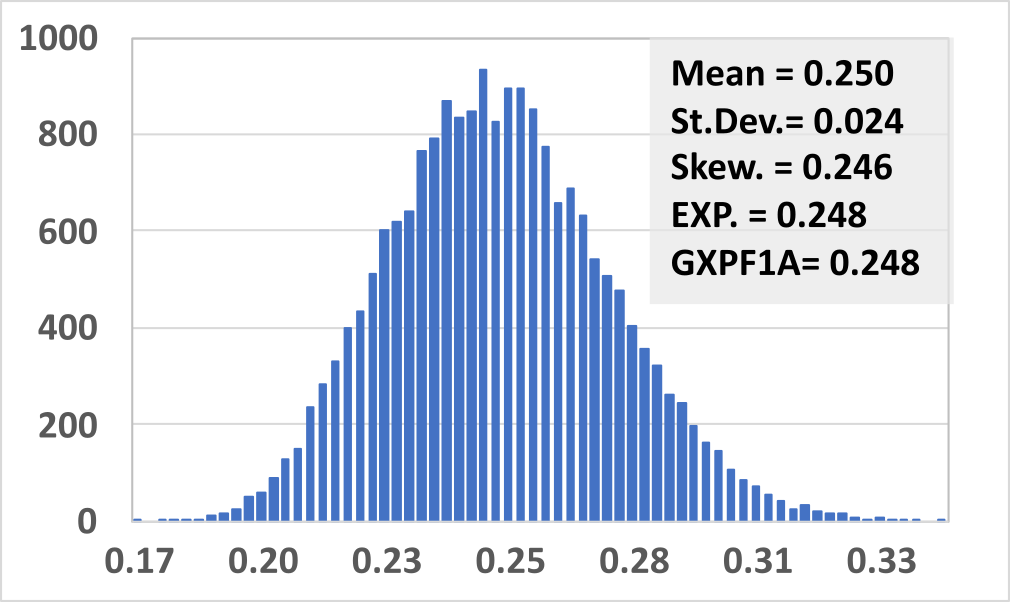}  
\\ \hline
$^{48}$Ti Occ(Pf5) & \includegraphics[width=0.32\linewidth]{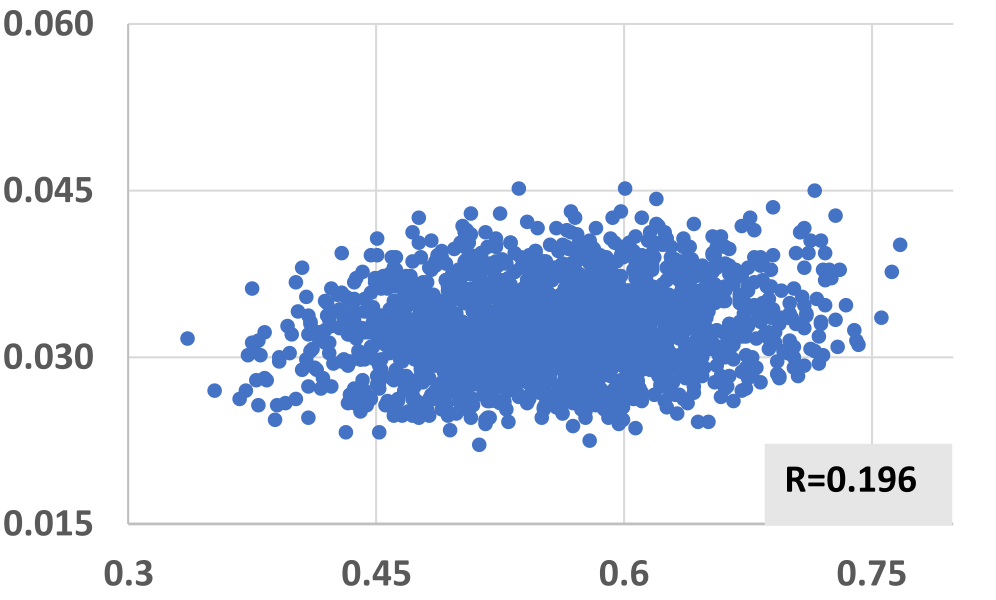} & \includegraphics[width=0.32\linewidth]{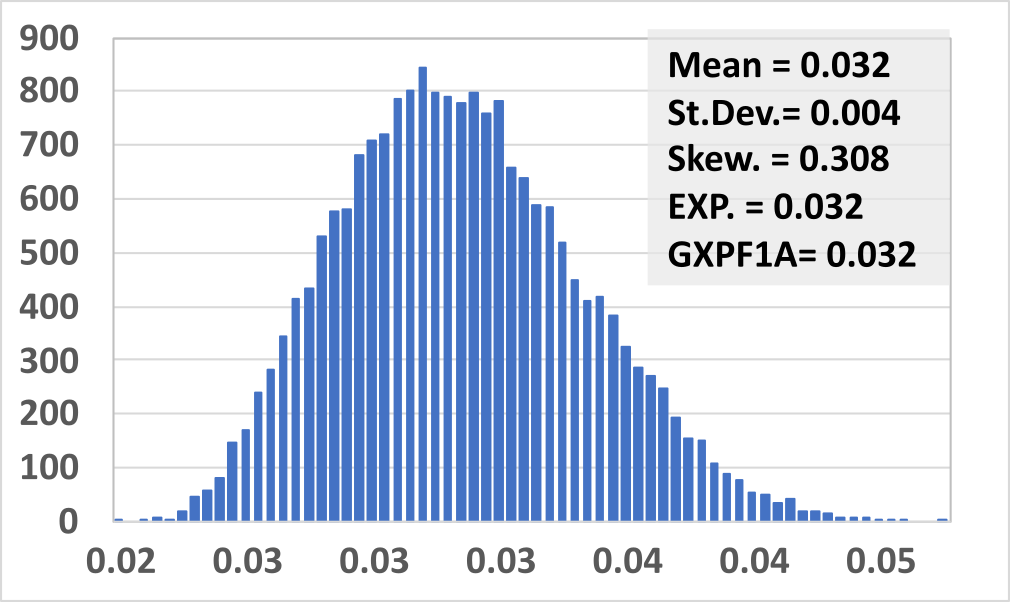}  
\\ \hline
$^{48}$Ti Occ(Pf7) & \includegraphics[width=0.32\linewidth]{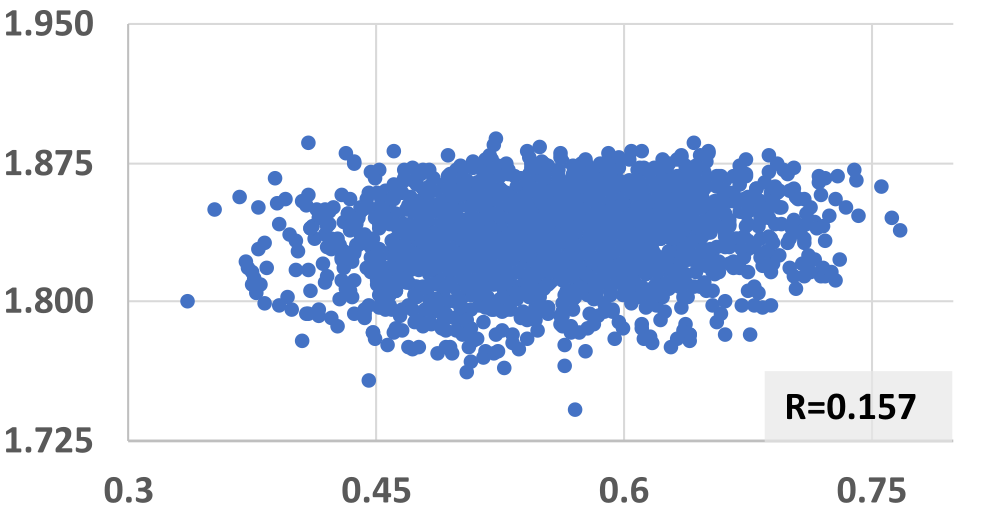} & \includegraphics[width=0.32\linewidth]{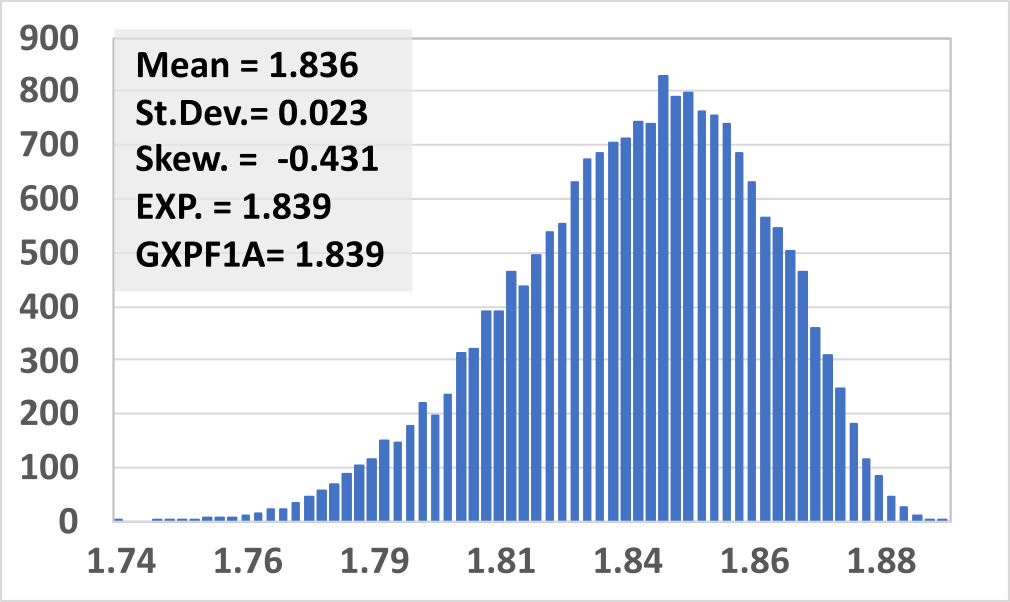}  
\\ \hline
$^{48}$Ti Occ(Pp1) & \includegraphics[width=0.32\linewidth]{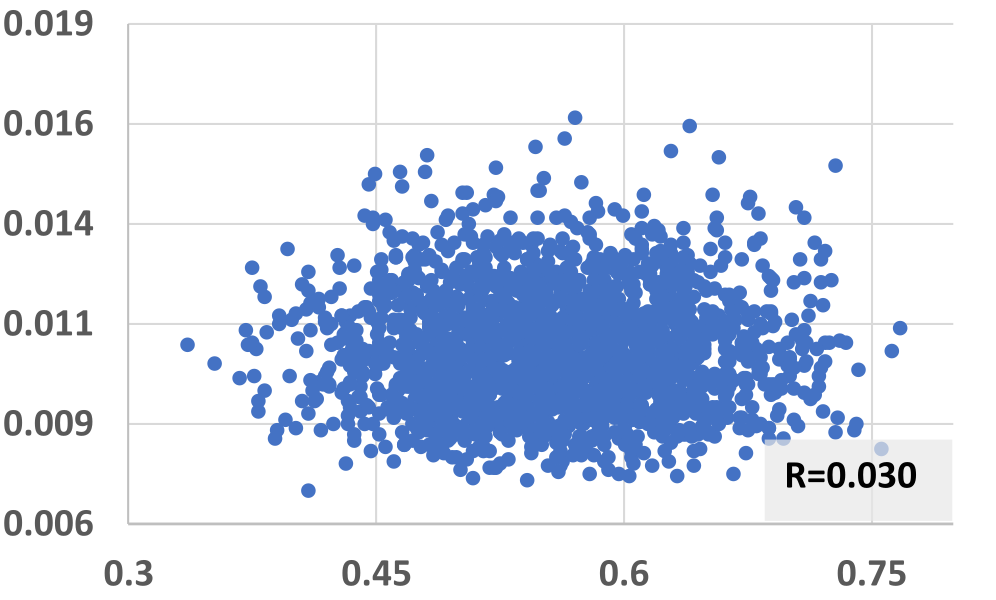} & \includegraphics[width=0.32\linewidth]{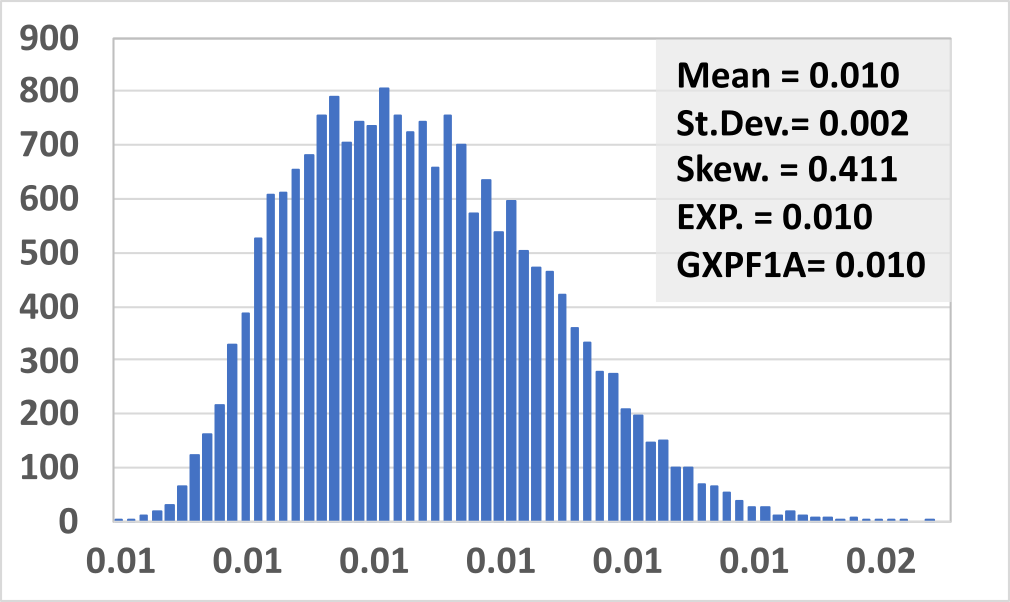}  
\\ \hline
$^{48}$Ti Occ(Pp3) & \includegraphics[width=0.32\linewidth]{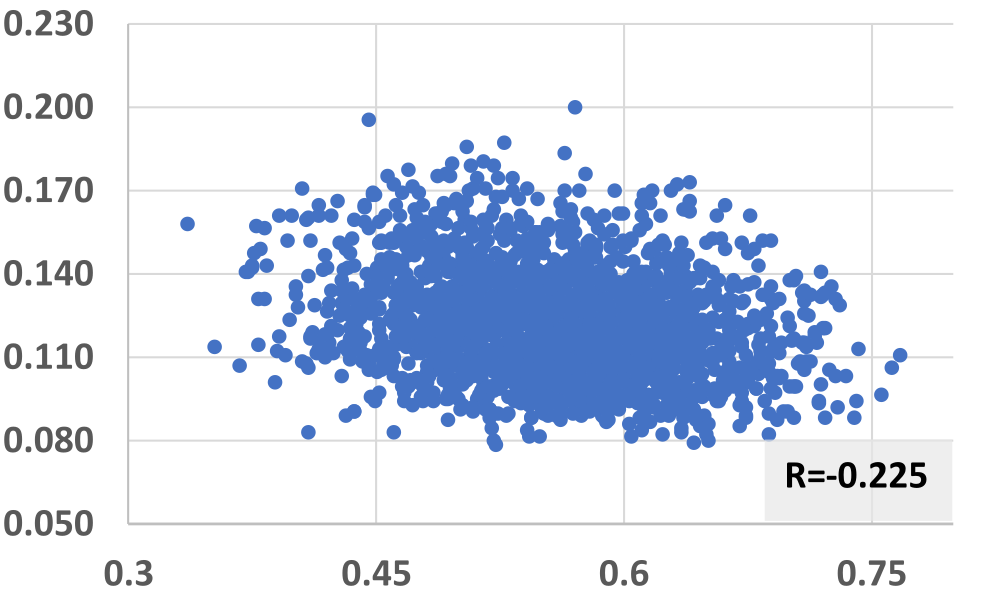} & \includegraphics[width=0.32\linewidth]{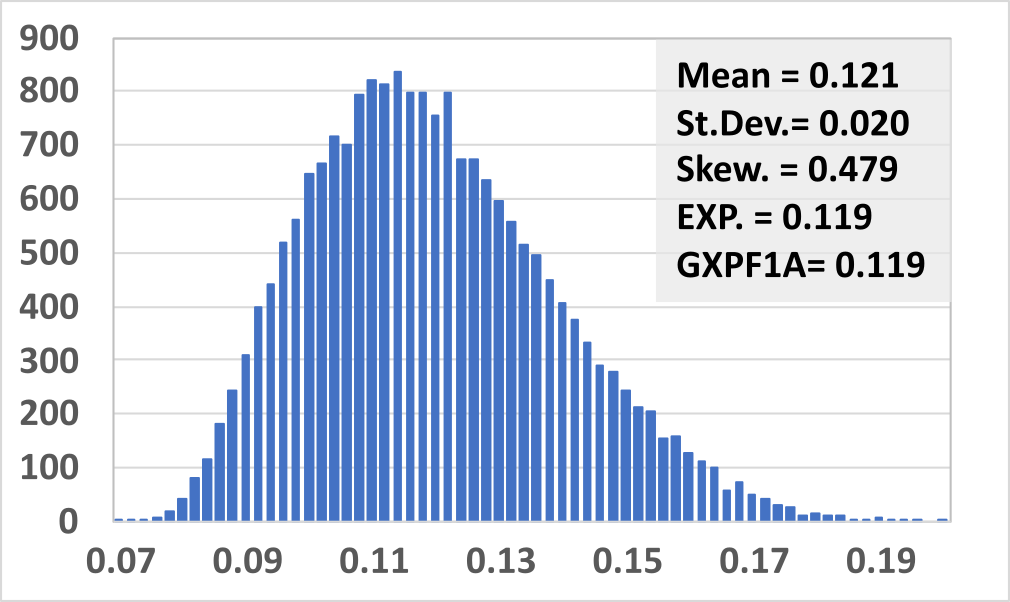}  
\\ \hline
\end{tabular}
 \caption{(Color online) Continuation of Table \ref{tab:parent-gxpf1a}. }
\label{tab:daughter-gxpf1a}
\end{table}

\begin{table}[htbp]
\centering
\begin{tabular}{c|c|c}
\hline
Observable & Correlation & PDF\\ \hline
$0\nu\beta\beta$ NME & \includegraphics[width=0.32\linewidth]{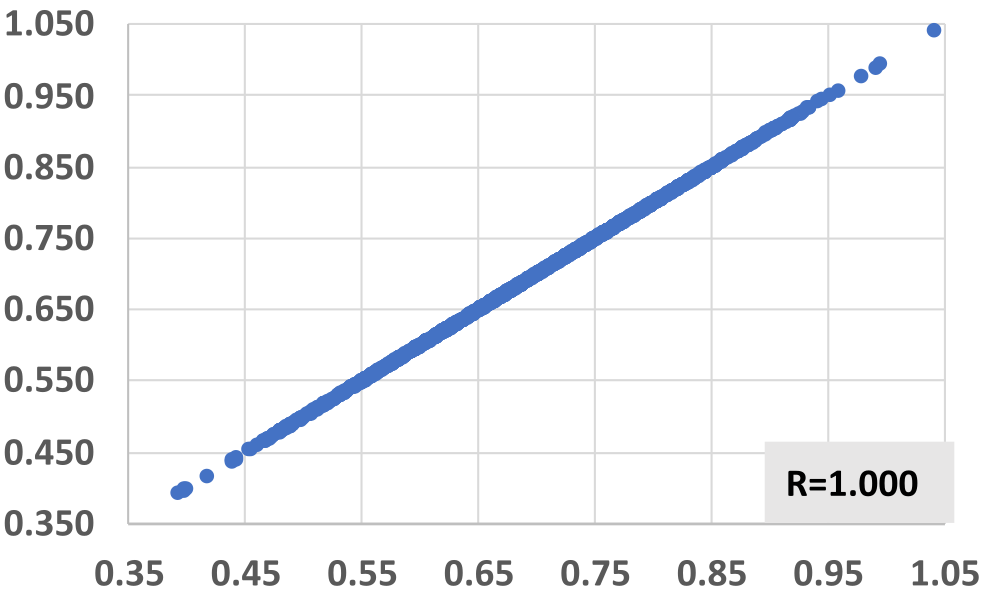} & \includegraphics[width=0.32\linewidth]{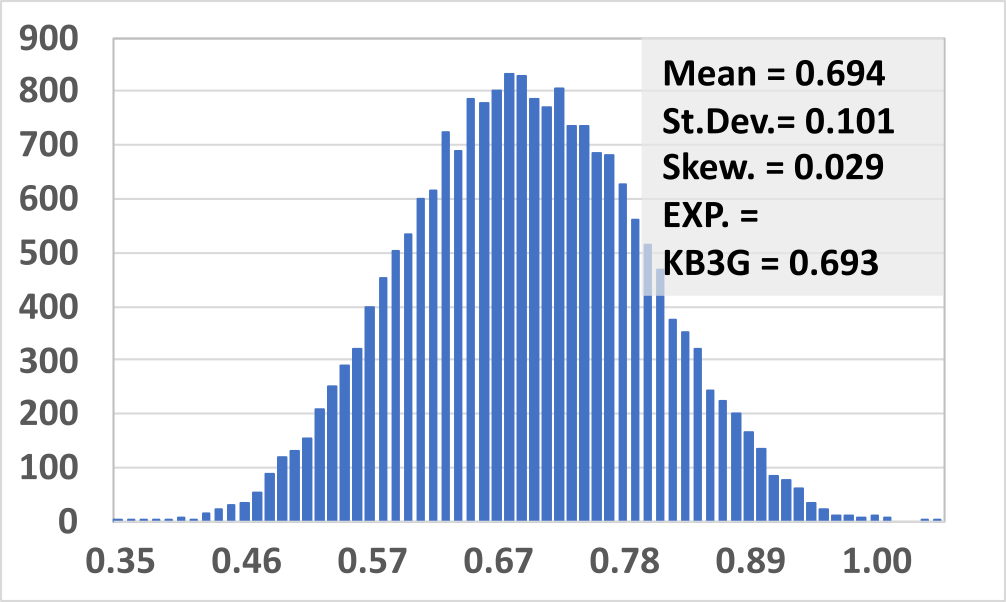} \\
\hline
$2\nu\beta\beta$ NME & \includegraphics[width=0.32\linewidth]{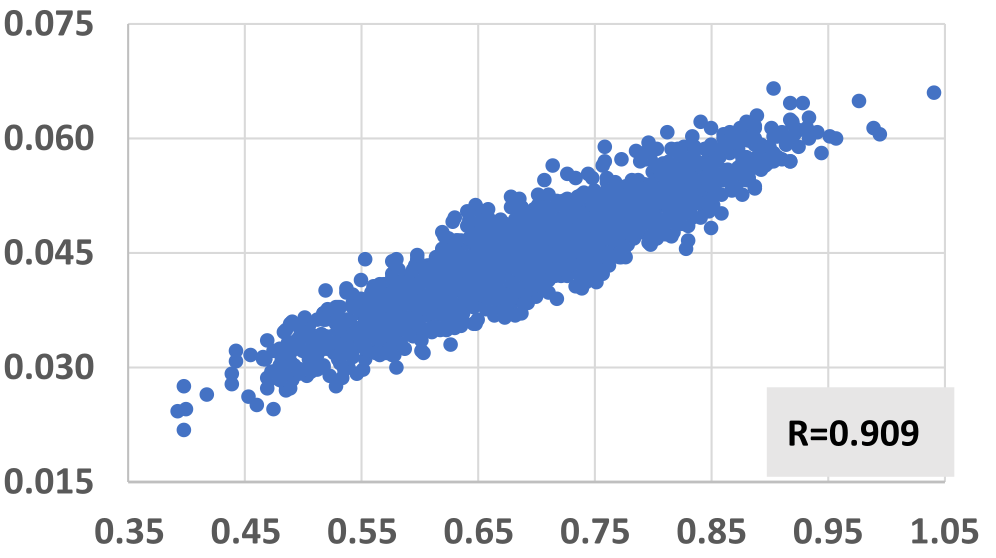} & \includegraphics[width=0.32\linewidth]{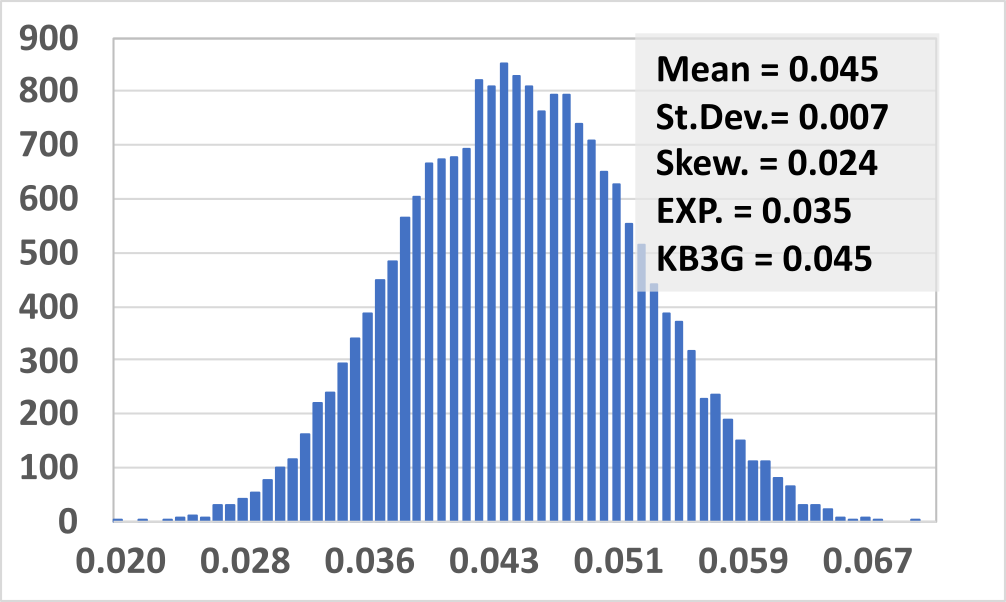}  
\\ \hline
$^{48}$Ca $\rightarrow$ $^{48}$Sc GT & \includegraphics[width=0.32\linewidth]{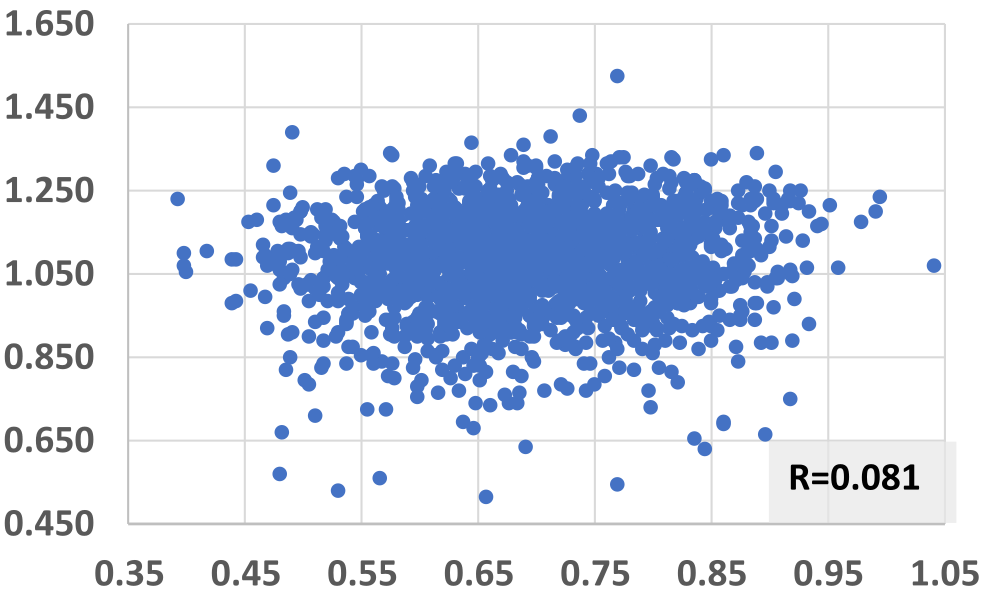} & \includegraphics[width=0.32\linewidth]{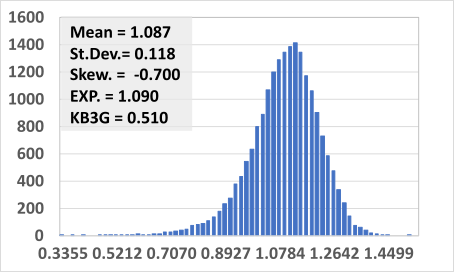}  
\\ \hline
$^{48}$Ti $\rightarrow$ $^{48}$Sc GT & \includegraphics[width=0.32\linewidth]{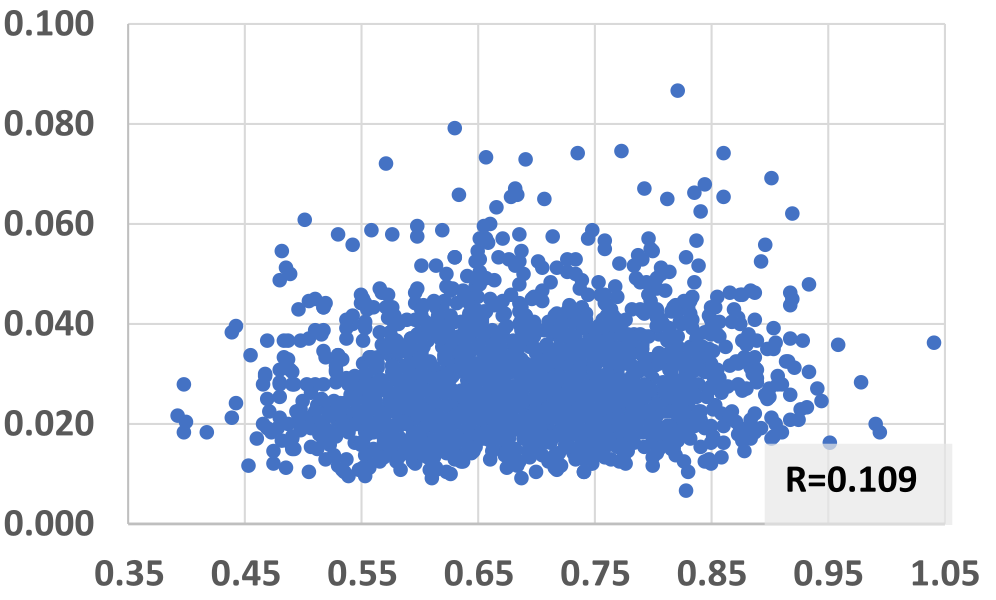} & \includegraphics[width=0.32\linewidth]{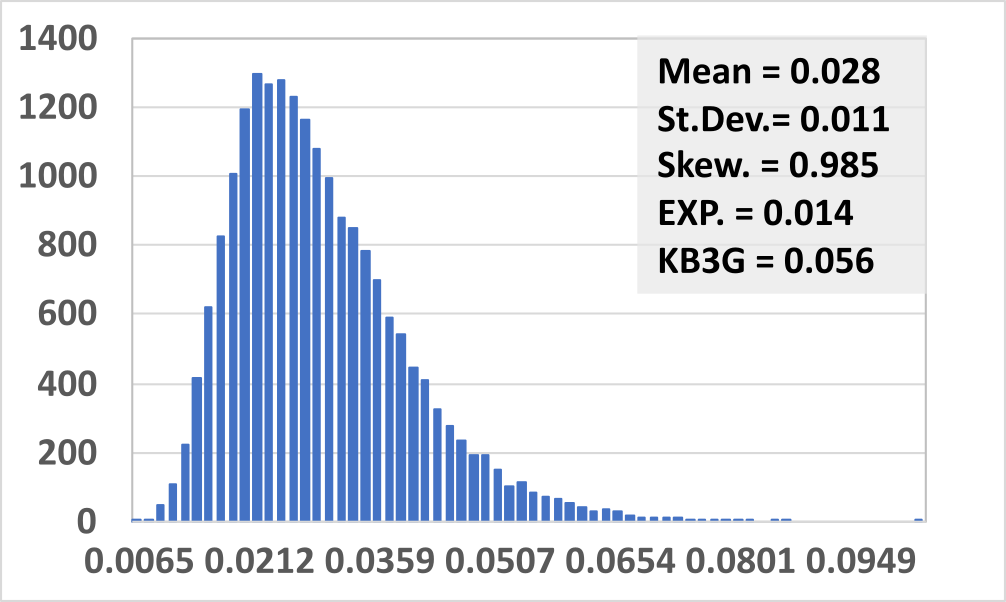}  
\\ \hline
$^{48}$Ca B(E2)$(\uparrow)$ & \includegraphics[width=0.32\linewidth]{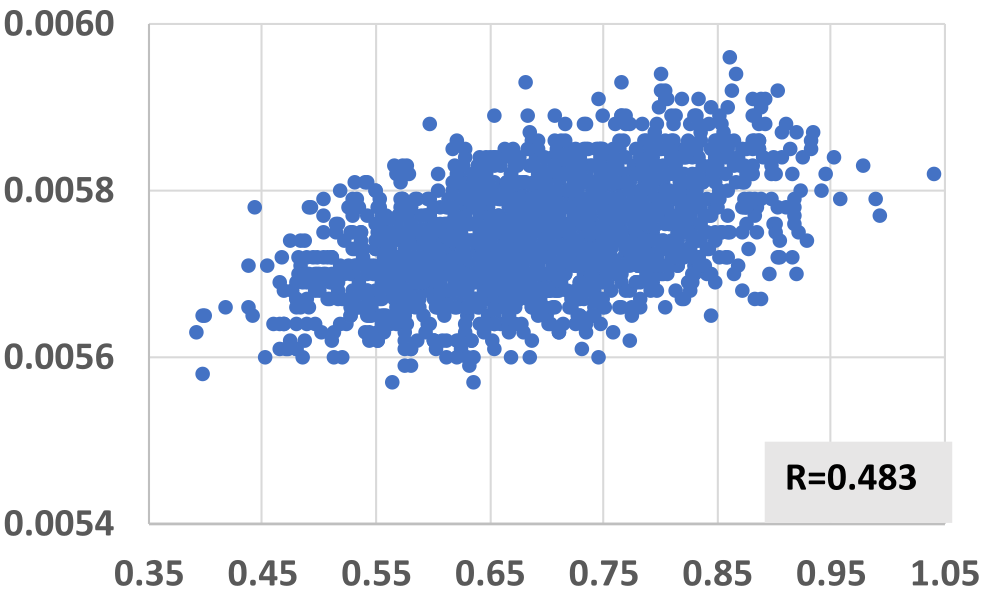} & \includegraphics[width=0.32\linewidth]{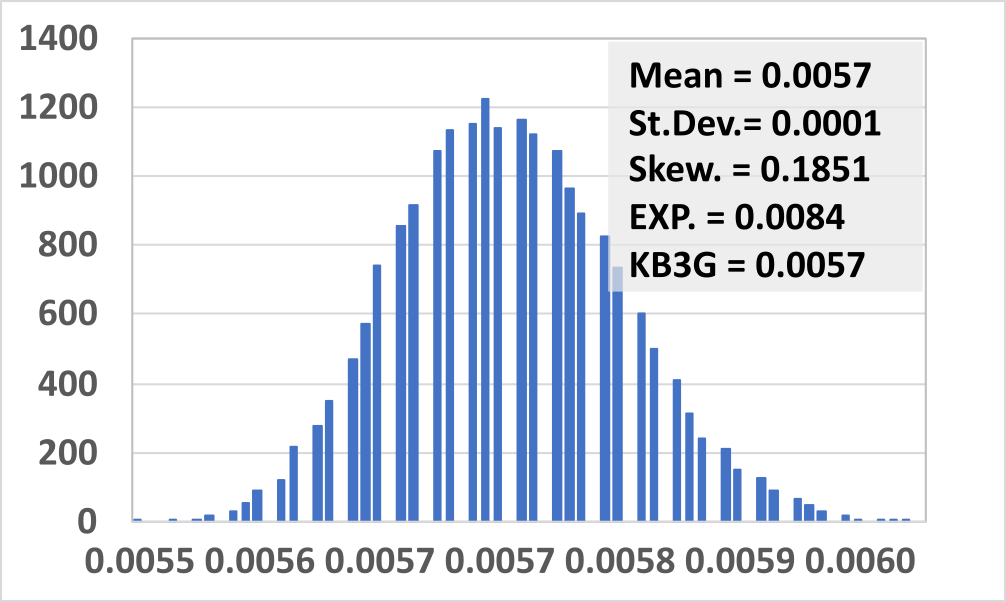} 
\\ \hline
$^{48}$Ca $2^{+}$ & \includegraphics[width=0.32\linewidth]{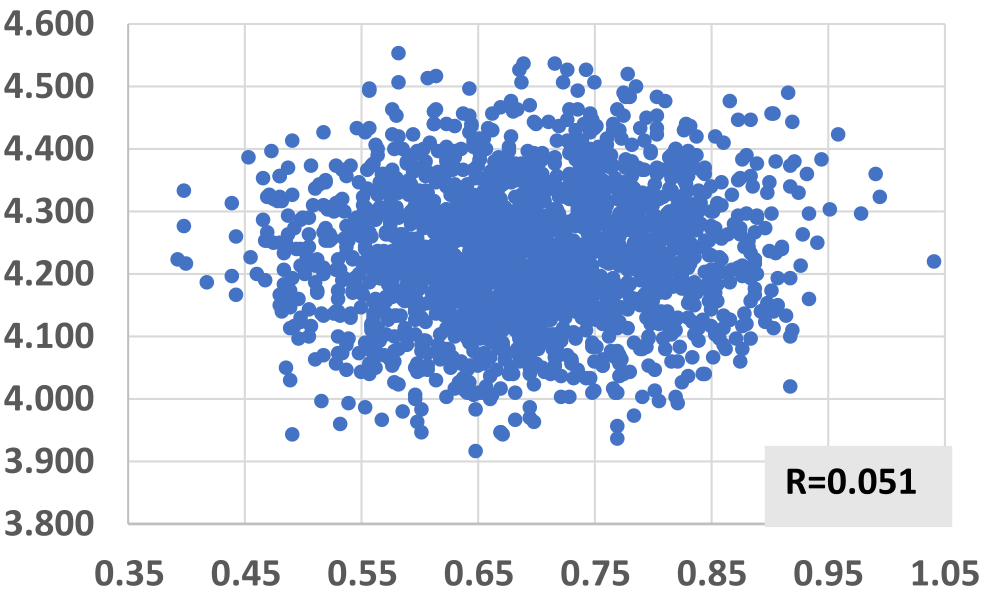} & \includegraphics[width=0.32\linewidth]{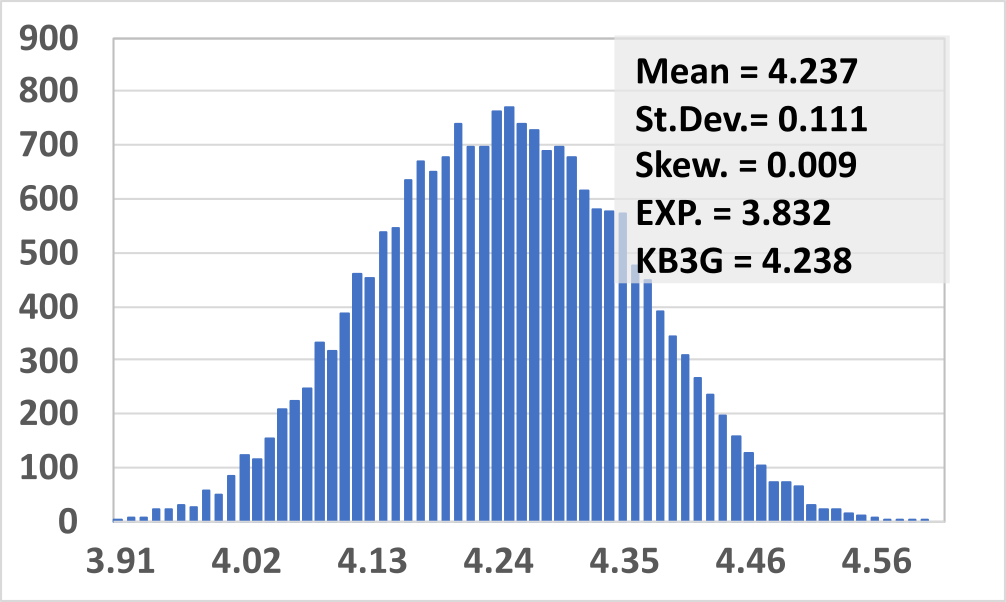}  
\\ \hline
$^{48}$Ca $4^{+}$ & \includegraphics[width=0.32\linewidth]{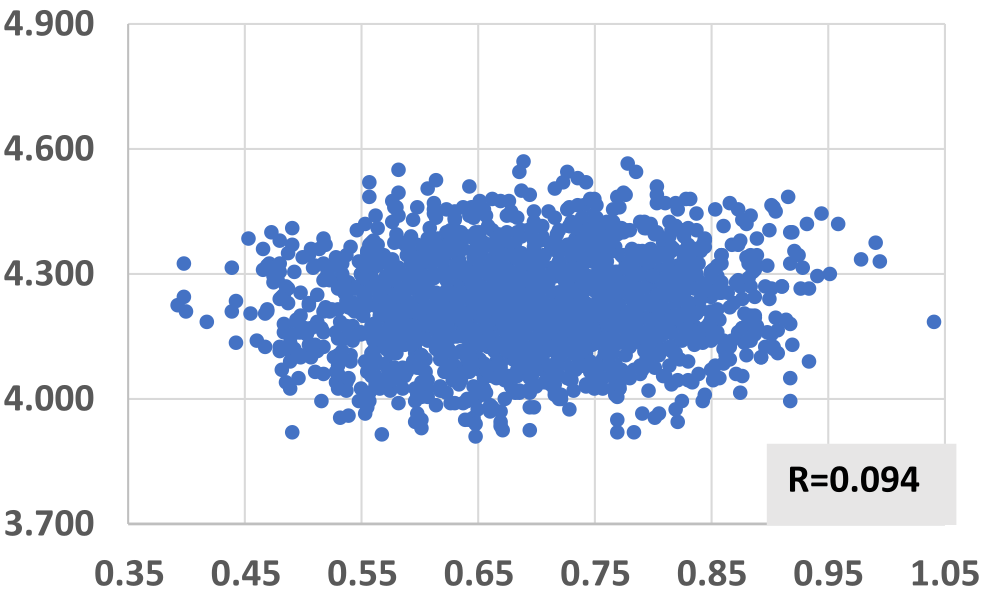} & \includegraphics[width=0.32\linewidth]{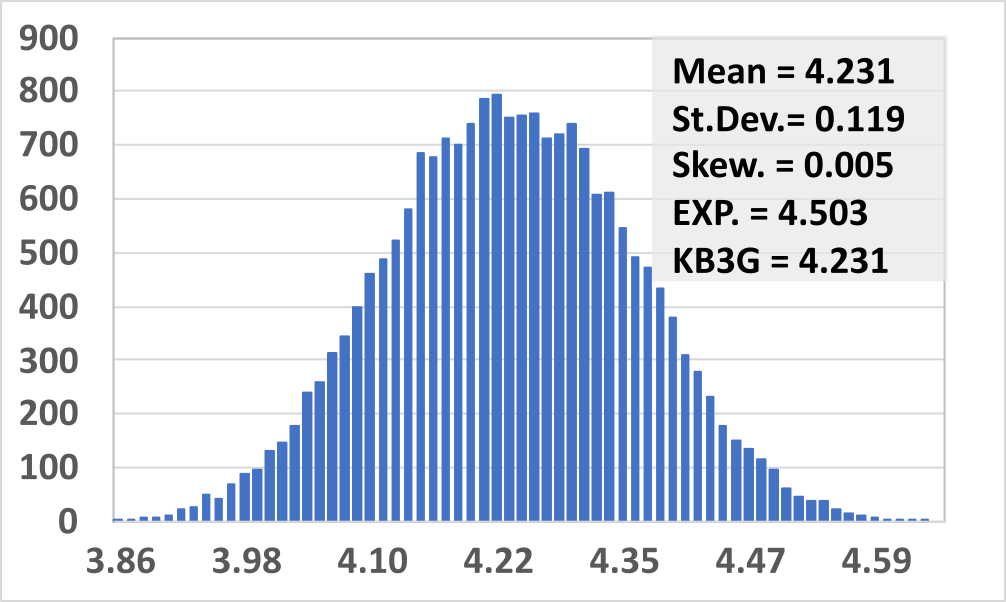}  
\\ \hline
$^{48}$Ca $6^{+}$ & \includegraphics[width=0.32\linewidth]{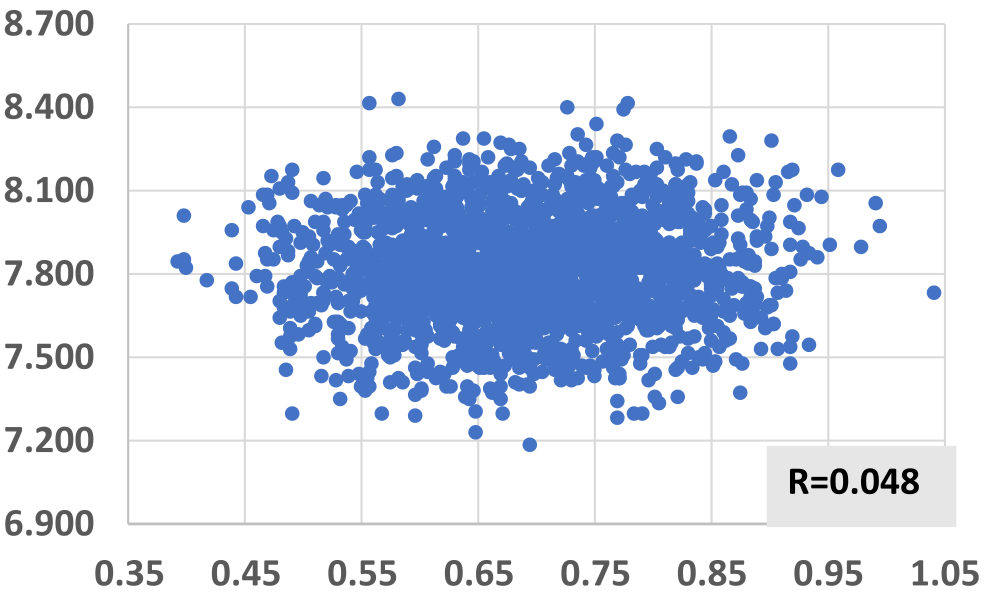} & \includegraphics[width=0.32\linewidth]{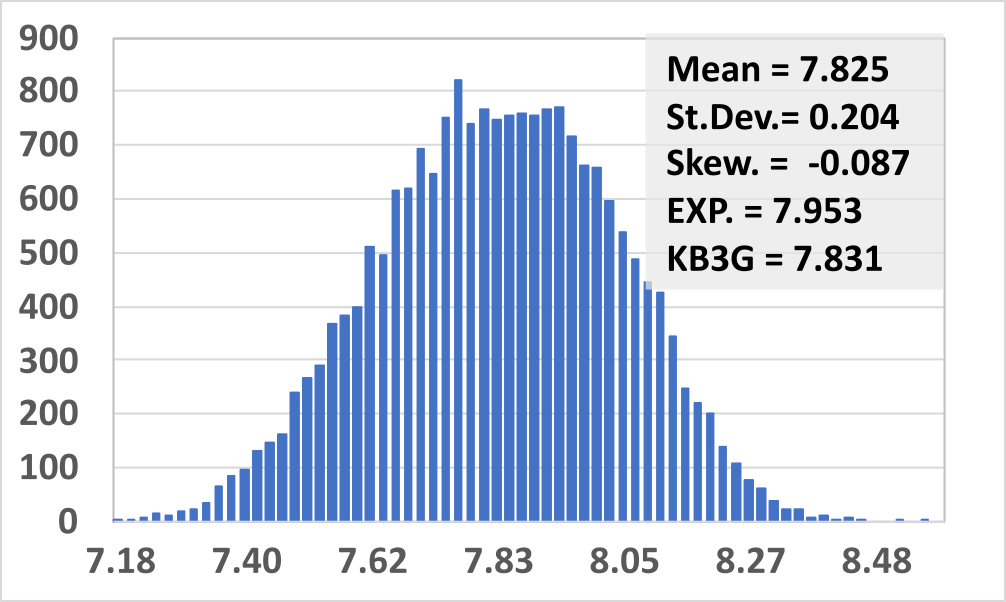}  
\\ \hline
$^{48}$Ca Occ(Nf5) & \includegraphics[width=0.32\linewidth]{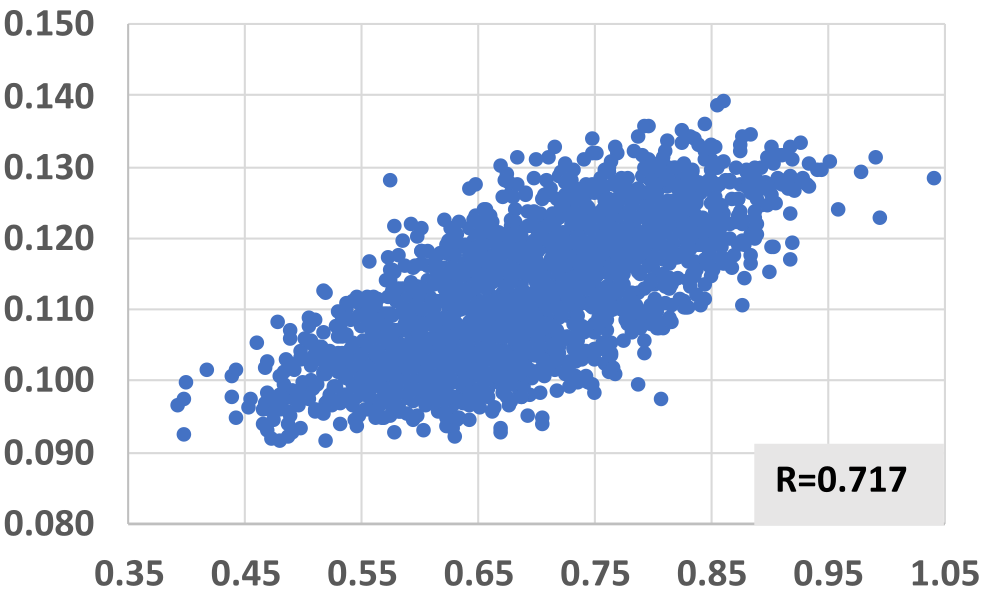} & \includegraphics[width=0.32\linewidth]{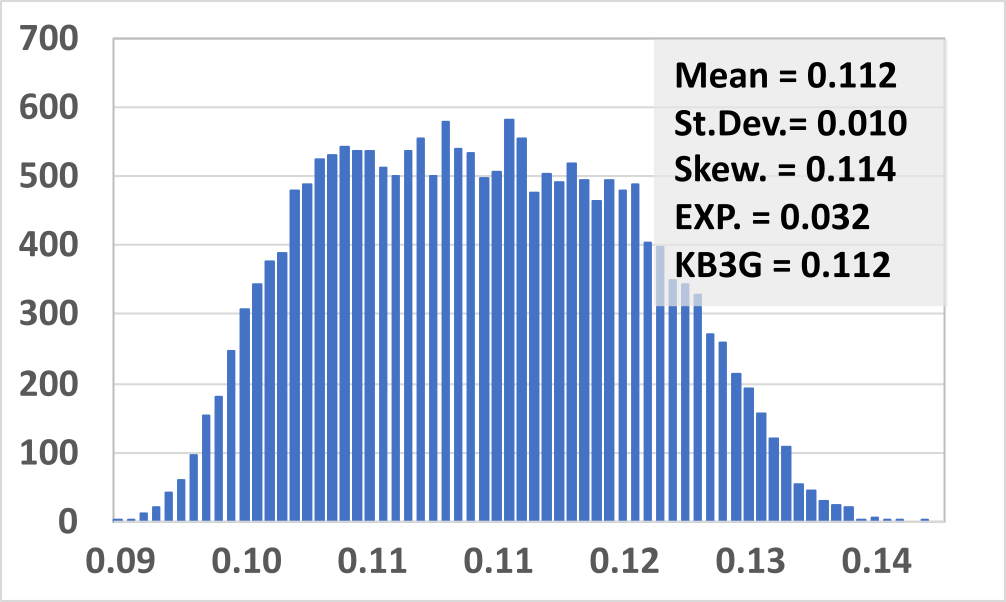}  
\\ \hline
$^{48}$Ca Occ(Nf7) & \includegraphics[width=0.32\linewidth]{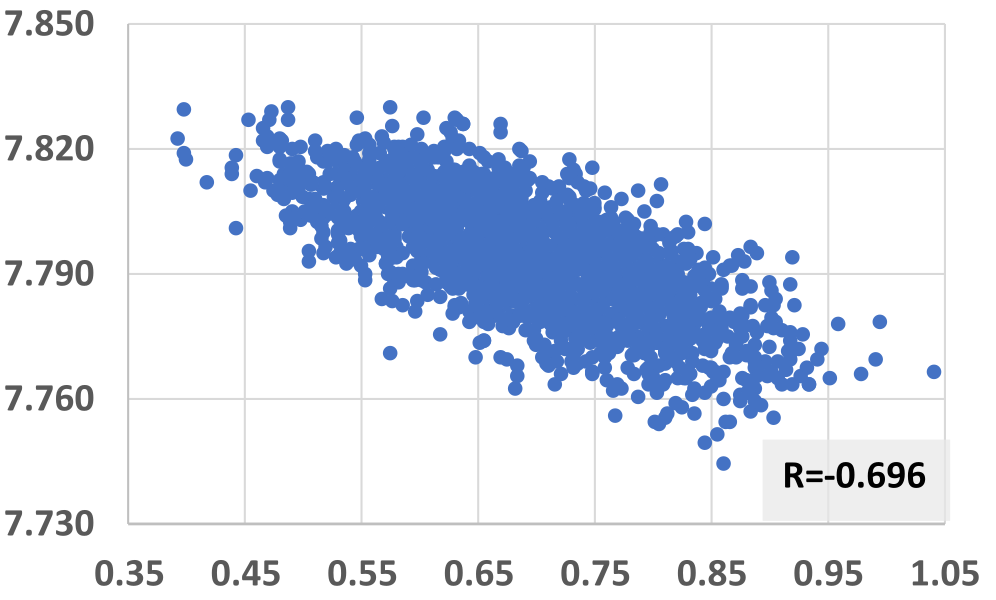} & \includegraphics[width=0.32\linewidth]{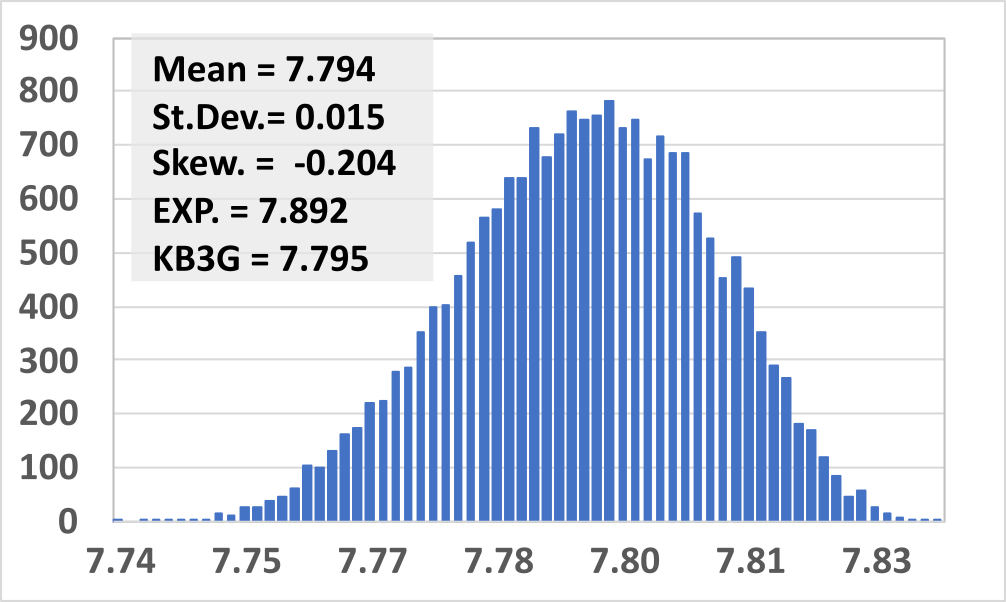}  
\\ \hline
$^{48}$Ca Occ(Np1) & \includegraphics[width=0.32\linewidth]{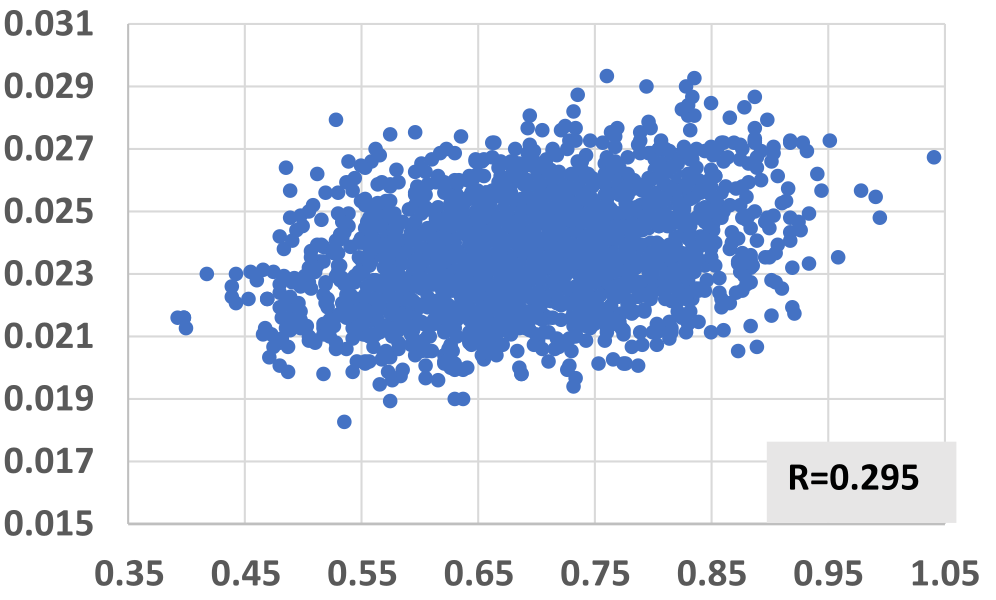} & \includegraphics[width=0.32\linewidth]{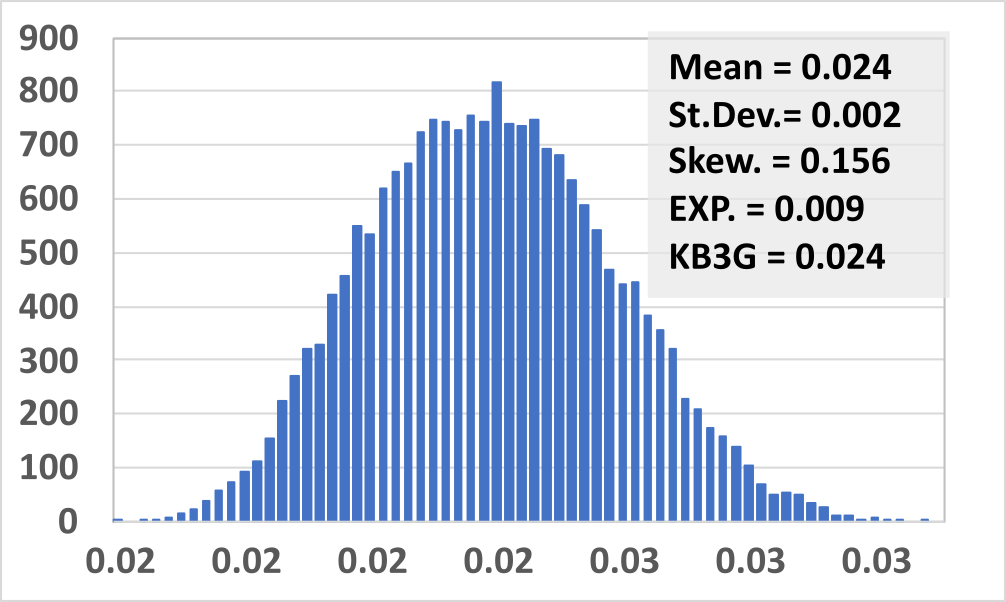}  
\\ \hline
$^{48}$Ca Occ(Np3) & \includegraphics[width=0.32\linewidth]{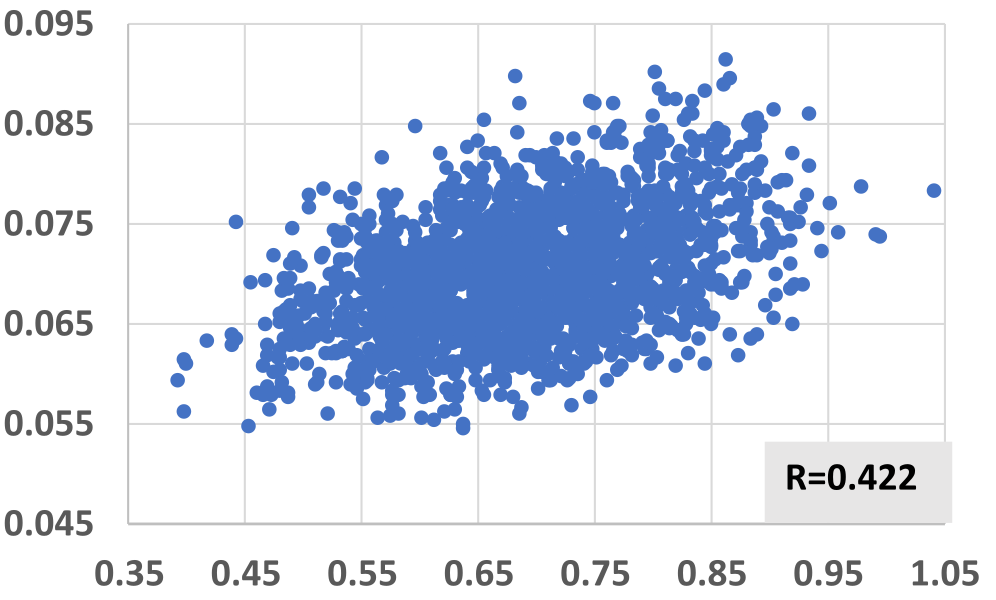} & \includegraphics[width=0.32\linewidth]{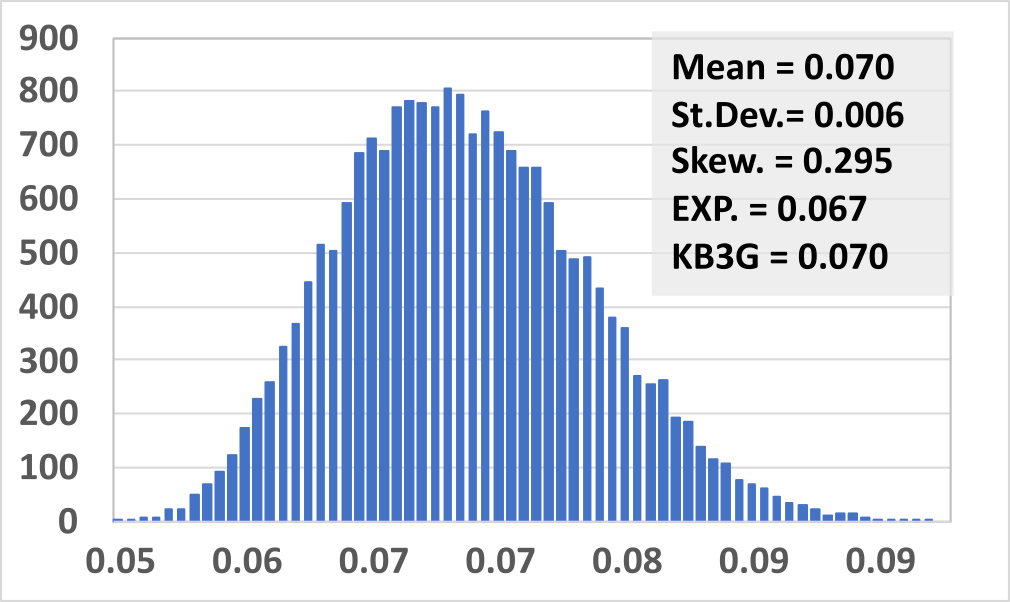}  
\\ \hline
\end{tabular}
 \caption{(Color online) Correlations scattered plots and PDFs for the KB3G starting Hamiltonian.}
\label{tab:parent-kb3g}
\end{table}

\begin{table}[htbp]
\centering
\begin{tabular}{c|c|c}
\hline
Observable & Correlation & PDF\\ \hline
$^{48}$Ti B(E2)$(\uparrow)$ & \includegraphics[width=0.32\linewidth]{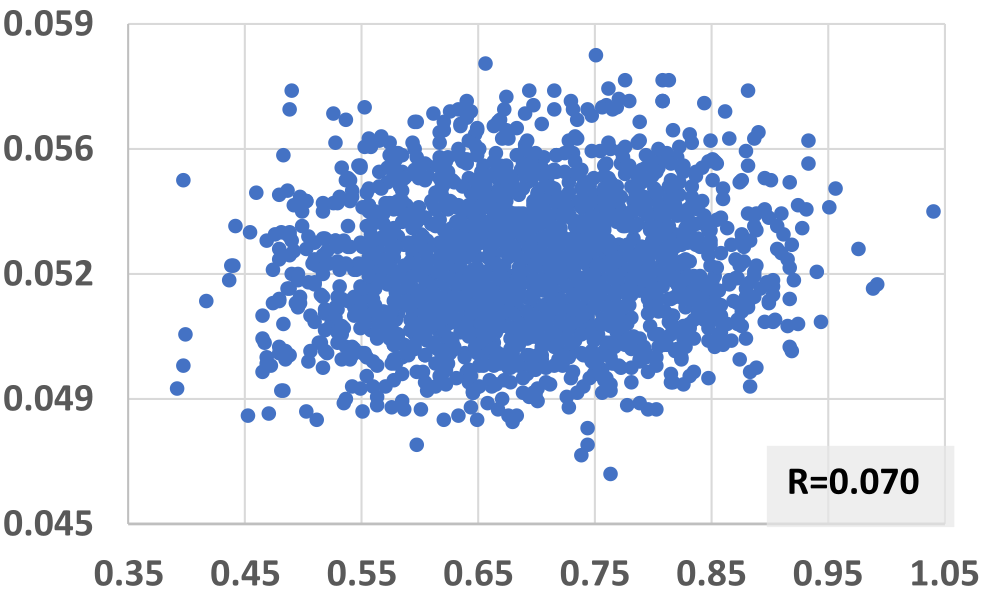} & \includegraphics[width=0.32\linewidth]{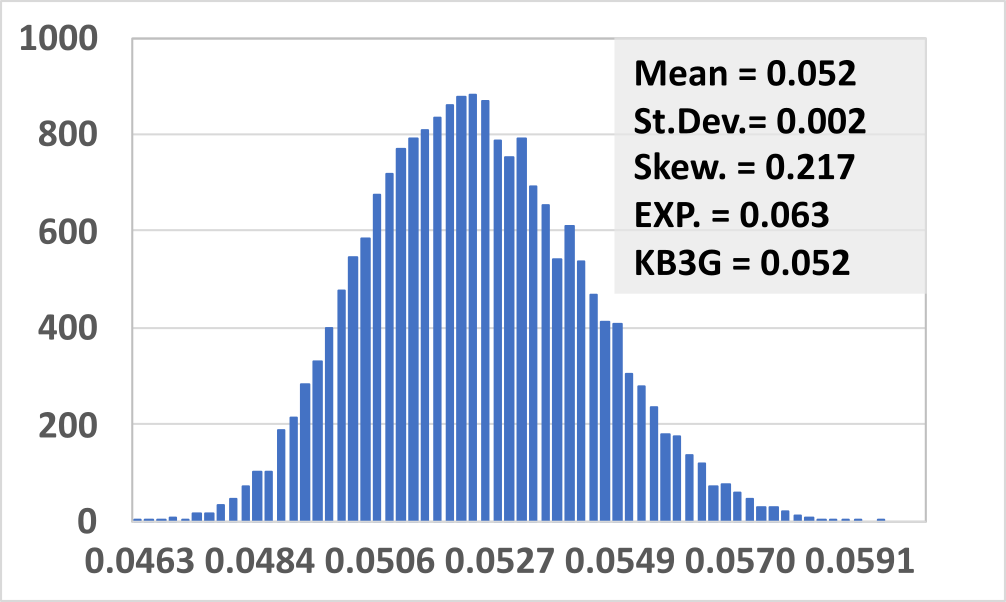} 
\\ \hline
$^{48}$Ti $2^{+}$ & \includegraphics[width=0.32\linewidth]{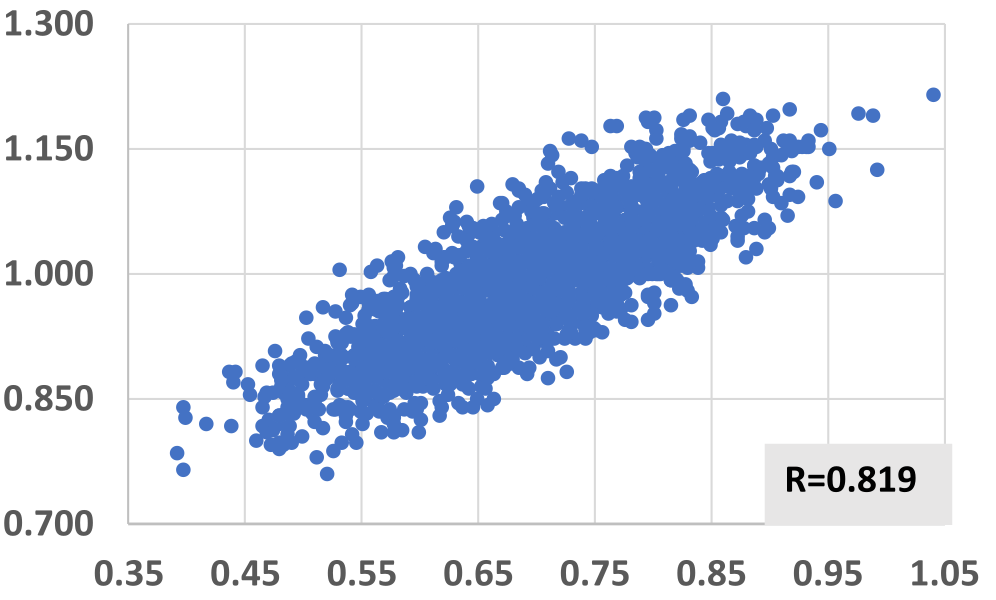} & 
 \includegraphics[width=0.32\linewidth]{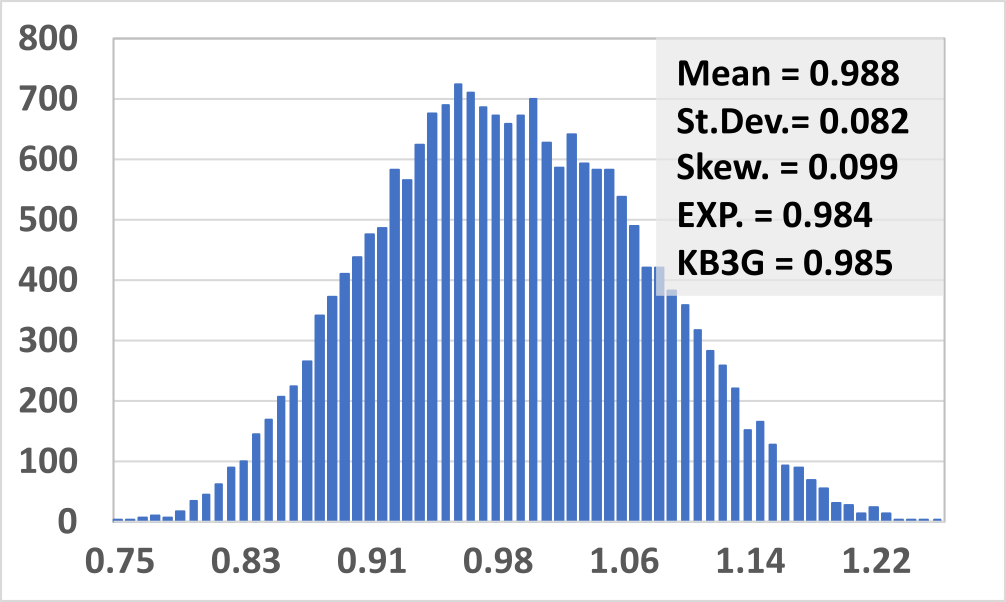} 
\\ \hline
$^{48}$Ti $4^{+}$ & \includegraphics[width=0.32\linewidth]{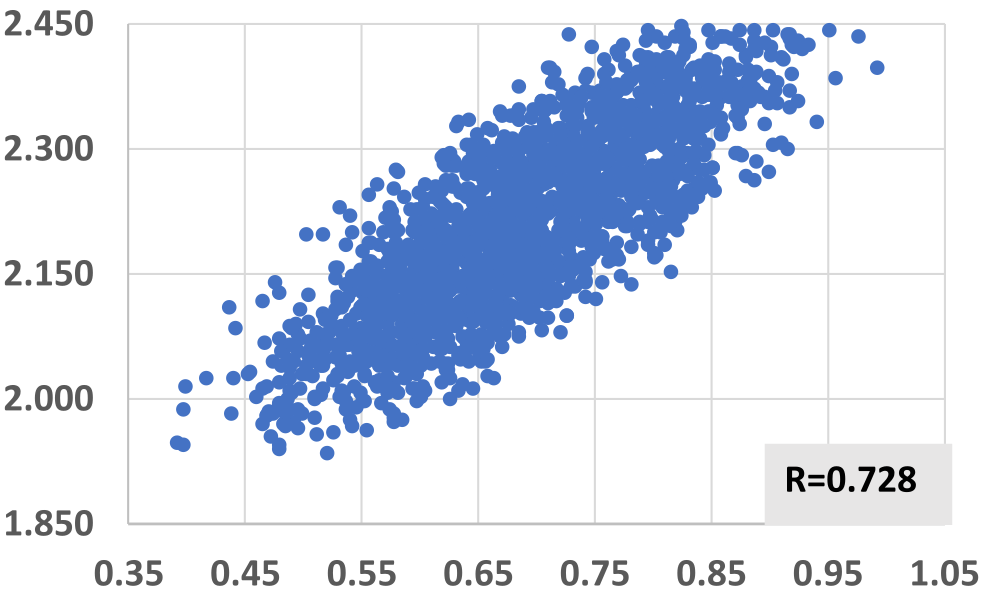} & \includegraphics[width=0.32\linewidth]{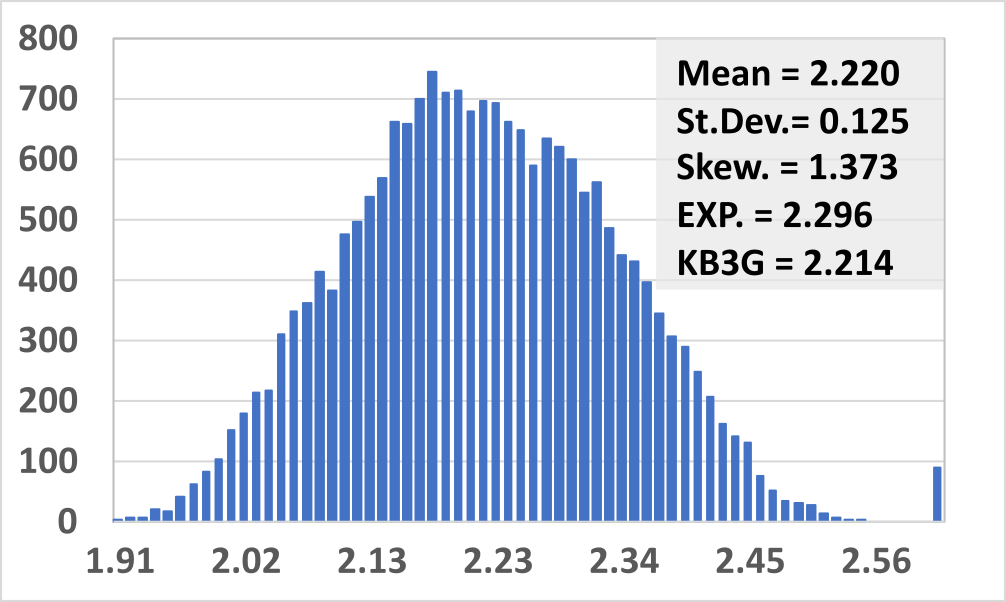}  
\\ \hline
$^{48}$Ti $6^{+}$ & \includegraphics[width=0.32\linewidth]{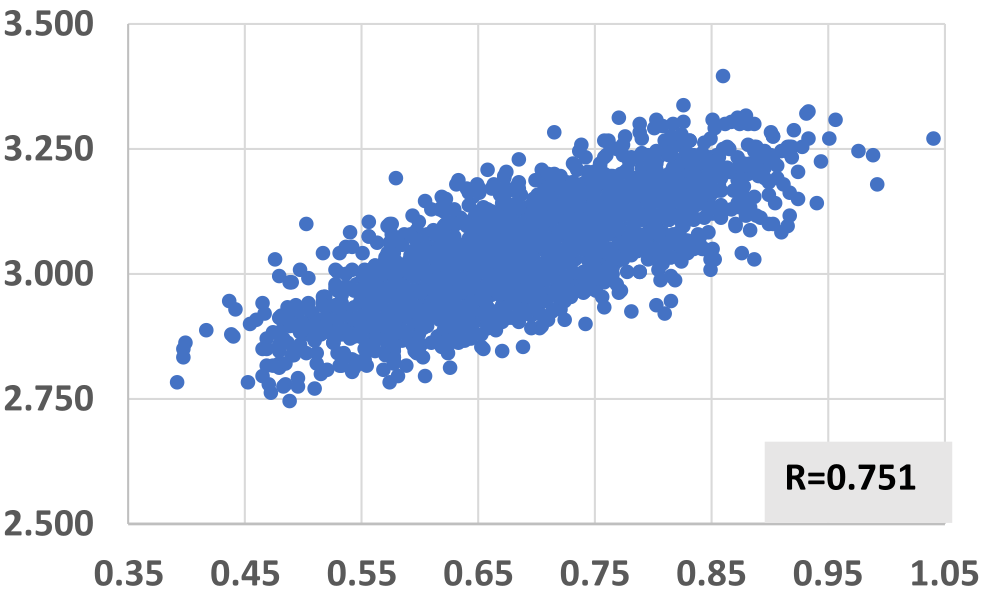} & \includegraphics[width=0.32\linewidth]{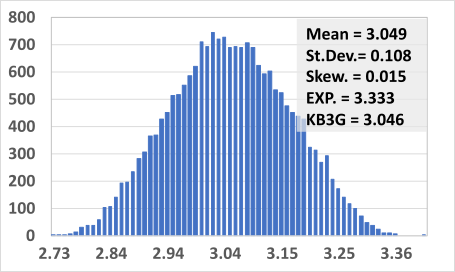}  
\\ \hline
$^{48}$Ti Occ(Nf5) & \includegraphics[width=0.32\linewidth]{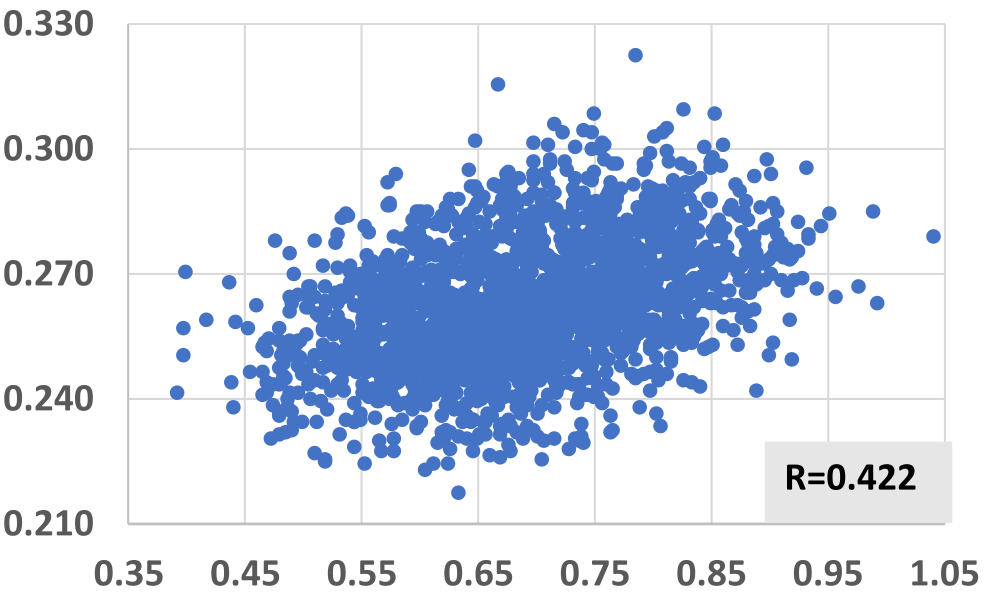} & \includegraphics[width=0.32\linewidth]{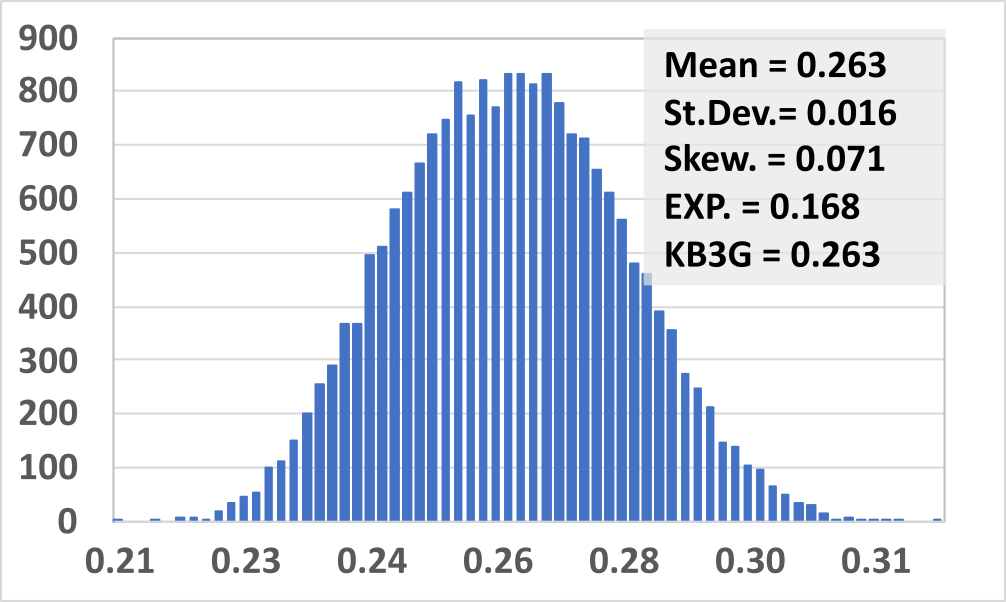}  
\\ \hline
$^{48}$Ti Occ(Nf7) & \includegraphics[width=0.32\linewidth]{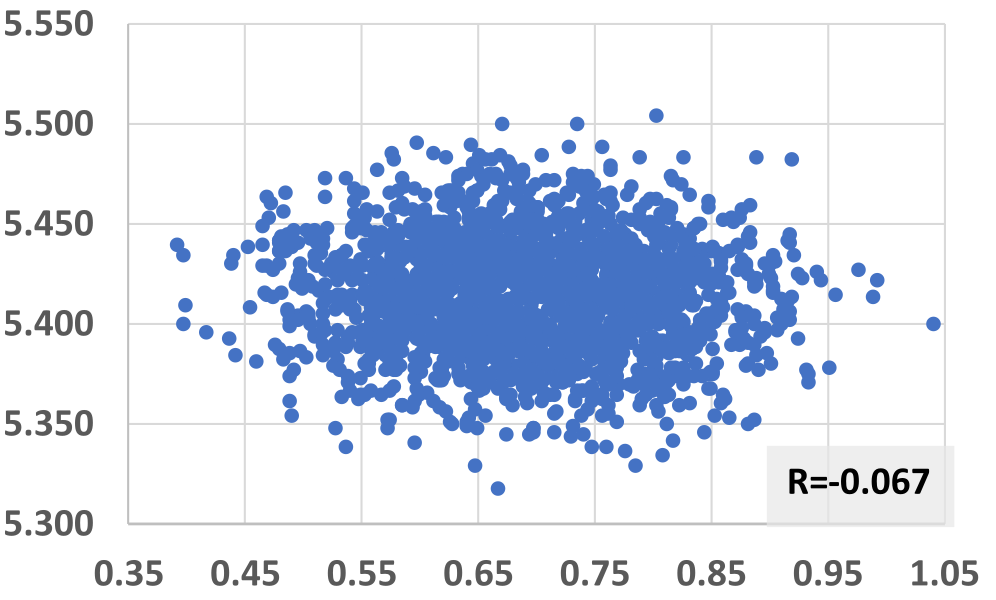} & \includegraphics[width=0.32\linewidth]{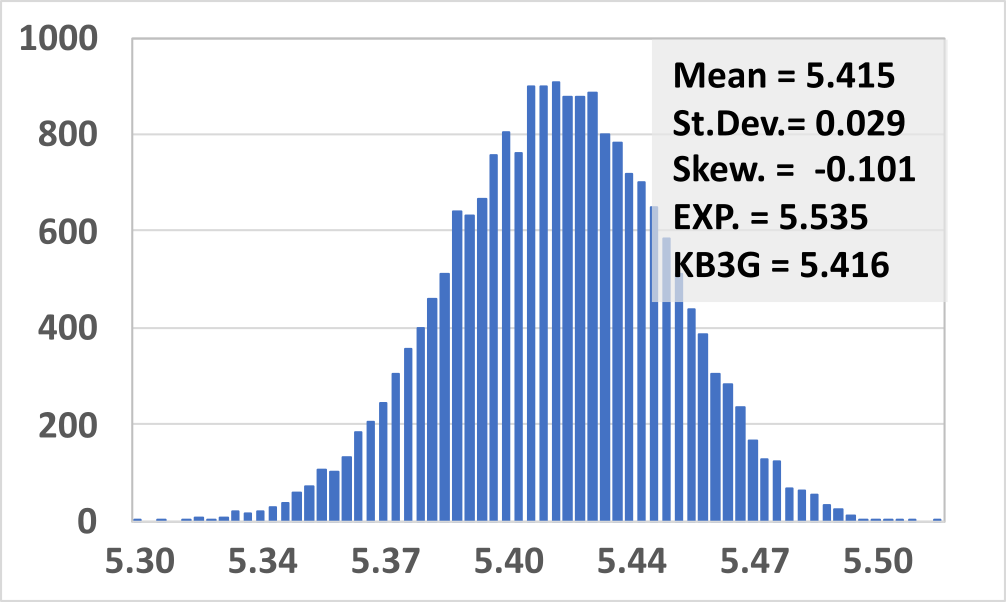}  
\\ \hline
$^{48}$Ti Occ(Np1) & \includegraphics[width=0.32\linewidth]{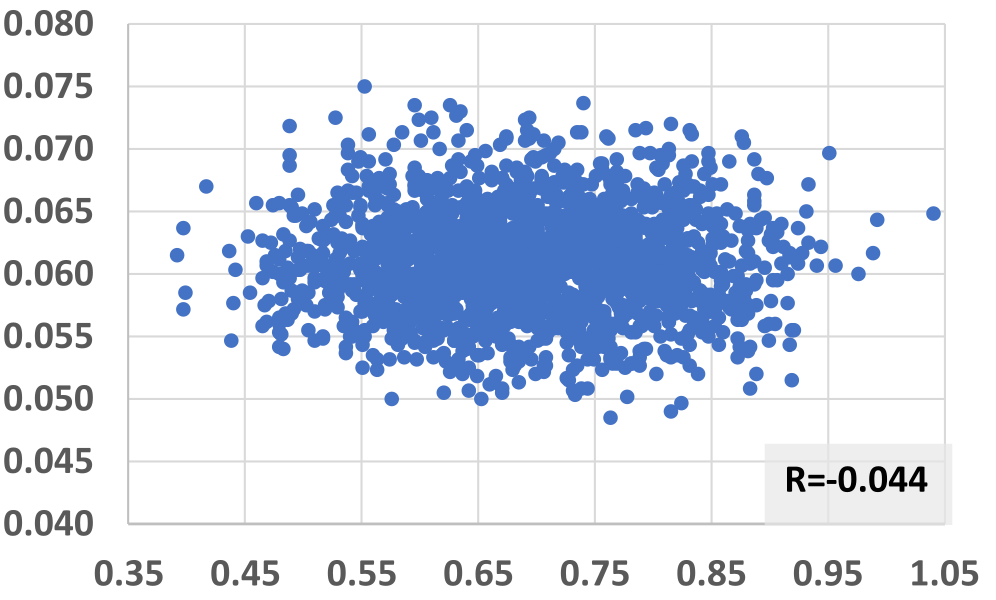} & \includegraphics[width=0.32\linewidth]{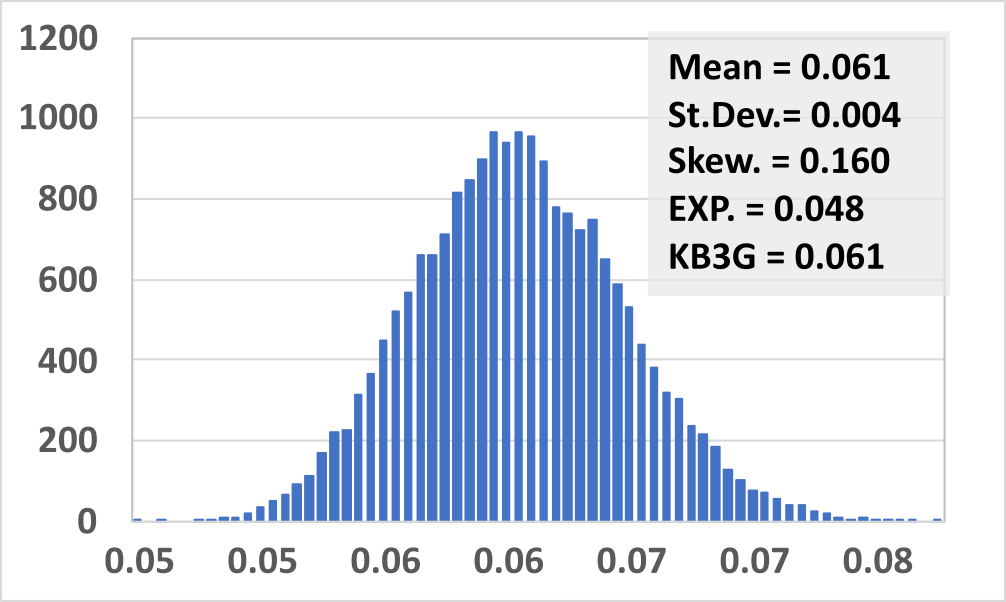}  
\\ \hline
$^{48}$Ti Occ(Np3) & \includegraphics[width=0.32\linewidth]{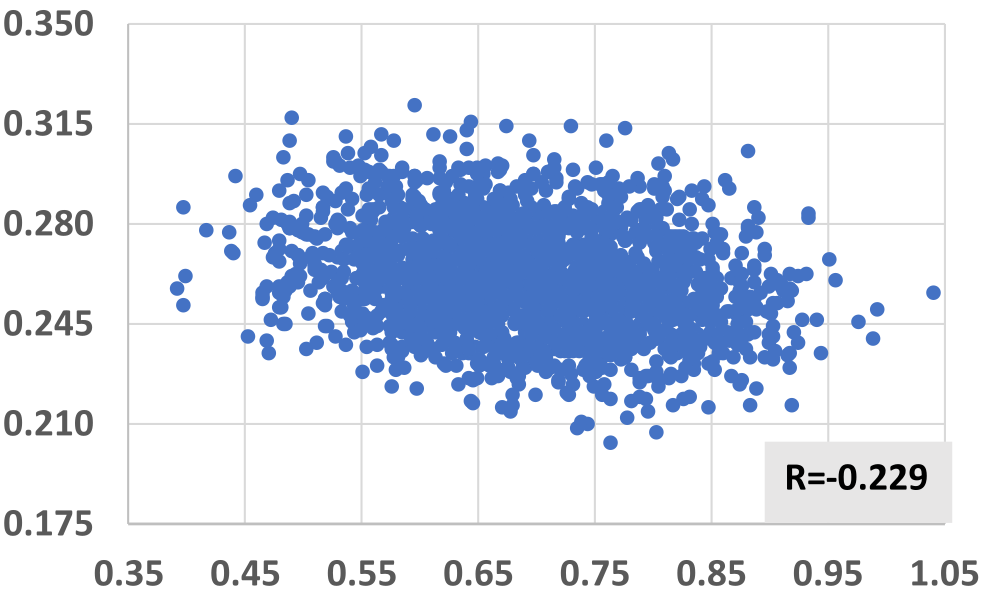} & \includegraphics[width=0.32\linewidth]{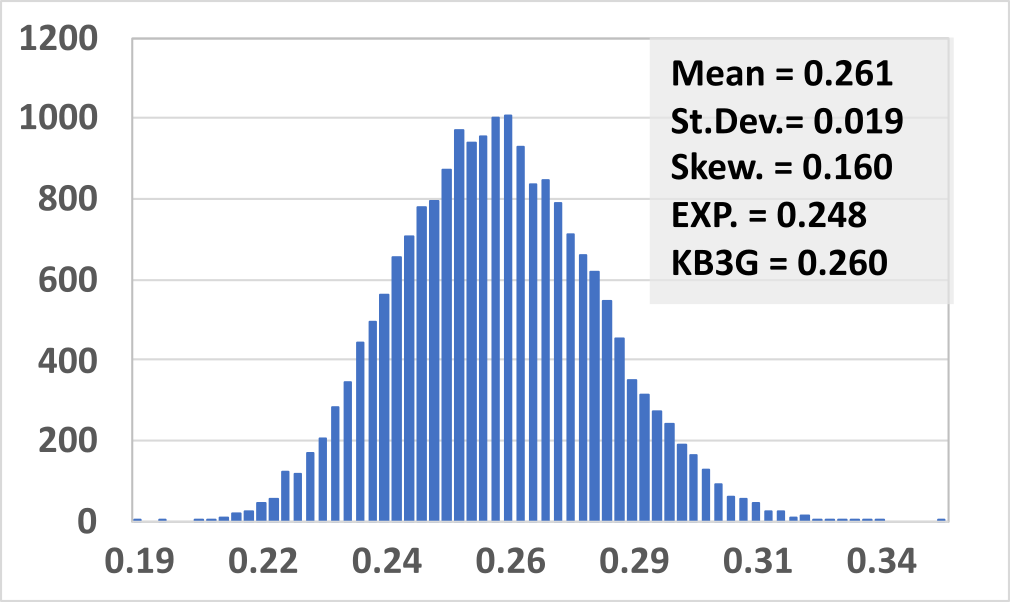}  
\\ \hline
$^{48}$Ti Occ(Pf5) & \includegraphics[width=0.32\linewidth]{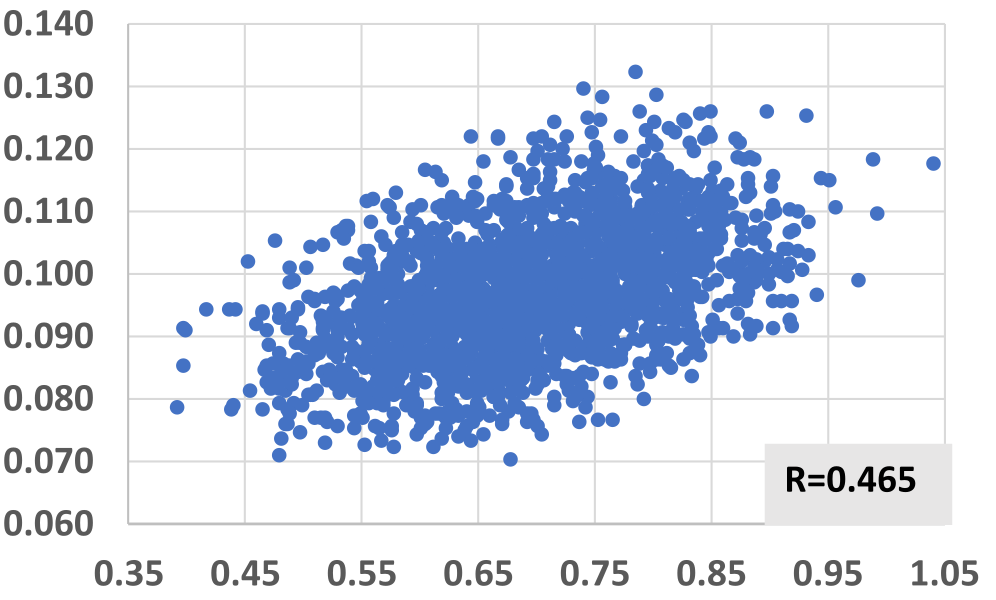} & \includegraphics[width=0.32\linewidth]{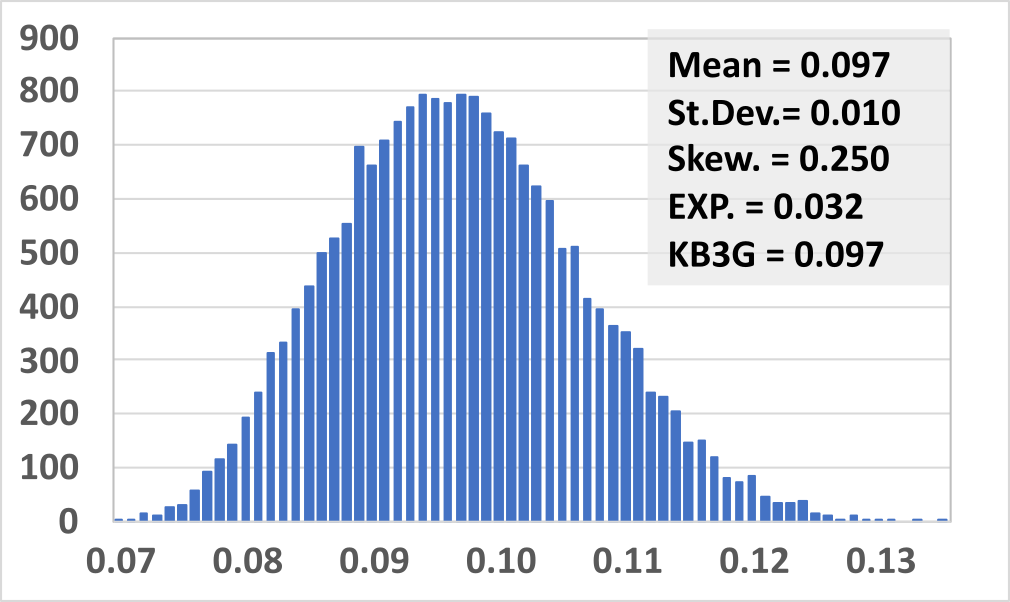}  
\\ \hline
$^{48}$Ti Occ(Pf7) & \includegraphics[width=0.32\linewidth]{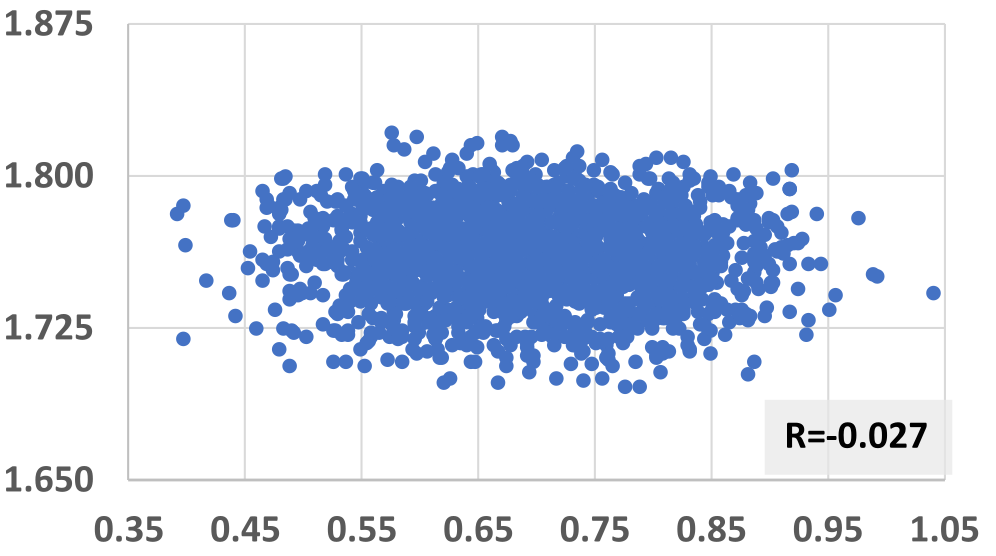} & \includegraphics[width=0.32\linewidth]{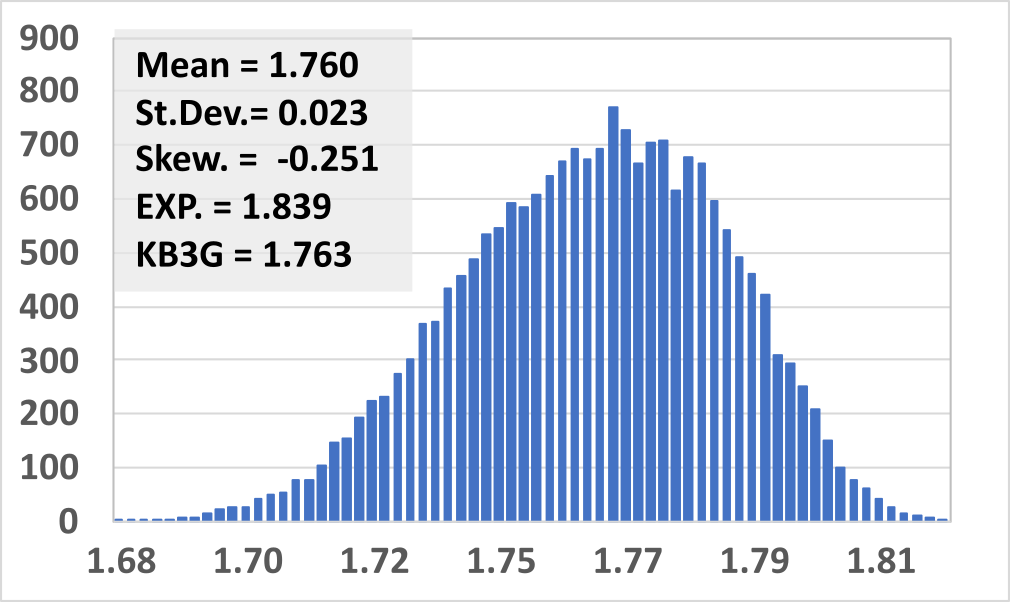}  
\\ \hline
$^{48}$Ti Occ(Pp1) & \includegraphics[width=0.32\linewidth]{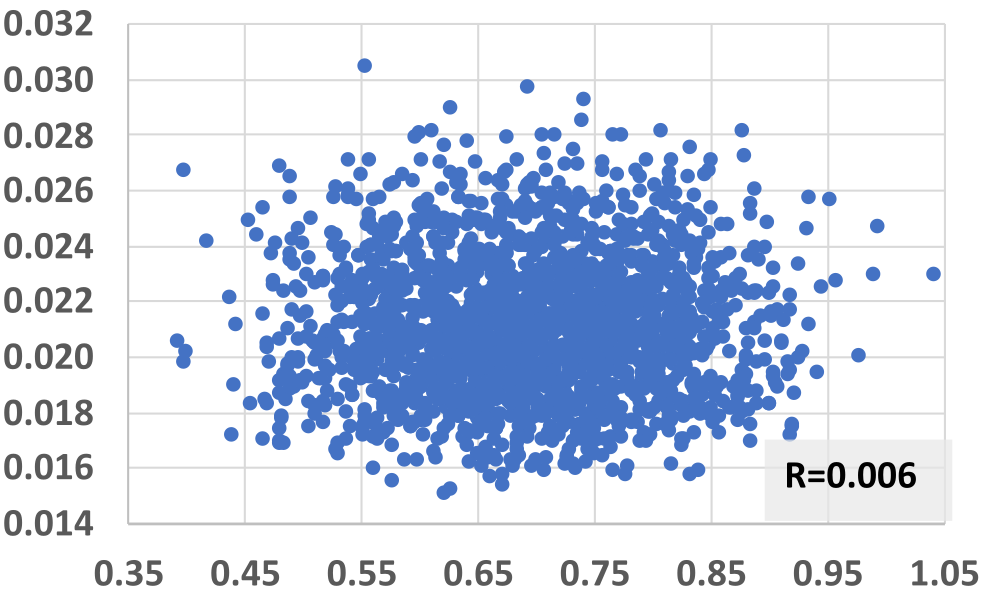} & \includegraphics[width=0.32\linewidth]{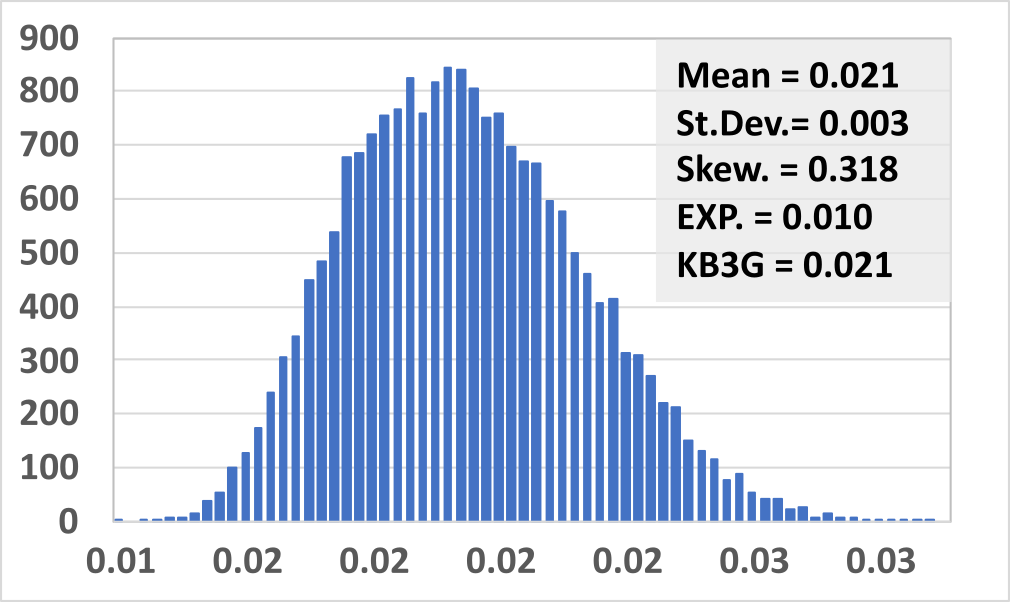}  
\\ \hline
$^{48}$Ti Occ(Pp3) & \includegraphics[width=0.32\linewidth]{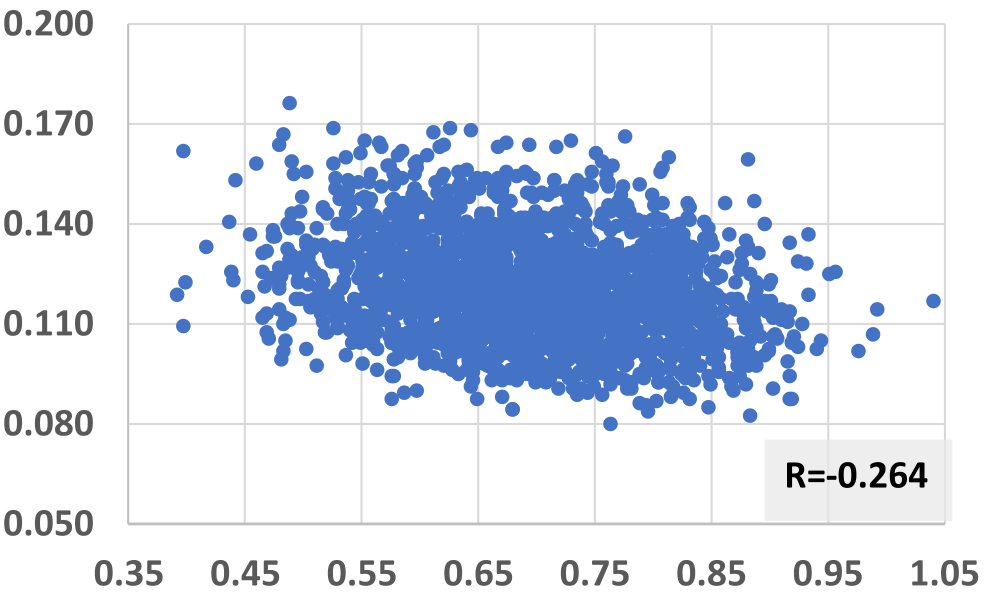} & \includegraphics[width=0.32\linewidth]{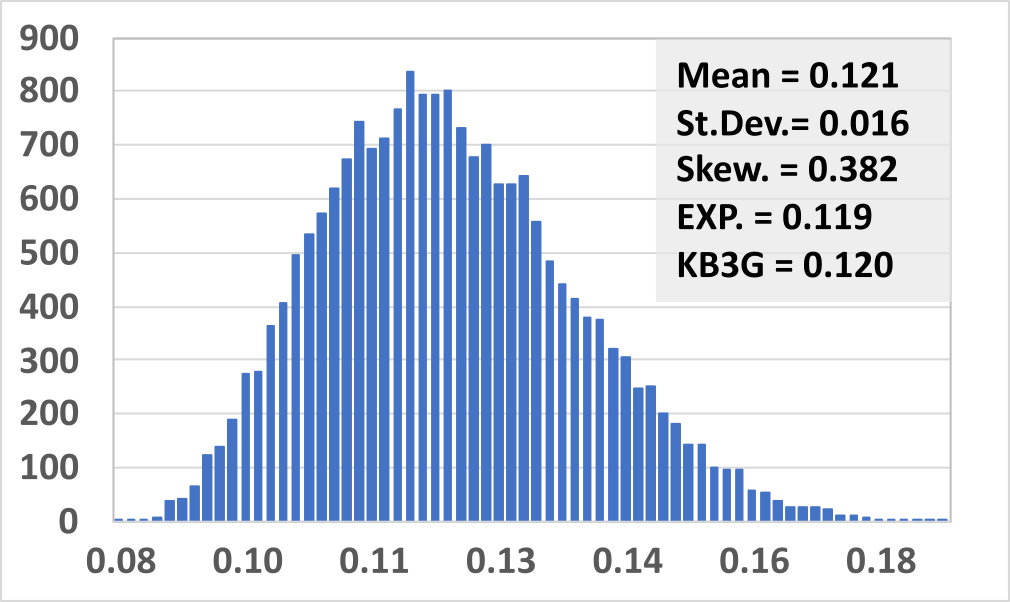}  
\\ \hline
\end{tabular}
 \caption{(Color online) Continuation of Table \ref{tab:parent-kb3g}. }
\label{tab:daughter-kb3g}
\end{table}
\vspace{0.5cm}
{\it Acknowledgements}. MH acknowledges support from the US Department of Energy grant DE-SC0022538 "Nuclear Astrophysics and Fundamental Symmetries". SS and AN acknowledge support by a grant of Romanian Ministry of Research, Innovation and Digitalization through the project CNCS – UEFISCDI number 99/2021 within PN-III-P4-ID-PCE-2020-2374.

\appendix*

\section{Gram-Charlier A series} \label{appendix}

In order to have a good representation for the PDF of the $0\nu\beta\beta$ NME, we consider small deviation from normal distribution via the Gram-Charlier A series \cite{Gram-Charlier}. This is given by:
\begin{equation} \label{GCeq}
\begin{split}
&  P(x) \approx \frac{1}{\sqrt{2\pi} \sigma} exp\left[-\frac{\left(x-\mu\right)^2}{2\sigma^2}\right] \\
\cdot & \left[ 1 + \frac{\mu_3}{3!} He_3((x-\mu)/\sigma) + \frac{\mu_4-3}{4!}He_4((x-\mu)/\sigma) \right]
\end{split}
\end{equation}

where $He_k(y)$ are the Chebyshev-Hermite polynomials, $He_3(y) = y^3 - 3y$ and $He_3(y) = y^4 - 6 y^2 + 3$, $\mu$ and $\sigma$ are the mean and variance of a probability density function (PDF), and $\mu_k$, with $k=3, 4$ are normalized moments of the same PDF, $P(x)$:

\begin{equation}
\mu_k = \int \left(\frac{x-\mu}{\sigma}\right)^k P(x) dx ,
\end{equation}

In practice, we use the sample moments, $\mu_3$ (skewnes) and $\mu_4 - 3$ (kurtosis), which in the limit of very large sample sizes become very close to the underlying moments.


\bibliographystyle{apsrev4-1}
\bibliography{bb-n}

\begin{thebibliography}{53}%
\makeatletter
\providecommand \@ifxundefined [1]{%
 \@ifx{#1\undefined}
}%
\providecommand \@ifnum [1]{%
 \ifnum #1\expandafter \@firstoftwo
 \else \expandafter \@secondoftwo
 \fi
}%
\providecommand \@ifx [1]{%
 \ifx #1\expandafter \@firstoftwo
 \else \expandafter \@secondoftwo
 \fi
}%
\providecommand \natexlab [1]{#1}%
\providecommand \enquote  [1]{``#1''}%
\providecommand \bibnamefont  [1]{#1}%
\providecommand \bibfnamefont [1]{#1}%
\providecommand \citenamefont [1]{#1}%
\providecommand \href@noop [0]{\@secondoftwo}%
\providecommand \href [0]{\begingroup \@sanitize@url \@href}%
\providecommand \@href[1]{\@@startlink{#1}\@@href}%
\providecommand \@@href[1]{\endgroup#1\@@endlink}%
\providecommand \@sanitize@url [0]{\catcode `\\12\catcode `\$12\catcode
  `\&12\catcode `\#12\catcode `\^12\catcode `\_12\catcode `\%12\relax}%
\providecommand \@@startlink[1]{}%
\providecommand \@@endlink[0]{}%
\providecommand \url  [0]{\begingroup\@sanitize@url \@url }%
\providecommand \@url [1]{\endgroup\@href {#1}{\urlprefix }}%
\providecommand \urlprefix  [0]{URL }%
\providecommand \Eprint [0]{\href }%
\providecommand \doibase [0]{http://dx.doi.org/}%
\providecommand \selectlanguage [0]{\@gobble}%
\providecommand \bibinfo  [0]{\@secondoftwo}%
\providecommand \bibfield  [0]{\@secondoftwo}%
\providecommand \translation [1]{[#1]}%
\providecommand \BibitemOpen [0]{}%
\providecommand \bibitemStop [0]{}%
\providecommand \bibitemNoStop [0]{.\EOS\space}%
\providecommand \EOS [0]{\spacefactor3000\relax}%
\providecommand \BibitemShut  [1]{\csname bibitem#1\endcsname}%
\let\auto@bib@innerbib\@empty
\bibitem [{\citenamefont {Avignone}\ \emph {et~al.}(2008)\citenamefont
  {Avignone}, \citenamefont {Elliott},\ and\ \citenamefont
  {Engel}}]{Avignone2008}%
  \BibitemOpen
  \bibfield  {author} {\bibinfo {author} {\bibfnamefont {F.~T.}\ \bibnamefont
  {Avignone}, \bibfnamefont {III}}, \bibinfo {author} {\bibfnamefont {S.~R.}\
  \bibnamefont {Elliott}}, \ and\ \bibinfo {author} {\bibfnamefont
  {J.}~\bibnamefont {Engel}},\ }\href {\doibase 10.1103/RevModPhys.80.481}
  {\bibfield  {journal} {\bibinfo  {journal} {Rev. Mod. Phys.}\ }\textbf
  {\bibinfo {volume} {80}},\ \bibinfo {pages} {481} (\bibinfo {year}
  {2008})}\BibitemShut {NoStop}%
\bibitem [{\citenamefont {Vergados}\ \emph {et~al.}(2012)\citenamefont
  {Vergados}, \citenamefont {Ejiri},\ and\ \citenamefont
  {Simkovic}}]{Vergados2012}%
  \BibitemOpen
  \bibfield  {author} {\bibinfo {author} {\bibfnamefont {J.~D.}\ \bibnamefont
  {Vergados}}, \bibinfo {author} {\bibfnamefont {H.}~\bibnamefont {Ejiri}}, \
  and\ \bibinfo {author} {\bibfnamefont {F.}~\bibnamefont {Simkovic}},\ }\href
  {\doibase 10.1088/0034-4885/75/10/106301} {\bibfield  {journal} {\bibinfo
  {journal} {Rep. Prog. Phys.}\ }\textbf {\bibinfo {volume} {75}},\ \bibinfo
  {pages} {106301} (\bibinfo {year} {2012})}\BibitemShut {NoStop}%
\bibitem [{\citenamefont {Doi}\ \emph {et~al.}(1985)\citenamefont {Doi},
  \citenamefont {Kotani},\ and\ \citenamefont {Takasugi}}]{Doi1985}%
  \BibitemOpen
  \bibfield  {author} {\bibinfo {author} {\bibfnamefont {M.}~\bibnamefont
  {Doi}}, \bibinfo {author} {\bibfnamefont {T.}~\bibnamefont {Kotani}}, \ and\
  \bibinfo {author} {\bibfnamefont {E.}~\bibnamefont {Takasugi}},\ }\href
  {\doibase 10.1143/PTPS.83.1} {\bibfield  {journal} {\bibinfo  {journal}
  {Prog. Theor. Phys. Suppl.}\ }\textbf {\bibinfo {volume} {83}},\ \bibinfo
  {pages} {1} (\bibinfo {year} {1985})}\BibitemShut {NoStop}%
\bibitem [{\citenamefont {Rodejohann}(2012)}]{Rodejohann2012}%
  \BibitemOpen
  \bibfield  {author} {\bibinfo {author} {\bibfnamefont {W.}~\bibnamefont
  {Rodejohann}},\ }\href {\doibase 10.1088/0954-3899/39/12/124008} {\bibfield
  {journal} {\bibinfo  {journal} {J. Phys. G}\ }\textbf {\bibinfo {volume}
  {39}},\ \bibinfo {pages} {124008} (\bibinfo {year} {2012})}\BibitemShut
  {NoStop}%
\bibitem [{\citenamefont {Deppisch}\ \emph {et~al.}(2012)\citenamefont
  {Deppisch}, \citenamefont {Hirsch},\ and\ \citenamefont
  {Pas}}]{Deppisch2012}%
  \BibitemOpen
  \bibfield  {author} {\bibinfo {author} {\bibfnamefont {F.~F.}\ \bibnamefont
  {Deppisch}}, \bibinfo {author} {\bibfnamefont {M.}~\bibnamefont {Hirsch}}, \
  and\ \bibinfo {author} {\bibfnamefont {H.}~\bibnamefont {Pas}},\ }\href
  {\doibase 10.1088/0954-3899/39/12/124007} {\bibfield  {journal} {\bibinfo
  {journal} {J. Phys. G}\ }\textbf {\bibinfo {volume} {39}},\ \bibinfo {pages}
  {124007} (\bibinfo {year} {2012})}\BibitemShut {NoStop}%
\bibitem [{\citenamefont {Horoi}\ and\ \citenamefont
  {Neacsu}(2016{\natexlab{a}})}]{HoroiNeacsu2016prd}%
  \BibitemOpen
  \bibfield  {author} {\bibinfo {author} {\bibfnamefont {M.}~\bibnamefont
  {Horoi}}\ and\ \bibinfo {author} {\bibfnamefont {A.}~\bibnamefont {Neacsu}},\
  }\href {\doibase 10.1103/PhysRevD.93.113014} {\bibfield  {journal} {\bibinfo
  {journal} {Phys. Rev. D}\ }\textbf {\bibinfo {volume} {93}},\ \bibinfo
  {pages} {113014} (\bibinfo {year} {2016}{\natexlab{a}})},\ \Eprint
  {http://arxiv.org/abs/arXiv:1511.00670 [hep-ph]} {arXiv:arXiv:1511.00670
  [hep-ph] [hep-ph]} \BibitemShut {NoStop}%
\bibitem [{\citenamefont {Neacsu}\ and\ \citenamefont
  {Horoi}(2016)}]{Neacsu2016ahep-dist}%
  \BibitemOpen
  \bibfield  {author} {\bibinfo {author} {\bibfnamefont {A.}~\bibnamefont
  {Neacsu}}\ and\ \bibinfo {author} {\bibfnamefont {M.}~\bibnamefont {Horoi}},\
  }\href@noop {} {\bibfield  {journal} {\bibinfo  {journal} {Advances in High
  Energy Physics}\ }\textbf {\bibinfo {volume} {2016}} (\bibinfo {year}
  {2016})}\BibitemShut {NoStop}%
\bibitem [{\citenamefont {Ahmed}\ \emph {et~al.}(2017)\citenamefont {Ahmed},
  \citenamefont {Neacsu},\ and\ \citenamefont {Horoi}}]{Ahmed2017}%
  \BibitemOpen
  \bibfield  {author} {\bibinfo {author} {\bibfnamefont {F.}~\bibnamefont
  {Ahmed}}, \bibinfo {author} {\bibfnamefont {A.}~\bibnamefont {Neacsu}}, \
  and\ \bibinfo {author} {\bibfnamefont {M.}~\bibnamefont {Horoi}},\
  }\href@noop {} {\bibfield  {journal} {\bibinfo  {journal} {Physics Letters
  B}\ }\textbf {\bibinfo {volume} {769}},\ \bibinfo {pages} {299} (\bibinfo
  {year} {2017})}\BibitemShut {NoStop}%
\bibitem [{\citenamefont {Engel}\ and\ \citenamefont
  {Men{\'{e}}ndez}(2017)}]{EngelMenendez2017}%
  \BibitemOpen
  \bibfield  {author} {\bibinfo {author} {\bibfnamefont {J.}~\bibnamefont
  {Engel}}\ and\ \bibinfo {author} {\bibfnamefont {J.}~\bibnamefont
  {Men{\'{e}}ndez}},\ }\href {\doibase 10.1088/1361-6633/aa5bc5} {\bibfield
  {journal} {\bibinfo  {journal} {Reports on Progress in Physics}\ }\textbf
  {\bibinfo {volume} {80}},\ \bibinfo {pages} {046301} (\bibinfo {year}
  {2017})}\BibitemShut {NoStop}%
\bibitem [{\citenamefont {Kotila}\ and\ \citenamefont
  {Iachello}(2012)}]{Kotila2012}%
  \BibitemOpen
  \bibfield  {author} {\bibinfo {author} {\bibfnamefont {J.}~\bibnamefont
  {Kotila}}\ and\ \bibinfo {author} {\bibfnamefont {F.}~\bibnamefont
  {Iachello}},\ }\href {\doibase 10.1103/PhysRevC.85.034316} {\bibfield
  {journal} {\bibinfo  {journal} {Phys. Rev. C}\ }\textbf {\bibinfo {volume}
  {85}},\ \bibinfo {pages} {034316} (\bibinfo {year} {2012})}\BibitemShut
  {NoStop}%
\bibitem [{\citenamefont {Stoica}\ and\ \citenamefont
  {Mirea}(2013)}]{StoicaMirea2013}%
  \BibitemOpen
  \bibfield  {author} {\bibinfo {author} {\bibfnamefont {S.}~\bibnamefont
  {Stoica}}\ and\ \bibinfo {author} {\bibfnamefont {M.}~\bibnamefont {Mirea}},\
  }\href {\doibase 10.1103/PhysRevC.88.037303} {\bibfield  {journal} {\bibinfo
  {journal} {Phys. Rev. C}\ }\textbf {\bibinfo {volume} {88}},\ \bibinfo
  {pages} {037303} (\bibinfo {year} {2013})}\BibitemShut {NoStop}%
\bibitem [{\citenamefont {Mirea}\ \emph {et~al.}(2015)\citenamefont {Mirea},
  \citenamefont {Pahomi},\ and\ \citenamefont {Stoica}}]{MireaPahomi2015}%
  \BibitemOpen
  \bibfield  {author} {\bibinfo {author} {\bibfnamefont {M.}~\bibnamefont
  {Mirea}}, \bibinfo {author} {\bibfnamefont {T.}~\bibnamefont {Pahomi}}, \
  and\ \bibinfo {author} {\bibfnamefont {S.}~\bibnamefont {Stoica}},\
  }\href@noop {} {\bibfield  {journal} {\bibinfo  {journal} {Rom. Rep. Phys.}\
  }\textbf {\bibinfo {volume} {67}},\ \bibinfo {pages} {872} (\bibinfo {year}
  {2015})}\BibitemShut {NoStop}%
\bibitem [{\citenamefont {Caurier}\ \emph {et~al.}(1990)\citenamefont
  {Caurier}, \citenamefont {Poves},\ and\ \citenamefont {Zuker}}]{Caurier1990}%
  \BibitemOpen
  \bibfield  {author} {\bibinfo {author} {\bibfnamefont {E.}~\bibnamefont
  {Caurier}}, \bibinfo {author} {\bibfnamefont {A.}~\bibnamefont {Poves}}, \
  and\ \bibinfo {author} {\bibfnamefont {A.~P.}\ \bibnamefont {Zuker}},\ }\href
  {\doibase 10.1016/0370-2693(90)91071-I} {\bibfield  {journal} {\bibinfo
  {journal} {Phys. Lett. B}\ }\textbf {\bibinfo {volume} {252}},\ \bibinfo
  {pages} {13} (\bibinfo {year} {1990})}\BibitemShut {NoStop}%
\bibitem [{\citenamefont {Caurier}\ \emph {et~al.}(1996)\citenamefont
  {Caurier}, \citenamefont {Nowacki}, \citenamefont {Poves},\ and\
  \citenamefont {Retamosa}}]{Caurier1996}%
  \BibitemOpen
  \bibfield  {author} {\bibinfo {author} {\bibfnamefont {E.}~\bibnamefont
  {Caurier}}, \bibinfo {author} {\bibfnamefont {F.}~\bibnamefont {Nowacki}},
  \bibinfo {author} {\bibfnamefont {A.}~\bibnamefont {Poves}}, \ and\ \bibinfo
  {author} {\bibfnamefont {J.}~\bibnamefont {Retamosa}},\ }\href {\doibase
  10.1103/PhysRevLett.77.1954} {\bibfield  {journal} {\bibinfo  {journal}
  {Phys. Rev. Lett.}\ }\textbf {\bibinfo {volume} {77}},\ \bibinfo {pages}
  {1954} (\bibinfo {year} {1996})}\BibitemShut {NoStop}%
\bibitem [{\citenamefont {Caurier}\ \emph {et~al.}(2005)\citenamefont
  {Caurier}, \citenamefont {Martinez-Pinedo}, \citenamefont {Nowacki},
  \citenamefont {Poves},\ and\ \citenamefont {Zuker}}]{Caurier2005}%
  \BibitemOpen
  \bibfield  {author} {\bibinfo {author} {\bibfnamefont {E.}~\bibnamefont
  {Caurier}}, \bibinfo {author} {\bibfnamefont {G.}~\bibnamefont
  {Martinez-Pinedo}}, \bibinfo {author} {\bibfnamefont {F.}~\bibnamefont
  {Nowacki}}, \bibinfo {author} {\bibfnamefont {A.}~\bibnamefont {Poves}}, \
  and\ \bibinfo {author} {\bibfnamefont {A.~P.}\ \bibnamefont {Zuker}},\ }\href
  {\doibase 10.1103/RevModPhys.77.427} {\bibfield  {journal} {\bibinfo
  {journal} {Rev. Mod. Phys.}\ }\textbf {\bibinfo {volume} {77}},\ \bibinfo
  {pages} {427} (\bibinfo {year} {2005})}\BibitemShut {NoStop}%
\bibitem [{\citenamefont {Horoi}\ \emph {et~al.}(2007)\citenamefont {Horoi},
  \citenamefont {Stoica},\ and\ \citenamefont {Brown}}]{HoroiStoicaBrown2007}%
  \BibitemOpen
  \bibfield  {author} {\bibinfo {author} {\bibfnamefont {M.}~\bibnamefont
  {Horoi}}, \bibinfo {author} {\bibfnamefont {S.}~\bibnamefont {Stoica}}, \
  and\ \bibinfo {author} {\bibfnamefont {B.~A.}\ \bibnamefont {Brown}},\ }\href
  {\doibase 10.1103/PhysRevC.75.034303} {\bibfield  {journal} {\bibinfo
  {journal} {Phys. Rev. C}\ }\textbf {\bibinfo {volume} {75}},\ \bibinfo
  {pages} {034303} (\bibinfo {year} {2007})}\BibitemShut {NoStop}%
\bibitem [{\citenamefont {Horoi}\ and\ \citenamefont
  {Stoica}(2010)}]{HoroiStoica2010}%
  \BibitemOpen
  \bibfield  {author} {\bibinfo {author} {\bibfnamefont {M.}~\bibnamefont
  {Horoi}}\ and\ \bibinfo {author} {\bibfnamefont {S.}~\bibnamefont {Stoica}},\
  }\href {\doibase 10.1103/PhysRevC.81.024321} {\bibfield  {journal} {\bibinfo
  {journal} {Phys. Rev. C}\ }\textbf {\bibinfo {volume} {81}},\ \bibinfo
  {pages} {024321} (\bibinfo {year} {2010})}\BibitemShut {NoStop}%
\bibitem [{\citenamefont {Horoi}(2013)}]{Horoi2013}%
  \BibitemOpen
  \bibfield  {author} {\bibinfo {author} {\bibfnamefont {M.}~\bibnamefont
  {Horoi}},\ }\href {\doibase 10.1103/PhysRevC.87.014320} {\bibfield  {journal}
  {\bibinfo  {journal} {Phys. Rev. C}\ }\textbf {\bibinfo {volume} {87}},\
  \bibinfo {pages} {014320} (\bibinfo {year} {2013})}\BibitemShut {NoStop}%
\bibitem [{\citenamefont {Horoi}\ and\ \citenamefont
  {Brown}(2013)}]{HoroiBrown2013}%
  \BibitemOpen
  \bibfield  {author} {\bibinfo {author} {\bibfnamefont {M.}~\bibnamefont
  {Horoi}}\ and\ \bibinfo {author} {\bibfnamefont {B.~A.}\ \bibnamefont
  {Brown}},\ }\href {\doibase 10.1103/PhysRevLett.110.222502} {\bibfield
  {journal} {\bibinfo  {journal} {Phys. Rev. Lett.}\ }\textbf {\bibinfo
  {volume} {110}},\ \bibinfo {pages} {222502} (\bibinfo {year}
  {2013})}\BibitemShut {NoStop}%
\bibitem [{\citenamefont {Sen'kov}\ and\ \citenamefont
  {Horoi}(2014)}]{SenkovHoroi2014}%
  \BibitemOpen
  \bibfield  {author} {\bibinfo {author} {\bibfnamefont {R.~A.}\ \bibnamefont
  {Sen'kov}}\ and\ \bibinfo {author} {\bibfnamefont {M.}~\bibnamefont
  {Horoi}},\ }\href {\doibase 10.1103/PhysRevC.90.051301} {\bibfield  {journal}
  {\bibinfo  {journal} {Phys. Rev. C}\ }\textbf {\bibinfo {volume} {90}},\
  \bibinfo {pages} {{051301(R)}} (\bibinfo {year} {2014})}\BibitemShut
  {NoStop}%
\bibitem [{\citenamefont {Neacsu}\ and\ \citenamefont
  {Horoi}(2015)}]{NeacsuHoroi2015}%
  \BibitemOpen
  \bibfield  {author} {\bibinfo {author} {\bibfnamefont {A.}~\bibnamefont
  {Neacsu}}\ and\ \bibinfo {author} {\bibfnamefont {M.}~\bibnamefont {Horoi}},\
  }\href {\doibase 10.1103/PhysRevC.91.024309} {\bibfield  {journal} {\bibinfo
  {journal} {Phys. Rev. C}\ }\textbf {\bibinfo {volume} {91}},\ \bibinfo
  {pages} {024309} (\bibinfo {year} {2015})}\BibitemShut {NoStop}%
\bibitem [{\citenamefont {Horoi}\ and\ \citenamefont
  {Neacsu}(2016{\natexlab{b}})}]{NeacsuHoroi2016}%
  \BibitemOpen
  \bibfield  {author} {\bibinfo {author} {\bibfnamefont {M.}~\bibnamefont
  {Horoi}}\ and\ \bibinfo {author} {\bibfnamefont {A.}~\bibnamefont {Neacsu}},\
  }\href {\doibase 10.1103/PhysRevC.93.024308} {\bibfield  {journal} {\bibinfo
  {journal} {Phys. Rev. C}\ }\textbf {\bibinfo {volume} {93}},\ \bibinfo
  {pages} {024308} (\bibinfo {year} {2016}{\natexlab{b}})}\BibitemShut
  {NoStop}%
\bibitem [{\citenamefont {Horoi}\ and\ \citenamefont
  {Neacsu}(2018)}]{18ho035502}%
  \BibitemOpen
  \bibfield  {author} {\bibinfo {author} {\bibfnamefont {M.}~\bibnamefont
  {Horoi}}\ and\ \bibinfo {author} {\bibfnamefont {A.}~\bibnamefont {Neacsu}},\
  }\href {\doibase 10.1103/PhysRevC.98.035502} {\bibfield  {journal} {\bibinfo
  {journal} {Phys. Rev. C}\ }\textbf {\bibinfo {volume} {98}},\ \bibinfo
  {pages} {035502} (\bibinfo {year} {2018})}\BibitemShut {NoStop}%
\bibitem [{\citenamefont {Suhonen}\ and\ \citenamefont
  {Civitarese}(1998)}]{SuhonenCivitarese1998}%
  \BibitemOpen
  \bibfield  {author} {\bibinfo {author} {\bibfnamefont {J.}~\bibnamefont
  {Suhonen}}\ and\ \bibinfo {author} {\bibfnamefont {O.}~\bibnamefont
  {Civitarese}},\ }\href {\doibase 10.1016/S0370-1573(97)00087-2} {\bibfield
  {journal} {\bibinfo  {journal} {Phys. Rep.}\ }\textbf {\bibinfo {volume}
  {300}},\ \bibinfo {pages} {123} (\bibinfo {year} {1998})}\BibitemShut
  {NoStop}%
\bibitem [{\citenamefont {Simkovic}\ \emph {et~al.}(1999)\citenamefont
  {Simkovic}, \citenamefont {Pantis}, \citenamefont {Vergados},\ and\
  \citenamefont {Faessler}}]{Simkovic1999}%
  \BibitemOpen
  \bibfield  {author} {\bibinfo {author} {\bibfnamefont {F.}~\bibnamefont
  {Simkovic}}, \bibinfo {author} {\bibfnamefont {G.}~\bibnamefont {Pantis}},
  \bibinfo {author} {\bibfnamefont {J.~D.}\ \bibnamefont {Vergados}}, \ and\
  \bibinfo {author} {\bibfnamefont {A.}~\bibnamefont {Faessler}},\ }\href
  {\doibase 10.1103/PhysRevC.60.055502} {\bibfield  {journal} {\bibinfo
  {journal} {Phys. Rev. C}\ }\textbf {\bibinfo {volume} {60}},\ \bibinfo
  {pages} {055502} (\bibinfo {year} {1999})}\BibitemShut {NoStop}%
\bibitem [{\citenamefont {Stoica}\ and\ \citenamefont
  {Klapdor-Kleingrothaus}(2001)}]{Stoica2001}%
  \BibitemOpen
  \bibfield  {author} {\bibinfo {author} {\bibfnamefont {S.}~\bibnamefont
  {Stoica}}\ and\ \bibinfo {author} {\bibfnamefont {H.}~\bibnamefont
  {Klapdor-Kleingrothaus}},\ }\href {\doibase 10.1016/S0375-9474(01)00988-5}
  {\bibfield  {journal} {\bibinfo  {journal} {Nucl. Phys. A}\ }\textbf
  {\bibinfo {volume} {694}},\ \bibinfo {pages} {269} (\bibinfo {year}
  {2001})}\BibitemShut {NoStop}%
\bibitem [{\citenamefont {Rodin}\ \emph {et~al.}(2006)\citenamefont {Rodin},
  \citenamefont {Faessler}, \citenamefont {Simkovic},\ and\ \citenamefont
  {Vogel}}]{Rodin2006}%
  \BibitemOpen
  \bibfield  {author} {\bibinfo {author} {\bibfnamefont {V.}~\bibnamefont
  {Rodin}}, \bibinfo {author} {\bibfnamefont {A.}~\bibnamefont {Faessler}},
  \bibinfo {author} {\bibfnamefont {F.}~\bibnamefont {Simkovic}}, \ and\
  \bibinfo {author} {\bibfnamefont {P.}~\bibnamefont {Vogel}},\ }\href
  {\doibase 10.1016/j.nuclphysa.2005.12.004} {\bibfield  {journal} {\bibinfo
  {journal} {Nucl. Phys. A}\ }\textbf {\bibinfo {volume} {766}},\ \bibinfo
  {pages} {107} (\bibinfo {year} {2006})}\BibitemShut {NoStop}%
\bibitem [{\citenamefont {Kortelainen}\ and\ \citenamefont
  {Suhonen}(2007)}]{KortelainenSuhonnen2007}%
  \BibitemOpen
  \bibfield  {author} {\bibinfo {author} {\bibfnamefont {M.}~\bibnamefont
  {Kortelainen}}\ and\ \bibinfo {author} {\bibfnamefont {J.}~\bibnamefont
  {Suhonen}},\ }\href {\doibase 10.1103/PhysRevC.75.051303} {\bibfield
  {journal} {\bibinfo  {journal} {Phys. Rev. C}\ }\textbf {\bibinfo {volume}
  {75}},\ \bibinfo {pages} {051303} (\bibinfo {year} {2007})}\BibitemShut
  {NoStop}%
\bibitem [{\citenamefont {Faessler}\ \emph {et~al.}(2012)\citenamefont
  {Faessler}, \citenamefont {Rodin},\ and\ \citenamefont
  {Simkovic}}]{Faessler2012}%
  \BibitemOpen
  \bibfield  {author} {\bibinfo {author} {\bibfnamefont {A.}~\bibnamefont
  {Faessler}}, \bibinfo {author} {\bibfnamefont {V.}~\bibnamefont {Rodin}}, \
  and\ \bibinfo {author} {\bibfnamefont {F.}~\bibnamefont {Simkovic}},\ }\href
  {\doibase 10.1088/0954-3899/39/12/124006} {\bibfield  {journal} {\bibinfo
  {journal} {J. Phys. G}\ }\textbf {\bibinfo {volume} {39}},\ \bibinfo {pages}
  {124006} (\bibinfo {year} {2012})}\BibitemShut {NoStop}%
\bibitem [{\citenamefont {Simkovic}\ \emph {et~al.}(2013)\citenamefont
  {Simkovic}, \citenamefont {Rodin}, \citenamefont {Faessler},\ and\
  \citenamefont {Vogel}}]{SimkovicRodin2013}%
  \BibitemOpen
  \bibfield  {author} {\bibinfo {author} {\bibfnamefont {F.}~\bibnamefont
  {Simkovic}}, \bibinfo {author} {\bibfnamefont {V.}~\bibnamefont {Rodin}},
  \bibinfo {author} {\bibfnamefont {A.}~\bibnamefont {Faessler}}, \ and\
  \bibinfo {author} {\bibfnamefont {P.}~\bibnamefont {Vogel}},\ }\href
  {\doibase 10.1103/PhysRevC.87.045501} {\bibfield  {journal} {\bibinfo
  {journal} {Phys. Rev. C}\ }\textbf {\bibinfo {volume} {87}},\ \bibinfo
  {pages} {045501} (\bibinfo {year} {2013})}\BibitemShut {NoStop}%
\bibitem [{\citenamefont {Barea}\ and\ \citenamefont
  {Iachello}(2009)}]{Barea2009}%
  \BibitemOpen
  \bibfield  {author} {\bibinfo {author} {\bibfnamefont {J.}~\bibnamefont
  {Barea}}\ and\ \bibinfo {author} {\bibfnamefont {F.}~\bibnamefont
  {Iachello}},\ }\href {\doibase 10.1103/PhysRevC.79.044301} {\bibfield
  {journal} {\bibinfo  {journal} {Phys. Rev. C}\ }\textbf {\bibinfo {volume}
  {79}},\ \bibinfo {pages} {044301} (\bibinfo {year} {2009})}\BibitemShut
  {NoStop}%
\bibitem [{\citenamefont {Barea}\ \emph {et~al.}(2013)\citenamefont {Barea},
  \citenamefont {Kotila},\ and\ \citenamefont {Iachello}}]{Barea2013}%
  \BibitemOpen
  \bibfield  {author} {\bibinfo {author} {\bibfnamefont {J.}~\bibnamefont
  {Barea}}, \bibinfo {author} {\bibfnamefont {J.}~\bibnamefont {Kotila}}, \
  and\ \bibinfo {author} {\bibfnamefont {F.}~\bibnamefont {Iachello}},\ }\href
  {\doibase 10.1103/PhysRevC.87.014315} {\bibfield  {journal} {\bibinfo
  {journal} {Phys. Rev. C}\ }\textbf {\bibinfo {volume} {87}},\ \bibinfo
  {pages} {014315} (\bibinfo {year} {2013})}\BibitemShut {NoStop}%
\bibitem [{\citenamefont {Rodriguez}\ and\ \citenamefont
  {Martinez-Pinedo}(2010)}]{Rodriguez2010}%
  \BibitemOpen
  \bibfield  {author} {\bibinfo {author} {\bibfnamefont {T.~R.}\ \bibnamefont
  {Rodriguez}}\ and\ \bibinfo {author} {\bibfnamefont {G.}~\bibnamefont
  {Martinez-Pinedo}},\ }\href {\doibase 10.1103/PhysRevLett.105.252503}
  {\bibfield  {journal} {\bibinfo  {journal} {Phys. Rev. Lett.}\ }\textbf
  {\bibinfo {volume} {105}},\ \bibinfo {pages} {252503} (\bibinfo {year}
  {2010})}\BibitemShut {NoStop}%
\bibitem [{\citenamefont {Rath}\ \emph {et~al.}(2013)\citenamefont {Rath},
  \citenamefont {Chandra}, \citenamefont {Chaturvedi}, \citenamefont {Lohani},
  \citenamefont {Raina},\ and\ \citenamefont {Hirsch}}]{Rath2013}%
  \BibitemOpen
  \bibfield  {author} {\bibinfo {author} {\bibfnamefont {P.~K.}\ \bibnamefont
  {Rath}}, \bibinfo {author} {\bibfnamefont {R.}~\bibnamefont {Chandra}},
  \bibinfo {author} {\bibfnamefont {K.}~\bibnamefont {Chaturvedi}}, \bibinfo
  {author} {\bibfnamefont {P.}~\bibnamefont {Lohani}}, \bibinfo {author}
  {\bibfnamefont {P.~K.}\ \bibnamefont {Raina}}, \ and\ \bibinfo {author}
  {\bibfnamefont {J.~G.}\ \bibnamefont {Hirsch}},\ }\href {\doibase
  10.1103/PhysRevC.88.064322} {\bibfield  {journal} {\bibinfo  {journal} {Phys.
  Rev. C}\ }\textbf {\bibinfo {volume} {88}},\ \bibinfo {pages} {064322}
  (\bibinfo {year} {2013})}\BibitemShut {NoStop}%
\bibitem [{\citenamefont {Novario}\ \emph {et~al.}(2021)\citenamefont
  {Novario}, \citenamefont {Gysbers}, \citenamefont {Engel}, \citenamefont
  {Hagen}, \citenamefont {Jansen}, \citenamefont {Morris}, \citenamefont
  {Navr{\'{a}}til}, \citenamefont {Papenbrock},\ and\ \citenamefont
  {Quaglioni}}]{Novario2021}%
  \BibitemOpen
  \bibfield  {author} {\bibinfo {author} {\bibfnamefont {S.}~\bibnamefont
  {Novario}}, \bibinfo {author} {\bibfnamefont {P.}~\bibnamefont {Gysbers}},
  \bibinfo {author} {\bibfnamefont {J.}~\bibnamefont {Engel}}, \bibinfo
  {author} {\bibfnamefont {G.}~\bibnamefont {Hagen}}, \bibinfo {author}
  {\bibfnamefont {G.~R.}\ \bibnamefont {Jansen}}, \bibinfo {author}
  {\bibfnamefont {T.~D.}\ \bibnamefont {Morris}}, \bibinfo {author}
  {\bibfnamefont {P.}~\bibnamefont {Navr{\'{a}}til}}, \bibinfo {author}
  {\bibfnamefont {T.}~\bibnamefont {Papenbrock}}, \ and\ \bibinfo {author}
  {\bibfnamefont {S.}~\bibnamefont {Quaglioni}},\ }\href {\doibase
  10.1103/PhysRevLett.126.182502} {\bibfield  {journal} {\bibinfo  {journal}
  {Phys. Rev. Lett.}\ }\textbf {\bibinfo {volume} {126}},\ \bibinfo {pages}
  {182502} (\bibinfo {year} {2021})}\BibitemShut {NoStop}%
\bibitem [{\citenamefont {Yao}\ \emph {et~al.}(2020)\citenamefont {Yao},
  \citenamefont {Bally}, \citenamefont {Engel}, \citenamefont {Wirth},
  \citenamefont {Rodr{\'{i}}guez},\ and\ \citenamefont {Hergert}}]{Yao2020}%
  \BibitemOpen
  \bibfield  {author} {\bibinfo {author} {\bibfnamefont {J.~M.}\ \bibnamefont
  {Yao}}, \bibinfo {author} {\bibfnamefont {B.}~\bibnamefont {Bally}}, \bibinfo
  {author} {\bibfnamefont {J.}~\bibnamefont {Engel}}, \bibinfo {author}
  {\bibfnamefont {R.}~\bibnamefont {Wirth}}, \bibinfo {author} {\bibfnamefont
  {T.~R.}\ \bibnamefont {Rodr{\'{i}}guez}}, \ and\ \bibinfo {author}
  {\bibfnamefont {H.}~\bibnamefont {Hergert}},\ }\href {\doibase
  10.1103/PhysRevLett.124.232501} {\bibfield  {journal} {\bibinfo  {journal}
  {Phys. Rev. Lett.}\ }\textbf {\bibinfo {volume} {124}},\ \bibinfo {pages}
  {232501} (\bibinfo {year} {2020})}\BibitemShut {NoStop}%
\bibitem [{\citenamefont {Belley}\ \emph {et~al.}(2021)\citenamefont {Belley},
  \citenamefont {Payne}, \citenamefont {Stroberg}, \citenamefont {Miyagi},\
  and\ \citenamefont {Holt}}]{Belley2021}%
  \BibitemOpen
  \bibfield  {author} {\bibinfo {author} {\bibfnamefont {A.}~\bibnamefont
  {Belley}}, \bibinfo {author} {\bibfnamefont {C.~G.}\ \bibnamefont {Payne}},
  \bibinfo {author} {\bibfnamefont {S.~R.}\ \bibnamefont {Stroberg}}, \bibinfo
  {author} {\bibfnamefont {T.}~\bibnamefont {Miyagi}}, \ and\ \bibinfo {author}
  {\bibfnamefont {J.~D.}\ \bibnamefont {Holt}},\ }\href {\doibase
  10.1103/PhysRevLett.126.042502} {\bibfield  {journal} {\bibinfo  {journal}
  {Phys. Rev. Lett.}\ }\textbf {\bibinfo {volume} {126}},\ \bibinfo {pages}
  {042502} (\bibinfo {year} {2021})}\BibitemShut {NoStop}%
\bibitem [{\citenamefont {Barabash}(2020)}]{Barabash2020}%
  \BibitemOpen
  \bibfield  {author} {\bibinfo {author} {\bibfnamefont {A.}~\bibnamefont
  {Barabash}},\ }\href {\doibase {10.3390/universe6100159}} {\bibfield
  {journal} {\bibinfo  {journal} {{UNIVERSE}}\ }\textbf {\bibinfo {volume}
  {{6}}},\ \bibinfo {pages} {{159}} (\bibinfo {year} {{2020}})}\BibitemShut
  {NoStop}%
\bibitem [{\citenamefont {Pritychenko}\ \emph {et~al.}(2016)\citenamefont
  {Pritychenko}, \citenamefont {Birch}, \citenamefont {Singh},\ and\
  \citenamefont {Horoi}}]{Pritychenko2016}%
  \BibitemOpen
  \bibfield  {author} {\bibinfo {author} {\bibfnamefont {B.}~\bibnamefont
  {Pritychenko}}, \bibinfo {author} {\bibfnamefont {M.}~\bibnamefont {Birch}},
  \bibinfo {author} {\bibfnamefont {B.}~\bibnamefont {Singh}}, \ and\ \bibinfo
  {author} {\bibfnamefont {M.}~\bibnamefont {Horoi}},\ }\href {\doibase
  10.1016/j.adt.2015.10.001} {\bibfield  {journal} {\bibinfo  {journal} {ATOMIC
  DATA AND NUCLEAR DATA TABLES}\ }\textbf {\bibinfo {volume} {107}},\ \bibinfo
  {pages} {1} (\bibinfo {year} {2016})}\BibitemShut {NoStop}%
\bibitem [{\citenamefont {Chen}(2022)}]{NNDC_A48-2022}%
  \BibitemOpen
  \bibfield  {author} {\bibinfo {author} {\bibfnamefont {J.}~\bibnamefont
  {Chen}},\ }\href {\doibase https://doi.org/10.1016/j.nds.2021.12.001}
  {\bibfield  {journal} {\bibinfo  {journal} {Nuclear Data Sheets}\ }\textbf
  {\bibinfo {volume} {179}},\ \bibinfo {pages} {1} (\bibinfo {year}
  {2022})}\BibitemShut {NoStop}%
\bibitem [{\citenamefont {Grewe}\ \emph {et~al.}(2007)\citenamefont {Grewe},
  \citenamefont {Frekers}, \citenamefont {Rakers}, \citenamefont {Adachi},
  \citenamefont {B\"aumer}, \citenamefont {Botha}, \citenamefont {Dohmann},
  \citenamefont {Fujita}, \citenamefont {Fujita}, \citenamefont {Hatanaka},
  \citenamefont {Nakanishi}, \citenamefont {Negret}, \citenamefont {Neveling},
  \citenamefont {Popescu}, \citenamefont {Sakemi}, \citenamefont {Shimbara},
  \citenamefont {Shimizu}, \citenamefont {Smit}, \citenamefont {Tameshige},
  \citenamefont {Tamii}, \citenamefont {Thies}, \citenamefont {Brentano},
  \citenamefont {Yosoi},\ and\ \citenamefont {Zegers}}]{GreweGT2007}%
  \BibitemOpen
  \bibfield  {author} {\bibinfo {author} {\bibfnamefont {E.-W.}\ \bibnamefont
  {Grewe}}, \bibinfo {author} {\bibfnamefont {D.}~\bibnamefont {Frekers}},
  \bibinfo {author} {\bibfnamefont {S.}~\bibnamefont {Rakers}}, \bibinfo
  {author} {\bibfnamefont {T.}~\bibnamefont {Adachi}}, \bibinfo {author}
  {\bibfnamefont {C.}~\bibnamefont {B\"aumer}}, \bibinfo {author}
  {\bibfnamefont {N.~T.}\ \bibnamefont {Botha}}, \bibinfo {author}
  {\bibfnamefont {H.}~\bibnamefont {Dohmann}}, \bibinfo {author} {\bibfnamefont
  {H.}~\bibnamefont {Fujita}}, \bibinfo {author} {\bibfnamefont
  {Y.}~\bibnamefont {Fujita}}, \bibinfo {author} {\bibfnamefont
  {K.}~\bibnamefont {Hatanaka}}, \bibinfo {author} {\bibfnamefont
  {K.}~\bibnamefont {Nakanishi}}, \bibinfo {author} {\bibfnamefont
  {A.}~\bibnamefont {Negret}}, \bibinfo {author} {\bibfnamefont
  {R.}~\bibnamefont {Neveling}}, \bibinfo {author} {\bibfnamefont
  {L.}~\bibnamefont {Popescu}}, \bibinfo {author} {\bibfnamefont
  {Y.}~\bibnamefont {Sakemi}}, \bibinfo {author} {\bibfnamefont
  {Y.}~\bibnamefont {Shimbara}}, \bibinfo {author} {\bibfnamefont
  {Y.}~\bibnamefont {Shimizu}}, \bibinfo {author} {\bibfnamefont {F.~D.}\
  \bibnamefont {Smit}}, \bibinfo {author} {\bibfnamefont {Y.}~\bibnamefont
  {Tameshige}}, \bibinfo {author} {\bibfnamefont {A.}~\bibnamefont {Tamii}},
  \bibinfo {author} {\bibfnamefont {J.}~\bibnamefont {Thies}}, \bibinfo
  {author} {\bibfnamefont {P.~v.}\ \bibnamefont {Brentano}}, \bibinfo {author}
  {\bibfnamefont {M.}~\bibnamefont {Yosoi}}, \ and\ \bibinfo {author}
  {\bibfnamefont {R.~G.~T.}\ \bibnamefont {Zegers}},\ }\href {\doibase
  10.1103/PhysRevC.76.054307} {\bibfield  {journal} {\bibinfo  {journal} {Phys.
  Rev. C}\ }\textbf {\bibinfo {volume} {76}},\ \bibinfo {pages} {054307}
  (\bibinfo {year} {2007})}\BibitemShut {NoStop}%
\bibitem [{\citenamefont {Richter}\ \emph {et~al.}(1991)\citenamefont
  {Richter}, \citenamefont {van~der Merwe}, \citenamefont {Julies},\ and\
  \citenamefont {Brown}}]{fpd6}%
  \BibitemOpen
  \bibfield  {author} {\bibinfo {author} {\bibfnamefont {A.}~\bibnamefont
  {Richter}}, \bibinfo {author} {\bibfnamefont {M.~G.}\ \bibnamefont {van~der
  Merwe}}, \bibinfo {author} {\bibfnamefont {R.~E.}\ \bibnamefont {Julies}}, \
  and\ \bibinfo {author} {\bibfnamefont {B.~A.}\ \bibnamefont {Brown}},\
  }\href@noop {} {\bibfield  {journal} {\bibinfo  {journal} {Nucl. Phys. A}\
  }\textbf {\bibinfo {volume} {523}},\ \bibinfo {pages} {325} (\bibinfo {year}
  {1991})}\BibitemShut {NoStop}%
\bibitem [{\citenamefont {Honma}\ \emph {et~al.}(2004)\citenamefont {Honma},
  \citenamefont {Otsuka}, \citenamefont {Brown},\ and\ \citenamefont
  {Mizusaki}}]{Honma2004}%
  \BibitemOpen
  \bibfield  {author} {\bibinfo {author} {\bibfnamefont {M.}~\bibnamefont
  {Honma}}, \bibinfo {author} {\bibfnamefont {T.}~\bibnamefont {Otsuka}},
  \bibinfo {author} {\bibfnamefont {B.~A.}\ \bibnamefont {Brown}}, \ and\
  \bibinfo {author} {\bibfnamefont {T.}~\bibnamefont {Mizusaki}},\ }\href
  {\doibase 10.1103/PhysRevC.69.034335} {\bibfield  {journal} {\bibinfo
  {journal} {Phys. Rev. C}\ }\textbf {\bibinfo {volume} {69}},\ \bibinfo
  {pages} {034335} (\bibinfo {year} {2004})}\BibitemShut {NoStop}%
\bibitem [{\citenamefont {Honma}\ \emph {et~al.}(2005)\citenamefont {Honma},
  \citenamefont {Otsuka}, \citenamefont {Brown},\ and\ \citenamefont
  {Mizusaki}}]{Honma2005}%
  \BibitemOpen
  \bibfield  {author} {\bibinfo {author} {\bibfnamefont {M.}~\bibnamefont
  {Honma}}, \bibinfo {author} {\bibfnamefont {T.}~\bibnamefont {Otsuka}},
  \bibinfo {author} {\bibfnamefont {B.~A.}\ \bibnamefont {Brown}}, \ and\
  \bibinfo {author} {\bibfnamefont {T.}~\bibnamefont {Mizusaki}},\ }\href@noop
  {} {\bibfield  {journal} {\bibinfo  {journal} {Eur. Phys. J. A}\ }\textbf
  {\bibinfo {volume} {25 Suppl. 1}},\ \bibinfo {pages} {499} (\bibinfo {year}
  {2005})}\BibitemShut {NoStop}%
\bibitem [{\citenamefont {Brown}\ and\ \citenamefont
  {Richter}(2006)}]{BrownRichter2006}%
  \BibitemOpen
  \bibfield  {author} {\bibinfo {author} {\bibfnamefont {B.~A.}\ \bibnamefont
  {Brown}}\ and\ \bibinfo {author} {\bibfnamefont {W.~A.}\ \bibnamefont
  {Richter}},\ }\href@noop {} {\bibfield  {journal} {\bibinfo  {journal} {Phys.
  Rev. C}\ }\textbf {\bibinfo {volume} {74}},\ \bibinfo {pages} {034315}
  (\bibinfo {year} {2006})}\BibitemShut {NoStop}%
\bibitem [{\citenamefont {Schiffer}\ \emph {et~al.}(2008)\citenamefont
  {Schiffer}, \citenamefont {Freeman}, \citenamefont {Clark}, \citenamefont
  {Deibel}, \citenamefont {Fitzpatrick}, \citenamefont {Gros}, \citenamefont
  {Heinz}, \citenamefont {Hirata}, \citenamefont {Jiang}, \citenamefont {Kay}
  \emph {et~al.}}]{Schiffer2008}%
  \BibitemOpen
  \bibfield  {author} {\bibinfo {author} {\bibfnamefont {J.~P.}\ \bibnamefont
  {Schiffer}}, \bibinfo {author} {\bibfnamefont {S.~J.}\ \bibnamefont
  {Freeman}}, \bibinfo {author} {\bibfnamefont {J.~A.}\ \bibnamefont {Clark}},
  \bibinfo {author} {\bibfnamefont {C.}~\bibnamefont {Deibel}}, \bibinfo
  {author} {\bibfnamefont {C.~R.}\ \bibnamefont {Fitzpatrick}}, \bibinfo
  {author} {\bibfnamefont {S.}~\bibnamefont {Gros}}, \bibinfo {author}
  {\bibfnamefont {A.}~\bibnamefont {Heinz}}, \bibinfo {author} {\bibfnamefont
  {D.}~\bibnamefont {Hirata}}, \bibinfo {author} {\bibfnamefont {C.~L.}\
  \bibnamefont {Jiang}}, \bibinfo {author} {\bibfnamefont {B.~P.}\ \bibnamefont
  {Kay}},  \emph {et~al.},\ }\href {\doibase 10.1103/PhysRevLett.100.112501}
  {\bibfield  {journal} {\bibinfo  {journal} {Phys. Rev. Lett.}\ }\textbf
  {\bibinfo {volume} {100}},\ \bibinfo {pages} {112501} (\bibinfo {year}
  {2008})}\BibitemShut {NoStop}%
\bibitem [{\citenamefont {Kay}\ \emph {et~al.}(2009)\citenamefont {Kay},
  \citenamefont {Schiffer}, \citenamefont {Freeman}, \citenamefont {Adachi},
  \citenamefont {Clark}, \citenamefont {Deibel}, \citenamefont {Fujita},
  \citenamefont {Fujita}, \citenamefont {Grabmayr}, \citenamefont {Hatanaka}
  \emph {et~al.}}]{Kay2009}%
  \BibitemOpen
  \bibfield  {author} {\bibinfo {author} {\bibfnamefont {B.~P.}\ \bibnamefont
  {Kay}}, \bibinfo {author} {\bibfnamefont {J.~P.}\ \bibnamefont {Schiffer}},
  \bibinfo {author} {\bibfnamefont {S.~J.}\ \bibnamefont {Freeman}}, \bibinfo
  {author} {\bibfnamefont {T.}~\bibnamefont {Adachi}}, \bibinfo {author}
  {\bibfnamefont {J.~A.}\ \bibnamefont {Clark}}, \bibinfo {author}
  {\bibfnamefont {C.~M.}\ \bibnamefont {Deibel}}, \bibinfo {author}
  {\bibfnamefont {H.}~\bibnamefont {Fujita}}, \bibinfo {author} {\bibfnamefont
  {Y.}~\bibnamefont {Fujita}}, \bibinfo {author} {\bibfnamefont
  {P.}~\bibnamefont {Grabmayr}}, \bibinfo {author} {\bibfnamefont
  {K.}~\bibnamefont {Hatanaka}},  \emph {et~al.},\ }\href {\doibase
  10.1103/PhysRevC.79.021301} {\bibfield  {journal} {\bibinfo  {journal} {Phys.
  Rev. C}\ }\textbf {\bibinfo {volume} {79}},\ \bibinfo {pages} {021301}
  (\bibinfo {year} {2009})}\BibitemShut {NoStop}%
\bibitem [{\citenamefont {Kay}\ \emph {et~al.}(2013)\citenamefont {Kay},
  \citenamefont {Bloxham}, \citenamefont {McAllister}, \citenamefont {Clark},
  \citenamefont {Deibel}, \citenamefont {Freedman}, \citenamefont {Freeman},
  \citenamefont {Han}, \citenamefont {Howard}, \citenamefont {Mitchell} \emph
  {et~al.}}]{Kay2013}%
  \BibitemOpen
  \bibfield  {author} {\bibinfo {author} {\bibfnamefont {B.~P.}\ \bibnamefont
  {Kay}}, \bibinfo {author} {\bibfnamefont {T.}~\bibnamefont {Bloxham}},
  \bibinfo {author} {\bibfnamefont {S.~A.}\ \bibnamefont {McAllister}},
  \bibinfo {author} {\bibfnamefont {J.~A.}\ \bibnamefont {Clark}}, \bibinfo
  {author} {\bibfnamefont {C.~M.}\ \bibnamefont {Deibel}}, \bibinfo {author}
  {\bibfnamefont {S.~J.}\ \bibnamefont {Freedman}}, \bibinfo {author}
  {\bibfnamefont {S.~J.}\ \bibnamefont {Freeman}}, \bibinfo {author}
  {\bibfnamefont {K.}~\bibnamefont {Han}}, \bibinfo {author} {\bibfnamefont
  {A.~M.}\ \bibnamefont {Howard}}, \bibinfo {author} {\bibfnamefont {A.~J.}\
  \bibnamefont {Mitchell}},  \emph {et~al.},\ }\href {\doibase
  10.1103/PhysRevC.87.011302} {\bibfield  {journal} {\bibinfo  {journal} {Phys.
  Rev. C}\ }\textbf {\bibinfo {volume} {87}},\ \bibinfo {pages} {011302}
  (\bibinfo {year} {2013})}\BibitemShut {NoStop}%
\bibitem [{\citenamefont {Fox}\ \emph {et~al.}(2020)\citenamefont {Fox},
  \citenamefont {Johnson},\ and\ \citenamefont {Perez}}]{PhysRevC.101.054308}%
  \BibitemOpen
  \bibfield  {author} {\bibinfo {author} {\bibfnamefont {J.~M.~R.}\
  \bibnamefont {Fox}}, \bibinfo {author} {\bibfnamefont {C.~W.}\ \bibnamefont
  {Johnson}}, \ and\ \bibinfo {author} {\bibfnamefont {R.~N.}\ \bibnamefont
  {Perez}},\ }\href {\doibase 10.1103/PhysRevC.101.054308} {\bibfield
  {journal} {\bibinfo  {journal} {Phys. Rev. C}\ }\textbf {\bibinfo {volume}
  {101}},\ \bibinfo {pages} {054308} (\bibinfo {year} {2020})}\BibitemShut
  {NoStop}%
\bibitem [{\citenamefont {Miller}\ and\ \citenamefont
  {Spencer}(1976)}]{MillerSpencer1976}%
  \BibitemOpen
  \bibfield  {author} {\bibinfo {author} {\bibfnamefont {G.}~\bibnamefont
  {Miller}}\ and\ \bibinfo {author} {\bibfnamefont {J.~E.}\ \bibnamefont
  {Spencer}},\ }\href@noop {} {\bibfield  {journal} {\bibinfo  {journal} {Ann.
  Phys. (NY)}\ }\textbf {\bibinfo {volume} {100}},\ \bibinfo {pages} {562}
  (\bibinfo {year} {1976})}\BibitemShut {NoStop}%
\bibitem [{\citenamefont {\ifmmode~\check{S}\else \v{S}\fi{}imkovic}\ \emph
  {et~al.}(2009)\citenamefont {\ifmmode~\check{S}\else \v{S}\fi{}imkovic},
  \citenamefont {Faessler}, \citenamefont {M\"uther}, \citenamefont {Rodin},\
  and\ \citenamefont {Stauf}}]{PhysRevC.79.055501}%
  \BibitemOpen
  \bibfield  {author} {\bibinfo {author} {\bibfnamefont {F.}~\bibnamefont
  {\ifmmode~\check{S}\else \v{S}\fi{}imkovic}}, \bibinfo {author}
  {\bibfnamefont {A.}~\bibnamefont {Faessler}}, \bibinfo {author}
  {\bibfnamefont {H.}~\bibnamefont {M\"uther}}, \bibinfo {author}
  {\bibfnamefont {V.}~\bibnamefont {Rodin}}, \ and\ \bibinfo {author}
  {\bibfnamefont {M.}~\bibnamefont {Stauf}},\ }\href {\doibase
  10.1103/PhysRevC.79.055501} {\bibfield  {journal} {\bibinfo  {journal} {Phys.
  Rev. C}\ }\textbf {\bibinfo {volume} {79}},\ \bibinfo {pages} {055501}
  (\bibinfo {year} {2009})}\BibitemShut {NoStop}%
\bibitem [{\citenamefont {Horoi}(2016)}]{HoroiAPS2016}%
  \BibitemOpen
  \bibfield  {author} {\bibinfo {author} {\bibfnamefont {M.}~\bibnamefont
  {Horoi}},\ }\href@noop {} {\bibfield  {journal} {\bibinfo  {journal} {Bull.
  Amer. Phys. Soc.}\ }\textbf {\bibinfo {volume} {61}},\ \bibinfo {pages} {123}
  (\bibinfo {year} {2016})}\BibitemShut {NoStop}%
\bibitem [{\citenamefont {Cramer}(1957)}]{Gram-Charlier}%
  \BibitemOpen
  \bibfield  {author} {\bibinfo {author} {\bibfnamefont {H.}~\bibnamefont
  {Cramer}},\ }\href@noop {} {\emph {\bibinfo {title} {Mathematical Methods of
  Statistics}}}\ (\bibinfo  {publisher} {Princeton University Press},\ \bibinfo
  {address} {Princeton},\ \bibinfo {year} {1957})\BibitemShut {NoStop}%
\end{thebibliography}%


\end{document}